\pdfoutput=1
\documentclass[11pt]{article}

\usepackage[a4paper,margin=1in]{geometry}
\usepackage{authblk}
\usepackage{graphicx}
\usepackage{multirow}
\usepackage{amsmath,amssymb,amsfonts}
\usepackage{amsthm}
\usepackage{mathrsfs}
\usepackage{xcolor}
\usepackage{textcomp}
\usepackage{booktabs}
\usepackage{algorithm}
\usepackage{float}
\usepackage{algorithmicx}
\usepackage{algpseudocode}
\usepackage{listings}
\usepackage{tabularx}
\usepackage{placeins}
\usepackage{needspace}
\usepackage{braket}
\usepackage{colortbl}
\usepackage{url}
\usepackage{cite}
\usepackage{microtype}
\usepackage[hidelinks]{hyperref}

\graphicspath{{./Figures/}}
\theoremstyle{definition}

\theoremstyle{remark}

\title{\textbf{QARIMA: A Quantum Approach to Classical Time Series Analysis}}

\author[1]{Nishikanta Mohanty\thanks{Corresponding author: nishikanta.m.mohanty@alumni.uts.edu.au}}
\author[2,3]{Bikash K. Behera}
\author[4]{Badshah Mukherjee}
\author[5]{Pravat Dash}
\author[3]{Giuseppe Sergioli}
\author[3,6]{Roberto Giuntini}

\affil[1]{Centre for Quantum Software and Information, University of Technology Sydney, Sydney, NSW 2007, Australia\\\texttt{nishikanta.m.mohanty@alumni.uts.edu.au}}
\affil[2]{Bikash's Quantum (OPC) Pvt. Ltd., Mohanpur, West Bengal 741246, India\\\texttt{bikas.riki@gmail.com}}
\affil[3]{Universit\`a degli Studi di Cagliari, Cagliari 09123, Italy\\\texttt{giuseppe.sergioli@gmail.com; giuntini@unica.it}}
\affil[4]{Independent Researcher, Dubai 9262, United Arab Emirates\\\texttt{badshah.mukherjee@outlook.com}}
\affil[5]{Independent Researcher, Bangalore 560068, India\\\texttt{pravat.dash@outlook.com}}
\affil[6]{Institute for Advanced Study, Technische Universit\"at M\"unchen, Garching b. M\"unchen 85748, Germany}

\date{}

\begin{document}
\maketitle

\begin{abstract}
We present QARIMA, a quantum state-similarity-based reconstruction of the classical ARIMA modelling pipeline. Rather than using a quantum circuit as a standalone forecaster, QARIMA preserves ARIMA's interpretable forecasting structure while reformulating its core building blocks through analogous quantum-compatible modules. The framework integrates quantum differencing assessment, QACF/QPACF lag discovery, compact-swap-test state projection, swap-test/VQC-based AR and MA coefficient estimation, and weak-lag refinement within a single ARIMA forecasting workflow. QACF and QPACF serve the functional roles of ACF and PACF for MA and AR lag discovery, but construct lag relevance through quantum measurement geometry rather than direct classical correlation. Given screened candidate orders $(p,d,q)$, AR and MA coefficients are estimated through state-alignment losses incorporating cosine alignment, entropy regularization, phase correction, and norm control. We evaluate QARIMA across environmental, climatic, industrial, and weather time-series datasets using rolling-origin out-of-sample testing against automated classical ARIMA baselines, with performance assessed through MSE, MAPE, and Diebold--Mariano tests. The results show that quantum state-similarity modules can produce competitive and, in several cases, improved forecasting behaviour while preserving ARIMA's transparency, modularity, and interpretability. QARIMA therefore establishes a distinct pathway for quantum-enhanced statistical forecasting: its modules remain functionally analogous to classical ARIMA subroutines, but are not algebraic replicas; they serve the same modelling roles through state overlap, projection, and measurement-driven parameter estimation.
\end{abstract}

\noindent\textbf{Keywords:} Compact swap test; variational quantum circuits; quantum optimization; ARIMA; quantum autocorrelation function; quantum partial autocorrelation function; forecast error metrics; Diebold--Mariano test.

\vspace{0.75em}
\noindent\textit{Funding statement: This work received no external funding.}

\section{Introduction}\label{sec1}

Time-series forecasting is a central problem in statistical modelling and machine learning, with applications across economics, energy, healthcare, climate research, retail, and industrial decision systems. The AutoRegressive Integrated Moving Average (ARIMA) framework \cite{box2015time} remains one of the most widely used approaches for univariate forecasting because it combines interpretability, modularity, and a transparent decomposition of temporal structure into autoregressive (AR), differencing (I), and moving-average (MA) components. Its strength lies not only in the final forecasting equation, but also in the modelling architecture that leads to it: differencing is used to control non-stationarity, ACF and PACF diagnostics support lag discovery, AR and MA estimation determine model parameters, and residual analysis evaluates remaining dependence.

Despite this interpretability, classical ARIMA model construction depends on statistical diagnostics and estimation procedures based on correlation, partial correlation, residual behaviour, likelihood, least-squares fitting, and information criteria such as AIC and BIC \cite{akaike1974new}. These tools are effective in many settings, but their order-selection behaviour can become sensitive when the series is noisy, short, nonlinear, quasi-periodic, or structurally changing. In such cases, alternative ways of representing lag relevance, residual dependence, and coefficient alignment may lead to different candidate models while preserving the ARIMA forecasting structure.

Quantum machine learning provides a natural basis for such an alternative representation. In particular, amplitude/state encoding, compact swap-test measurement, quantum state overlap/similarity, phase-corrected projection, and variational quantum circuit (VQC) parameterization allow lag vectors, residual windows, and coefficient vectors to be compared through quantum-compatible state similarity rather than only through classical correlation or regression alignment \cite{schuld2021_QML_NonMarkovian,cerezo2021variational,lloyd2020quantum}. The compact swap test provides a measurement-derived similarity mechanism between encoded quantum-compatible states, making it suitable for constructing quantum analogues of dependence diagnostics and coefficient-estimation objectives.

We introduce QARIMA, a quantum state-similarity-based reconstruction of the classical ARIMA modelling pipeline. Rather than using a quantum circuit as a standalone forecasting model, QARIMA preserves the ARIMA forecasting equation and reconstructs its core statistical building blocks through analogous quantum-compatible modules. Differencing assessment, autocorrelation analysis, partial-autocorrelation analysis, autoregressive coefficient estimation, moving-average residual modelling, and weak-lag refinement are reformulated through compact-swap-test measurement, quantum state overlap/similarity, phase-corrected projection, and shallow VQC-style coefficient refinement.

Within this framework, QACF and QPACF act as quantum state-similarity diagnostics for lag discovery. They serve the functional roles of classical ACF and PACF within the ARIMA pipeline, but construct lag relevance through quantum measurement geometry rather than direct classical correlation. Candidate AR and MA orders are screened through these quantum diagnostics, after which AR and MA coefficients are estimated using swap-test-derived state-alignment losses with cosine alignment, entropy regularization, phase correction, and norm control. The VQC component is used as a shallow, hardware-compatible coefficient-refinement mechanism embedded inside the ARIMA structure, not as an independent forecasting engine.

This design preserves ARIMA’s interpretability while changing the mechanism through which order discovery and parameter estimation are performed. The resulting framework integrates quantum differencing assessment, QACF/QPACF lag discovery, compact-swap-test state projection, swap-test/VQC-based AR and MA coefficient estimation, and weak-lag refinement within a single interpretable forecasting workflow. In this sense, QARIMA does not algebraically replicate classical ARIMA subroutines; it provides functionally analogous quantum state-similarity modules that serve the same modelling roles through state overlap, projection, and measurement-driven parameter estimation. The empirical evaluation investigates whether this reconstruction yields useful forecasting behaviour across multiple real-world datasets while retaining the transparency and modularity of ARIMA.

\FloatBarrier
\subsection{Related Work and Novelty}
\label{sec:related_work}

Quantum time-series forecasting has recently emerged as an active research area spanning variational quantum circuits, quantum recurrent neural networks, quantum reservoir computing, and hybrid quantum-classical forecasting architectures. In many of these approaches, quantum circuits are employed as direct forecasting models, where historical observations are encoded into parameterized quantum circuits and measured observables are used to generate future predictions \cite{emmanoulopoulos2022qmlfinance,kaushik2022onestep,dimitrijevs2026hybridforecasting}. These studies demonstrate the feasibility of quantum circuits for temporal regression tasks, but they generally treat forecasting as a direct prediction problem rather than reconstructing the statistical modelling workflow of established forecasting frameworks.

A second class of approaches extends deep-learning-based sequence models using quantum components. Examples include quantum long short-term memory (LSTM) networks, quantum recurrent neural networks (RNN), and quantum segment recurrent neural networks \cite{chen2022qlstm,li2023qrnn,moon2025qsegrnn}. In these methods, quantum circuits augment recurrent architectures through quantum-enhanced gating, state representations, or sequence-learning mechanisms. While such approaches provide quantum-enhanced temporal modelling capabilities, they remain fundamentally neural forecasting architectures and do not explicitly preserve the statistical structure of classical forecasting methodologies.

Quantum reservoir computing has also been explored for time-series prediction by exploiting the memory and nonlinear dynamics of quantum systems \cite{kutvonen2020qrc,mujal2023qrc,kobayashi2024feedbackqrc}. Here, forecasting is achieved through the evolution of a quantum reservoir followed by a trained readout layer. These methods demonstrate the ability of quantum systems to capture temporal dependencies, but forecasting remains driven by reservoir dynamics rather than by interpretable statistical modelling stages. More recently, researchers have investigated long-horizon, multivariate, and application-specific quantum forecasting models \cite{chittoor2025qultsf}. Benchmarking studies have further emphasized the importance of evaluating quantum forecasting methods against strong classical baselines under controlled validation protocols \cite{jones2024benchmarking}. Collectively, these studies establish the viability of quantum-enhanced forecasting, but they largely focus on replacing, augmenting, or benchmarking forecasting engines rather than reformulating the internal statistical building blocks of an established forecasting pipeline.

\begin{figure}[htbp]
\centering
\includegraphics[width=0.85\linewidth]{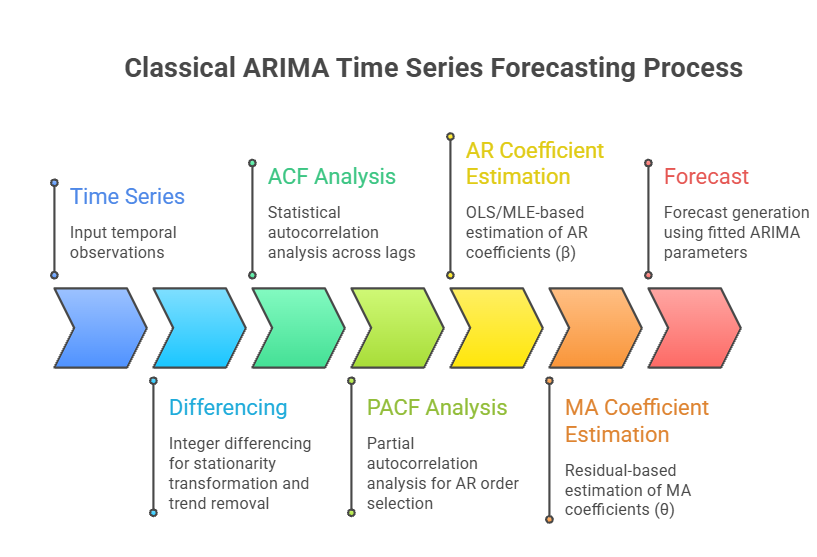}
\caption{Classical ARIMA forecasting workflow. The process consists of stationarity transformation, autocorrelation and partial-autocorrelation analysis for lag discovery, estimation of autoregressive and moving-average coefficients, and forecast generation using the fitted ARIMA model.}
\label{fig:classical_arima_workflow}
\end{figure}

\begin{figure}[htbp]
\centering
\includegraphics[width=0.85\linewidth]{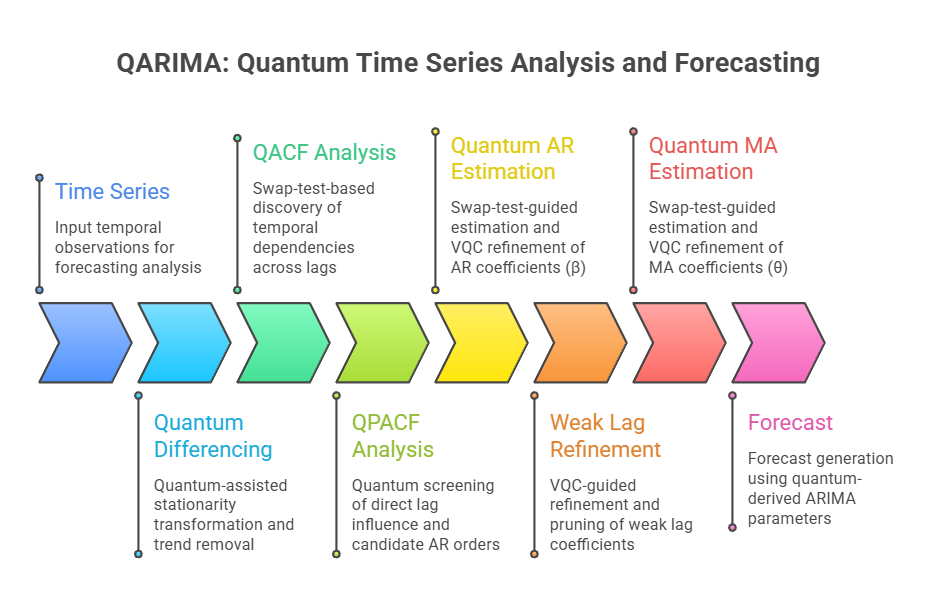}
\caption{QARIMA forecasting workflow. The major stages of classical ARIMA are reformulated using quantum-compatible primitives including quantum differencing, swap-test-driven QACF/QPACF lag discovery, quantum-assisted coefficient estimation, and VQC-guided weak-lag refinement while preserving the ARIMA forecasting structure.}
\label{fig:qarima_workflow}
\end{figure}

In contrast, QARIMA occupies a distinct position within this literature. Rather than introducing a quantum circuit as a standalone forecasting engine, the proposed framework preserves the ARIMA forecasting equation and reconstructs the major modelling stages of ARIMA through analogous quantum state-similarity modules. Classical differencing assessment, ACF, PACF, AR estimation, MA residual modelling, and weak-lag refinement are assigned quantum-compatible counterparts based on compact swap-test measurement, quantum state overlap/similarity, phase-corrected projection, and shallow VQC-style coefficient refinement. Figures~\ref{fig:classical_arima_workflow} and~\ref{fig:qarima_workflow} illustrate the conceptual distinction between the classical ARIMA pipeline and the proposed QARIMA formulation.

Classical ARIMA performs forecasting through a sequence of interpretable statistical operations: differencing is used to control non-stationarity, ACF and PACF analysis guide lag discovery, AR and MA estimation determine model parameters, and forecast generation proceeds through the fitted ARIMA equation. QARIMA preserves this architecture, but reconstructs the modelling path through quantum state-similarity mechanisms. Differencing assessment is formulated through quantum-compatible projection behaviour; temporal dependence is screened through Quantum Autocorrelation Function (QACF) and Quantum Partial Autocorrelation Function (QPACF) diagnostics; AR and MA coefficients are estimated through swap-test-derived state-alignment losses; and weak autoregressive relationships are refined through a VQC-guided coefficient-refinement stage. Importantly, these quantum modules are functionally analogous to classical ARIMA subroutines but are not algebraic replicas. They serve the same modelling roles through state overlap, projection, and measurement geometry while the final forecasting equation remains ARIMA-based.

\begin{table}[H]
\centering
\caption{Key distinctions between classical ARIMA and QARIMA.}
\begin{tabular}{p{3.5cm}p{4cm}p{4cm}}
\hline
Component & Classical ARIMA & QARIMA \\
\hline
Dependency Discovery & ACF/PACF & QACF/QPACF \\
AR Estimation & OLS/MLE & Swap-Test Estimation \\
Weak Lag Refinement & None & VQC-guided \\
MA Estimation & Residual-based & Swap-Test Estimation \\
Forecasting Engine & ARIMA Equation & ARIMA Equation \\
\hline
\end{tabular}
\end{table}
\FloatBarrier

An equally important distinction concerns the role of the variational quantum circuit. Existing quantum forecasting architectures commonly employ variational quantum circuits, quantum recurrent networks, or quantum reservoirs as the forecasting model itself. In such approaches, the quantum model directly maps historical observations to future predictions. QARIMA adopts a different design. The VQC is not used as an independent forecasting engine; rather, it serves as a shallow, hardware-compatible coefficient-refinement parameterization embedded inside the ARIMA pipeline. Candidate lag structures are first identified through quantum state-similarity diagnostics, after which AR and MA coefficients are refined through swap-test-derived state-alignment losses. Forecast generation continues to be performed through the ARIMA model structure using quantum-derived parameters.

Consequently, QARIMA is neither a direct quantum forecaster nor a quantum-enhanced recurrent architecture. It represents a quantum state-similarity reconstruction of the ARIMA methodology itself. To the best of our knowledge, this is the first end-to-end ARIMA-oriented quantum forecasting framework that systematically reformulates differencing assessment, dependency discovery, coefficient estimation, residual modelling, and lag refinement using compact swap-test measurement and VQC-style coefficient refinement while preserving the statistical interpretability and forecasting structure of classical ARIMA.

\begin{table}[H]
\centering
\caption{Positioning of QARIMA relative to related quantum forecasting approaches.}
\label{tab:related_work_summary}
\scriptsize
\setlength{\tabcolsep}{2pt}
\begin{tabular}{p{3.2cm} p{4.2cm} p{5.2cm}}
\hline
\textbf{Approach} & \textbf{Main idea} & \textbf{Difference from QARIMA} \\
\hline
VQC/PQC forecasting 
& Quantum circuits are used as direct regression or sliding-window predictors \cite{emmanoulopoulos2022qmlfinance,kaushik2022onestep,dimitrijevs2026hybridforecasting}.
& QARIMA preserves the ARIMA forecasting equation while reformulating its internal modelling stages through quantum state-similarity modules. \\
\hline
Quantum recurrent models 
& Quantum circuits augment RNN/LSTM-style sequence learning \cite{chen2022qlstm,li2023qrnn,moon2025qsegrnn}.
& QARIMA remains statistically interpretable and does not rely on hidden-state recurrent learning. \\
\hline
Quantum reservoir computing 
& Quantum dynamics provide temporal memory and nonlinear readout features \cite{kutvonen2020qrc,mujal2023qrc,kobayashi2024feedbackqrc}.
& QARIMA uses explicit QACF/QPACF and AR/MA estimation instead of reservoir dynamics. \\
\hline
Long-horizon and multivariate quantum forecasting 
& Hybrid quantum models target longer-range or multivariate forecasting applications \cite{chittoor2025qultsf}.
& QARIMA focuses on a quantum-compatible reconstruction of the ARIMA pipeline. \\
\hline
Quantum forecasting benchmarks 
& Quantum forecasting models are compared against classical baselines under controlled validation protocols \cite{jones2024benchmarking}.
& QARIMA follows this discipline through fixed VQC configurations, rolling-origin evaluation, MSE/MAPE reporting, and Diebold-Mariano testing. \\
\hline
\end{tabular}
\end{table}

\FloatBarrier
\section{Background}
\label{Background}

ARIMA (AutoRegressive Integrated Moving Average) models are among the most widely used approaches for univariate time-series forecasting. Originally developed within the Box-Jenkins framework \cite{Wilson_2016}, ARIMA combines differencing, autoregressive modelling, and moving-average error correction into a unified and interpretable forecasting framework. Given a time series ${y_t}_{t=1}^{N}$, an ARIMA$(p,d,q)$ model applies $d$ differencing operations to reduce non-stationary behaviour and then models the transformed series using autoregressive (AR) and moving-average (MA) components. The model is expressed as

\begin{eqnarray}
B_p(L)(1-L)^d y_t = \Theta_q(L)\varepsilon_t,
\end{eqnarray}

where $L$ denotes the lag operator, $B_p(L)=1-b_1L-\cdots-b_pL^p$ is the AR polynomial, $\Theta_q(L)=1+\theta_1L+\cdots+\theta_qL^q$ is the MA polynomial, and $\varepsilon_t$ represents white-noise innovations. After differencing, the ARIMA process can be written as

\begin{eqnarray}
y_t = \sum_{i=1}^{p} b_i y_{t-i}
+ \varepsilon_t
+ \sum_{j=1}^{q}\theta_j \varepsilon_{t-j},
\end{eqnarray}

indicating that the current observation is represented through a combination of past observations and past forecast errors. A central challenge in ARIMA modelling is the identification of the orders $(p,d,q)$ and the estimation of the corresponding AR and MA parameters. Classical approaches typically rely on differencing analysis, autocorrelation functions (ACF), partial autocorrelation functions (PACF), least-squares or likelihood-based estimation, residual diagnostics, and information-theoretic criteria such as AIC or BIC.

The proposed QARIMA framework retains this statistical architecture but reformulates the modelling path through quantum state-similarity modules. In QARIMA, the classical roles of differencing assessment, ACF/PACF-based lag discovery, AR coefficient estimation, MA residual modelling, and lag refinement are preserved, but the mechanisms used to perform these roles are reconstructed using compact swap-test measurement, quantum state overlap/similarity, phase-corrected projection, and shallow VQC-style coefficient refinement. Thus, the final forecasting equation remains ARIMA-based, while order discovery and parameter estimation are governed by quantum-compatible state-similarity and measurement-driven alignment procedures rather than purely classical correlation and regression machinery.

\FloatBarrier
\section{Methodology}\label{sec:methodology}

The proposed QARIMA methodology constructs an interpretable ARIMA-style forecasting model through a sequence of quantum state-similarity modules. The workflow preserves the classical ARIMA modelling architecture, but replaces its main diagnostic and estimation mechanisms with quantum-compatible counterparts based on compact swap-test measurement, state overlap/similarity, phase-corrected projection, and shallow VQC-based coefficient refinement.

The pipeline begins with the construction of lagged representations of the input series and the selection of the differencing order $d^\star$. Instead of applying a purely classical unit-root or stationarity diagnostic, QARIMA estimates $d^\star$ through a quantum state-similarity differencing assessment. Successive delayed and differenced representations are compared using compact swap-test projections, and the selected differencing order is determined from the projection loss together with stabilization or attenuation of the learned dependence parameter. This produces the transformed series $\Delta^{d^\star}y_t$ used in the subsequent AR and MA stages.

After differencing, temporal dependence is screened through QPACF and QACF. QPACF acts as the quantum state-similarity counterpart of PACF and is used to identify candidate autoregressive orders $p$. QACF acts as the quantum state-similarity counterpart of ACF and is applied
to the AR residual sequence to screen candidate moving-average orders $q$. Candidate residual lags whose QACF scores exceed the configurable empirical cutoff $\tau_{\mathrm{QACF}}$ are retained for MA-loss evaluation. If no lag passes this cutoff, the procedure reverts to the predefined bounded MA-order search range. These diagnostics preserve the functional roles of PACF and ACF within ARIMA, but construct lag relevance using compact swap-test-derived state similarity rather than direct classical correlation.

Given the QPACF-screened candidate AR orders, QARIMA estimates autoregressive coefficients using a VQC-based state-alignment loss. For each candidate $p$, a delayed AR design matrix is formed from $\Delta^{d^\star}y_t$, coefficients are initialized using ordinary least squares, and a shallow VQC refines the coefficient vector by minimizing a loss containing prediction error, cosine-alignment penalty, and entropy regularization. The best AR model is selected by comparing the refined AR losses across the QPACF-screened candidate set.

An optional weak-lag refinement stage is then applied around the selected AR anchor model. The anchor coefficients remain fixed, while a small number of weak higher-order lag coefficients are introduced and refined through a sparse VQC-based objective. This stage allows QARIMA to test whether weak lag information improves the AR state-alignment objective without replacing the selected anchor AR structure.

After AR fitting, residuals are computed from the QPACF-guided and VQC-refined AR models. These residuals serve both as diagnostic outputs for AR adequacy and as the input sequence for MA modelling. QACF is then applied to the AR residual sequence to compute aggregate
state-similarity scores between aligned residual vectors. Lags exceeding
$\tau_{\mathrm{QACF}}$ are retained as candidate MA orders, with fallback to
the bounded MA-order search range when the screened set is empty. For each candidate $q$, a delayed residual matrix is constructed, MA coefficients are initialized, and a quantum-regularized MA loss is minimized using compact swap-test similarity, phase-corrected projection, entropy regularization, and coefficient stabilization.

The final QARIMA/ARMA summary is produced by applying the MA estimation procedure to each retained AR model. Since each AR candidate order $p$ has its own residual sequence, the selected MA order is conditional on that AR residual structure and is denoted $q_p^\star$. The final model summary therefore records candidate configurations of the form $(p,d^\star,q_p^\star)$ together with the corresponding AR coefficients, MA coefficients, residual statistics, and minimized loss. Forecast generation remains ARIMA-based, using the quantum-derived orders and coefficient estimates.

The detailed procedures are presented in Section~\ref{Process&Algo}. The compact swap-test and state-preparation routines are given in Algorithms~\ref{alg:compact-swap-projection}, \ref{alg:prep-swaptest}, and \ref{alg:compact-swap-test-dot}. Quantum differencing is summarized in Algorithm~\ref{alg:estimate-d}, with delay-matrix construction and differenced-series generation given in Algorithms~\ref{alg:build-delay} and \ref{alg:generate-diff}. QACF and QPACF lag-screening are described in Algorithms~\ref{alg:q-acf} and \ref{alg:quantum-pacf}. AR loss estimation and QPACF-guided AR refinement are given in Algorithms~\ref{alg:quantum-ar-loss} and \ref{alg:vqc-p-estimation}. Weak-lag extension and progressive weak-lag refinement are described in Algorithms~\ref{alg:weak-lag-extension-vqc} and \ref{alg:progressive-weak-vqc}. AR residual evaluation is given in Algorithm~\ref{alg:residual-estimation}. MA loss estimation, MA order selection, and VQC-refined MA coefficient training are given in Algorithms~\ref{alg:quantum-ma-loss}, \ref{alg:ma-q-estimation}, and \ref{alg:vqc-ma-train}. The final ARMA summary across retained AR models is given in Algorithm~\ref{alg:run-ma-all}. Together, these components define an interpretable QARIMA pipeline in which quantum effects enter order discovery, residual modelling, weak-lag refinement, and AR/MA coefficient estimation, while the final forecasting structure remains ARIMA-based.

\FloatBarrier
\section{Processes and Algorithms}\label{Process&Algo}
In this section, we will detail the algorithms that are used to calculate Quantum ARIMA components for  parameter estimation. Along with these, we will describe some supporting mechanisms that are used to aid the major algorithms.

\FloatBarrier
\subsection{Quantum-Inspired State Similarity via Compact Swap Test}

To quantify the alignment between an input vector $\mathbf{x}_t$ and a parameter vector $\boldsymbol{\theta}$, QARIMA employs a compact swap-test-based state-similarity primitive~\cite{buhrman2001quantum,schuld2019quantum}. The purpose of this primitive is not to recover a signed classical inner product directly. Instead, the compact swap test estimates a measurement-derived overlap magnitude between encoded, normalized quantum-compatible states. This overlap is then used as a fidelity-style state-alignment component inside the QACF/QPACF screening rules and the AR/MA empirical-risk losses.

Given two real-valued vectors $\mathbf{x},\boldsymbol{\theta}\in\mathbb{R}^{n}$, we first construct the normalized auxiliary states:
\begin{eqnarray}
\phi &=& \left[\frac{\left\|\mathbf{x}\right\|}{\sqrt{Z}},-\frac{\left\|\boldsymbol{\theta}\right\|}{\sqrt{Z}}
\right],
\qquad
Z=\left\|\mathbf{x}\right\|^2 + \left\|\boldsymbol{\theta}\right\|^2 ,
\end{eqnarray}
and
\begin{eqnarray}
\psi &=& \frac{1}{\sqrt{2}}\left[\frac{x_1}{\left\|\mathbf{x}\right\|},
\frac{\theta_1}{\left\|\boldsymbol{\theta}\right\|},\ldots,\frac{x_n}{\left\|\mathbf{x}\right\|},
\frac{\theta_n}{\left\|\boldsymbol{\theta}\right\|}
\right].
\end{eqnarray}

These vectors are padded, when necessary, to match quantum register lengths that are powers of two. The compact swap test is then applied to the encoded states. Let $p_0^{\mathrm{meas}}$ denote the measured probability of obtaining outcome $0$ on the control qubit. For normalized encoded states, the swap-test measurement satisfies:
\begin{eqnarray}
p_0^{\mathrm{meas}} &=& \frac{1+\left|\left\langle\phi\middle|\psi\right\rangle\right|^2
}{2}.
\end{eqnarray}
The corresponding overlap-magnitude estimate is:
\begin{eqnarray}
\widehat{s}_{\mathrm{swap}} &=& \sqrt{2p_0^{\mathrm{meas}}-1}.
\end{eqnarray}

The quantity $\widehat{s}_{\mathrm{swap}}$ is interpreted as a non-negative quantum state-overlap score, not as a signed classical correlation coefficient. This distinction is central to the QARIMA construction. In the lag-screening stages, QACF and QPACF use $\widehat{s}_{\mathrm{swap}}$ to measure the strength of lagged state alignment. For this purpose, the magnitude of state similarity is sufficient for identifying candidate lag relevance. In the AR and MA coefficient-estimation stages, signed forecast effects are supplied by the trainable coefficient vectors, the observed target or residual errors, and the phase-corrected dot-product terms in the loss. Thus, the compact swap test contributes a measurement-driven alignment signal, while signed reconstruction is handled by the coefficient parameterization and prediction-error component of the empirical-risk objective.

We distinguish three related quantities throughout the paper. The quantity $p_0^{\mathrm{meas}}$ is the raw control-qubit measurement probability obtained from the compact swap test. The quantity $\widehat{s}_{\mathrm{swap}}$ is the derived overlap-magnitude score computed from $p_0^{\mathrm{meas}}$. The entropy terms used later in the AR and MA losses do not reuse $p_0^{\mathrm{meas}}$ directly; instead, they use a separate overlap-derived uncertainty probability, denoted $\pi_t$, which is defined locally in the AR and MA loss sections. Thus, $p_0^{\mathrm{meas}}$, $\widehat{s}_{\mathrm{swap}}$, and $\pi_t$ are related but not interchangeable.

Although the underlying time series is univariate, QACF and QPACF do not treat lag comparison as a comparison of isolated scalar observations alone. For a given lag $k$, QARIMA forms aligned lagged arrays from the current segment and its delayed counterpart, for example $\mathbf{y}^{(0)}_{k}=[y_{k+1},\ldots,y_N]$ and $\mathbf{y}^{(k)}=[y_1,\ldots,y_{N-k}]$, or corresponding lag-window representations when windowed diagnostics are used. These aligned arrays are encoded into normalized quantum-compatible states and compared through the compact swap-test projection. The resulting score therefore measures state alignment between lagged temporal representations, not merely overlap between two isolated scalar values.

Normalization during state preparation is used only for overlap estimation. It does not remove the scale information required for forecasting. In the AR and MA estimation stages, prediction magnitudes are reconstructed outside the compact swap-test measurement using the appropriate norm factors, such as $\|\mathbf{x}_t\|\,\|\mathbf{b}\|$ for autoregressive prediction and $\|\boldsymbol{\varepsilon}_t\|\,\|\boldsymbol{\theta}\|$ for moving-average residual correction. Directional and signed effects are supplied by the coefficient-vector orientation and by the cosine-projection reference used in the phase-corrected AR and MA prediction equations. Thus, QARIMA separates normalized state-alignment measurement from norm-based scale reconstruction.

The compact swap-test score is estimated using repeated circuit executions over a specified number of shots. Algorithm~\ref{alg:prep-swaptest} outlines the state-preparation routine, and Algorithm~\ref{alg:compact-swap-projection} details the compact swap-test computation used throughout the QARIMA pipeline.

\begin{algorithm}[H]
\caption{Compact Swap Test Projection}
\label{alg:compact-swap-projection}
\begin{algorithmic}[1]
\Require Input $\mathbf{x}, \boldsymbol{\theta} \in \mathbb{R}^n$, number of shots $S$
\State Compute $\phi, \psi \gets$ \Call{prep-swaptest}{$\mathbf{x}, \boldsymbol{\theta}$}
\State Normalize $\phi, \psi$ to unit $\ell_2$ norm
\State Pad $\psi$ to length $2^k$ for minimal $k$ such that $2^k \geq \text{len}(\psi)$
\State Construct quantum registers: one control qubit, one ancilla for $\phi$, and $k$ ancillas for $\psi$
\State Initialize $\phi$ into ancilla qubit, $\psi$ into multi-qubit register
\State Apply Hadamard to control qubit
\State Apply controlled-swap between $\phi$ and $\psi$
\State Apply Hadamard to control qubit
\State Measure control qubit $S$ times to estimate $p_0^{\mathrm{meas}}$
\State \Return state-overlap score $\widehat{s}_{\mathrm{swap}}\gets\sqrt{2p_0^{\mathrm{meas}}-1}$
\end{algorithmic}
\end{algorithm}
\FloatBarrier

\begin{figure}[htbp]
\centering
\includegraphics[width=0.72\linewidth]{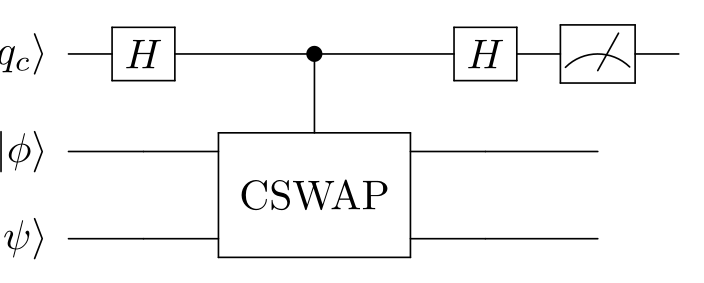}
\caption{Compact swap test estimating state similarity between encoded states \(\ket{\phi}\) and \(\ket{\psi}\). The control qubit is prepared in \(|0\rangle\), followed by Hadamard, controlled-swap, Hadamard, and measurement. The measured probability \(p_0^{\mathrm{meas}}\) is used to compute the overlap-magnitude score \(\widehat{s}_{\mathrm{swap}}=\sqrt{2p_0^{\mathrm{meas}}-1}\). Entropy regularization in the AR and MA losses uses a separate overlap-derived probability \(\pi_t\), not the raw measurement probability \(p_0^{\mathrm{meas}}\).}
\label{fig:qcircuit-swaptest}
\end{figure}

\FloatBarrier
\subsection{Quantum State-Similarity Differencing Assessment}
 
To determine the differencing order $d$, QARIMA preserves the functional objective of differencing in classical ARIMA: reducing persistent temporal dependence until an additional differencing operation provides limited further benefit. Classical implementations commonly identify this point through unit-root diagnostics such as the Augmented Dickey--Fuller test
\cite{Dickey_Fuller_1979}, autocorrelation decay, visual inspection, or information criteria. QARIMA instead constructs successive delayed and differenced representations and evaluates the remaining dependence through compact swap-test-based state projection. The proposed differencing stage is therefore not an algebraic reproduction of a classical unit-root test; rather, it provides a quantum state-similarity mechanism for assessing how temporal
dependence changes across successive differencing levels.

The process begins by constructing a delay matrix from the raw series (Algorithm~\ref{alg:build-delay}) and generating differenced variants up to a pre-defined maximum order $d_{\max}$ (Algorithm~\ref{alg:generate-diff}). For each differencing level $d$, QARIMA forms an aligned pair of transformed representations. At $d=0$, the level series is compared with its delayed representation. For $d \geq 1$, the current differenced state $\text{Del}_{d}$ is compared with the preceding transformed state $\text{Del}_{d-1}$ through the delayed matrix alignment. These two representations are encoded into quantum-compatible states and evaluated using a compact swap-test projection.

For each candidate differencing level $d$, a pair of projection parameters $(\alpha_d,\gamma_d)$ is estimated. The parameter $\alpha_d$ acts as an intercept-like projection component, while $\gamma_d$ captures the remaining state-aligned dependence between the current differentiated representation and the preceding delayed or differentiated state. Specifically, we define the swap-test-based projection as
\begin{eqnarray}
\hat{y}_t^{(d)} =
\left\langle
\psi(\mathbf{x}_t^{(d)})
\middle|
\psi(\boldsymbol{\theta}_d)
\right\rangle_{\mathrm{swap}},
\end{eqnarray}
where
\begin{eqnarray}
\mathbf{x}_t^{(d)} = [1, x_t^{(d)}] \ \text{and}\ \boldsymbol{\theta}_d &=& [\alpha_d, \gamma_d]
\end{eqnarray}
are encoded into quantum-compatible states using amplitude encoding. The notation $\langle\cdot|\cdot\rangle_{\mathrm{swap}}$ denotes the compact-swap-test-derived state overlap/similarity between the encoded representations. The objective at each differencing level is to minimize:
\begin{eqnarray}
\mathcal{L}_{d(\alpha_d, \gamma_d)} =
\frac{1}{N_d} \sum_{t}
\left(
\text{Del}_d(t) - \hat{y}_t^{(d)}
\right)^2.
\end{eqnarray}

For each candidate differencing level $d$, the parameters $(\alpha_d,\gamma_d)$ are estimated by minimizing $\mathcal{L}_d$ using a derivative-free optimizer such as COBYLA. The parameter $\gamma_d$ measures the remaining state-aligned dependence after applying differencing level $d$.The convergence of the numerical optimizer and the stabilization used for $d$-selection are distinct. Optimizer convergence determines whether $(\alpha_d,\gamma_d)$ has been adequately estimated for a fixed value of $d$. After obtaining the fitted values $\gamma_0,\gamma_1,\ldots,\gamma_d$, QARIMA examines their behaviour across successive differencing levels. Attenuation of $\gamma_d$, or stabilization of its recent fitted values within the threshold $\epsilon$, indicates that an additional differencing operation produces limited further reduction in state-aligned dependence. The selected order $d^\star$ is therefore determined from the projection loss together with the attenuation or stabilization of $\gamma_d$. This provides a measurement-driven stopping criterion for differencing depth and preserves the classical objective of avoiding both insufficient differencing and unnecessary over-differencing.The complete estimation routine is summarized in Algorithm~\ref{alg:estimate-d}.

This construction provides a distinct computational-modelling advantage over classical differencing diagnostics. Rather than reducing the decision to a scalar unit-root statistic or classical autocorrelation behaviour, QARIMA compares delayed and differentiated state representations through compact swap-test measurement. Consequently, $d$-selection is governed by the same state overlap, projection, and measurement geometry used later in QACF, QPACF, and AR/MA coefficient estimation. The differencing stage therefore becomes an integrated part of the quantum state-similarity pipeline rather than a separate classical preprocessing step.

\begin{algorithm}[H]
\caption{Quantum State-Similarity Estimation of Differencing Order $d$}
\label{alg:estimate-d}
\begin{algorithmic}[1]
\Require Time series $y$, maximum differencing depth $d_{\max}$, lag order $p$, convergence threshold $\epsilon$, patience $T$, maximum optimizer iterations $I_{\max}$, primary loss metric $\mathcal{L}$, random seed $s$

\State Construct delay matrix from $y$ using $p$ lags (Algorithm~\ref{alg:build-delay})
\State Generate differenced series up to $d_{\max}$ (Algorithm~\ref{alg:generate-diff})
\State Initialize $d \gets 0$ and $M_{\mathrm{log}} \gets \emptyset$

\State \textbf{Evaluate $d=0$:}
\State Set $x_t \gets \mathrm{lag}_1(t)$ and $y_t \gets y(t)$
\State Optimize $(\alpha_0,\gamma_0)$ using the projection loss $\mathcal{L}_0$
\State Compute $\hat{y}_t$ using the compact swap-test projection
\State Store $\alpha_0$, $\gamma_0$, and $\mathcal{L}_0$ in $M_{\mathrm{log}}$

\For{$d=1$ to $d_{\max}$}
\If{$\mathrm{Del}_{d}$ or $\mathrm{Del}_{d-1}$ is not defined}
\State \textbf{break}
\EndIf

\State Set $x_t \gets \mathrm{Del}_{d-1}(t)$ and $y_t \gets \mathrm{Del}_{d}(t)$
\State Initialize optimizer with seed $s+d$
\State Optimize $(\alpha_d,\gamma_d)$ using the projection loss $\mathcal{L}_d$
\State Compute $\hat{y}_t$ using the compact swap-test projection
\State Store $\alpha_d$, $\gamma_d$, and $\mathcal{L}_d$ in $M_{\mathrm{log}}$

\If{$d\geq T-1$}
    \State Set $\Gamma_d\gets \{\gamma_{d-T+1},\ldots,\gamma_d\}$
    \If{$\max(\Gamma_d)-\min(\Gamma_d)<\epsilon$}
        \State \textbf{break}
        \Comment{Limited change across recent differencing levels}
    \EndIf
\EndIf
\EndFor

\State Select $d^\star$ using the projection losses $\{\mathcal{L}_d\}$ together with attenuation or stabilization of the fitted sequence $\{\gamma_d\}$
\Return $d^*$, $\alpha_{d^*}$, $\gamma_{d^*}$
\end{algorithmic}
\end{algorithm}
\FloatBarrier

The estimation process is governed by several hyperparameters that control convergence behavior, optimization stability, and evaluation fidelity. These include the maximum differencing depth $d_{\max}$, maximum number of optimization iterations, convergence threshold for $\gamma$, a patience parameter that determines early stopping, and a tunable random seed for initialization. Additionally, the primary evaluation metric such as MSE or mean absolute error (MAE) can be selected to align with specific forecasting goals.

\begin{table}[H]
\centering
\caption{Hyperparameters for Estimating Differencing Order $d$}
\label{tab:d-hyperparams}
\small
\begin{tabularx}{\linewidth}{@{} l l X @{}}
\toprule
\textbf{Symbol} & \textbf{Name} & \textbf{Description} \\
\midrule
$d_{\max}$ & Max differencing depth & Maximum number of differencing levels tested. \\
$p$ & Lag order & Number of past lags used to construct the delay matrix. \\
$\epsilon$ & Convergence threshold & Threshold for detecting stabilization of the fitted $\gamma_d$ values across successive differencing levels. \\
$T$ & Patience & Number of recent differencing levels used to confirm stabilization of $\gamma_d$. \\
$I_{\max}$ & Max optimizer iterations & Maximum optimizer iterations used to estimate $(\alpha_d,\gamma_d)$ at each fixed differencing level. \\
$\mathcal{L}$ & Loss metric & Primary evaluation metric, such as MSE or MAE. \\
$s$ & Random seed & Seed used for reproducibility across differencing levels. \\
\bottomrule
\end{tabularx}
\end{table}
\FloatBarrier

\FloatBarrier
\subsection{Quantum-Inspired Autocorrelation Function (QACF)}

In the classical Box-Jenkins workflow, the autocorrelation function (ACF) quantifies dependence between $y_t$ and its lagged values $y_{t-k}$ without conditioning on intermediate lags, and is commonly used as a diagnostic for identifying moving-average (MA) structure and candidate MA orders $q$ in ARIMA models~\cite{box2015time,shumway2017time}. In QARIMA, this modelling role is preserved through a quantum state-similarity analogue, referred to as the Quantum-Inspired Autocorrelation Function (QACF). QACF is applied primarily to the residual sequence generated by the fitted AR component and is used to screen candidate MA orders, while QPACF is used separately for autoregressive-order screening.

The proposed QACF replaces the scalar sample-autocorrelation calculation with compact-swap-test-derived state similarity~\cite{buhrman2001quantum}. Let $\{\varepsilon_t\}_{t=1}^{N}$ denote the residual sequence obtained after AR modelling. For each candidate lag $k$, two aligned residual vectors are constructed:
\begin{eqnarray}
\boldsymbol{\varepsilon}^{(0)}_k &=& [\varepsilon_1,\varepsilon_2,\ldots,\varepsilon_{N-k}], \nonumber\\
\boldsymbol{\varepsilon}^{(k)}_k &=& [\varepsilon_{k+1},\varepsilon_{k+2},\ldots,\varepsilon_N].
\end{eqnarray}

These aligned vectors are passed through the common compact-swap-test
state-preparation routine, which produces normalized and padded
quantum-compatible representations. The compact swap test then evaluates
their state similarity through the measured control-qubit probability
$p_{0,k}^{\mathrm{meas}}$. The QACF score at lag $k$ is defined as
\begin{eqnarray}
\widehat{\rho}^{\,\mathrm{Q}}_k &=& \sqrt{2p_{0,k}^{\mathrm{meas}}-1}.
\end{eqnarray}

Equivalently, the score may be written as
\begin{eqnarray}
\widehat{\rho}^{\,\mathrm{Q}}_k &=& \mathrm{SwapSim}
\left(\boldsymbol{\varepsilon}^{(0)}_k,\boldsymbol{\varepsilon}^{(k)}_k\right),
\end{eqnarray}
where $\mathrm{SwapSim}(\cdot,\cdot)$ denotes the compact-swap-test
state-similarity operation.

The resulting QACF value measures the lag-dependent alignment between two complete residual-series representations. For each lag $k$, the compact swaptest compares the residual sequence with its aligned lag-shifted counterpart, so that changes in $\widehat{\rho}^{\,\mathrm{Q}}_k$ reflect changes in the persistence of residual structure across lags. QACF therefore exhibits autocorrelation-like behaviour at the series level, although the dependence is constructed through normalized quantum state overlap rather than through mean-centered covariance. It is not obtained by independently comparing
scalar residual pairs and averaging their scores; the complete aligned residual vectors are encoded and evaluated jointly.

The QACF scores are subsequently used to restrict the candidate MA-order
set through an operational similarity cutoff:
\begin{eqnarray}
\mathcal{Q}_{\mathrm{QACF}} &=& \left\{k\in\{1,\ldots,K\}:\widehat{\rho}^{\,\mathrm{Q}}_k
\geq \tau_{\mathrm{QACF}}\right\}.
\end{eqnarray}

In the present implementation, the default cutoff is
\begin{eqnarray}
\tau_{\mathrm{QACF}} &=& 0.15.
\end{eqnarray}

This value is a configurable empirical candidate-gating parameter rather than
a classical statistical significance boundary. If no lag satisfies the QACF
cutoff, the procedure reverts to the predefined bounded MA-order search
range. Each retained candidate order is subsequently evaluated through the
quantum-regularized MA loss, and the final MA order is selected from the
resulting loss comparison. Thus, QACF serves as a residual-dependence
screening mechanism within the QARIMA pipeline, while final MA-order
selection remains governed by the downstream coefficient-estimation and
loss-minimization stage.

\begin{algorithm}[H]
\caption{Quantum-Inspired ACF Estimation}
\label{alg:q-acf}
\begin{algorithmic}[1]
\Require AR residuals $\{\varepsilon_t\}_{t=1}^{N}$, maximum lag $K$, shots $S$, cutoff $\tau_{\mathrm{QACF}}$, fallback order set $\mathcal{Q}_0$
\Ensure QACF scores $\{\widehat{\rho}^{\,\mathrm{Q}}_k\}_{k=0}^{K}$, candidate MA orders $\mathcal{Q}_{\mathrm{cand}}$

\State Initialize $\mathrm{QACF}\gets[1.0]$

\For{$k=1$ to $K$}
    \State Set
    $\mathbf{x}_k\gets[\varepsilon_1,\ldots,\varepsilon_{N-k}]$
    and
    $\mathbf{y}_k\gets[\varepsilon_{k+1},\ldots,\varepsilon_N]$
    \State Compute
    $(\boldsymbol{\phi}_k,\boldsymbol{\psi}_k)
    \gets
    \mathrm{prep\_swaptest}(\mathbf{x}_k,\mathbf{y}_k)$
    \State Compute
    $\widehat{\rho}^{\,\mathrm{Q}}_k
    \gets
    \mathrm{CompactSwapProjection}
    (\boldsymbol{\phi}_k,\boldsymbol{\psi}_k,S)$
    \State Append $\widehat{\rho}^{\,\mathrm{Q}}_k$ to $\mathrm{QACF}$
\EndFor

\State Set
$\mathcal{Q}_{\mathrm{QACF}}
\gets
\{k:\widehat{\rho}^{\,\mathrm{Q}}_k\geq\tau_{\mathrm{QACF}}\}$

\If{$\mathcal{Q}_{\mathrm{QACF}}\neq\varnothing$}
    \State $\mathcal{Q}_{\mathrm{cand}}\gets\mathcal{Q}_{\mathrm{QACF}}$
\Else
    \State $\mathcal{Q}_{\mathrm{cand}}\gets\mathcal{Q}_0$
\EndIf

\State \Return $\mathrm{QACF}$, $\mathcal{Q}_{\mathrm{cand}}$
\end{algorithmic}
\end{algorithm}
\FloatBarrier

\begin{table}[H]
\centering
\caption{Hyperparameters for Quantum-Inspired ACF Estimation}
\label{tab:q-acf-hparams}
\resizebox{\columnwidth}{!}{%
\begin{tabular}{p{0.16\columnwidth} p{0.24\columnwidth} p{0.51\columnwidth}}
\toprule
\textbf{Symbol} & \textbf{Name} & \textbf{Description} \\
\midrule
$K$ & max\_lag & Maximum residual lag evaluated by QACF. \\
$S$ & shots & Number of compact swap-test measurements per lag. \\
$\tau_{\mathrm{QACF}}$ & acf\_threshold & Empirical cutoff for retaining candidate MA orders; default value $0.15$. \\
$\mathcal{Q}_0$ & fallback q range & Bounded MA-order set used when no QACF lag passes the cutoff. \\
$\widehat{\rho}^{\,\mathrm{Q}}_k$ & QACF score & State-similarity score between aligned residual vectors at lag $k$. \\
$\mathcal{Q}_{\mathrm{cand}}$ & candidate orders & MA-order set passed to the MA loss evaluation. \\
\bottomrule
\end{tabular}%
}
\end{table}
\FloatBarrier

In the classical Box--Jenkins workflow, ACF identifies lagged dependence by measuring the relationship between a residual series and its shifted counterpart. QACF preserves this series-level dependence analysis and the associated MA-order discovery role, but evaluates the relationship through compact-swap-test state overlap. Classical ACF measures signed,
mean-centered linear dependence, whereas QACF measures the magnitude of alignment between complete encoded residual-series representations at each lag. The resulting QACF profile therefore provides an autocorrelation-like description of residual persistence in quantum state-similarity space. Lags retained from this profile define candidate MA orders, which are then evaluated through residual modelling and quantum-regularized MA coefficient
estimation.

\FloatBarrier
\subsection{Quantum State-Similarity Partial Autocorrelation Function (QPACF)}

The partial autocorrelation function (PACF) is a fundamental diagnostic in classical time-series analysis for identifying autoregressive structure. In the Box--Jenkins workflow, PACF supports selection of the autoregressive order $p$ by assessing the relevance of individual lags after lower-order temporal effects have been considered~\cite{box2015time}. QARIMA introduces a Quantum State-Similarity Partial Autocorrelation Function, denoted QPACF,
as its quantum formulation of this AR-order discovery mechanism. QPACF produces a lag-indexed dependence profile for identifying candidate AR orders, while QACF is applied separately to the residual series for MA-order discovery.

QPACF reformulates the PACF diagnostic in quantum state-similarity space. Classical PACF quantifies lag relevance through residualized signed correlation. QPACF instead evaluates the alignment between encoded current and lagged temporal representations using compact-swap-test projection. The resulting quantity is not a generic similarity measure: it is indexed
explicitly by the temporal lag $k$, and its variation across lags describes how strongly the encoded temporal state retains alignment with its lagged counterpart. QPACF therefore captures autocorrelation-like AR dependence through quantum state overlap, projection, and measurement geometry.

For each candidate lag $k$, aligned temporal representations are formed from $y_t$ and $y_{t-k}$. The aligned values, or their corresponding lag-window representations, are encoded into quantum-compatible states and evaluated using the compact swap  test~\cite{buhrman2001quantum,schuld2019quantum}. This operation follows the compact-swap-projection and  state-preparation procedures described in Algorithms~\ref{alg:compact-swap-projection} and~\ref{alg:prep-swaptest}. For each aligned temporal pair, the compact
swap test produces the state-overlap score $\widehat{s}^{Q}_{t,k}$. The QPACF value at lag $k$ is then defined as

\begin{eqnarray}
\widehat{\rho}^{\,Q}_k
&=&
\frac{1}{N-k}
\sum_{t=k+1}^{N}
\widehat{s}^{Q}_{t,k}.
\end{eqnarray}

The quantity $\widehat{\rho}^{\,Q}_k$ measures lag-specific temporal-state alignment. When scalar lag values are encoded, it aggregates the state-similarity scores obtained from the aligned observations at lag $k$. When delayed-window or lag-matrix representations are encoded, the same construction measures alignment between richer temporal states. In both
cases, the sequence
\[
\left\{
\widehat{\rho}^{\,Q}_1,
\widehat{\rho}^{\,Q}_2,
\ldots,
\widehat{\rho}^{\,Q}_K
\right\}
\]
forms a quantum partial-lag dependence profile whose peaks identify candidate autoregressive structure.

The partial-lag assessment is completed within the QARIMA AR-order discovery process. QPACF first evaluates the direct state-alignment signal associated with each candidate lag. The retained lag structures are then evaluated jointly through the autoregressive design matrix and the VQC-refined AR loss. This second stage determines whether the QPACF-selected
structure retains predictive value when the candidate lags are represented together. Consequently, QPACF provides the quantum dependence signal for AR-order discovery, while the subsequent variational refinement determines the final order $p^\star$ and its associated coefficient vector.

After computing QPACF values for all lags up to a maximum lag $K$, a thresholding mechanism is applied to generate the candidate autoregressive lag set. Three screening strategies are implemented:
\begin{itemize}
\item a static threshold $\tau=1.96/\sqrt{N}$, used as a PACF-inspired reference band rather than as a formal classical PACF hypothesis test~\cite{shumway2017time};
\item a dynamic threshold based on a selected percentile of the QPACF
magnitudes; and
\item a standard-deviation threshold defined as $\mu+\sigma$, where $\mu$ and $\sigma$ are the mean and standard deviation of the QPACF magnitudes.
\end{itemize}

A fallback mechanism additionally retains lags whose QPACF values lie
between
\[
\tau_f=\beta\tau
\]
and the primary threshold $\tau$. This prevents moderately aligned temporal lags from being removed before variational evaluation. The retained lags are therefore QPACF-derived AR candidates rather than final coefficient decisions. Their joint predictive contribution is subsequently evaluated through VQC-based coefficient refinement and AR-loss comparison.

QPACF is thus a quantum state-similarity formulation of the PACF diagnostic, rather than an algebraic replication of the classical residualized correlation coefficient. Classical PACF and QPACF pursue the same modelling objective---identification of autoregressive lag structure through different dependence geometries. Classical PACF operates through
residualized linear correlation, whereas QPACF operates through lag-indexed quantum state alignment followed by joint variational evaluation of the selected AR structure.

\begin{algorithm}[H]
\caption{QPACF Partial-Lag Screening Using Compact Swap Test}
\label{alg:quantum-pacf}
\begin{algorithmic}[1]
\Require Time series $y_{1:N}$, maximum lag $K$, compact swap-test function $\texttt{SwapTest}$, shots $S$, threshold mode $T_m$, static threshold numerator $Z$, percentile level $\xi$, standard-deviation multiplier $c_{\sigma}$, fallback ratio $\beta$, fallback flag $f$
\Ensure QPACF scores $\{\widehat{\rho}^{\,Q}_k\}_{k=1}^{K}$, selected candidate AR lags $\mathcal{L}_{\mathrm{final}}$, threshold $\tau$

\State Initialize $\texttt{QPACF}\gets[\,]$

\For{$k=1$ to $K$}
    \If{$N\leq k+1$}
        \State Append $0$ to $\texttt{QPACF}$
        \State \textbf{continue}
    \EndIf

    \State Form aligned lag pairs $\{(y_t,y_{t-k})\}_{t=k+1}^{N}$
    \State Initialize accumulator $a_k\gets0$ and count $m_k\gets0$

    \For{$t=k+1$ to $N$}
        \State Compute swap-test state-similarity score
        $\widehat{s}^{Q}_{t,k}\gets\texttt{SwapTest}(y_t,y_{t-k},S)$
        \State Update $a_k\gets a_k+\widehat{s}^{Q}_{t,k}$
        \State Update $m_k\gets m_k+1$
    \EndFor

    \State Compute QPACF lag score
    $\widehat{\rho}^{\,Q}_k\gets a_k/m_k$
    \State Append $\widehat{\rho}^{\,Q}_k$ to $\texttt{QPACF}$
\EndFor

\State Compute magnitudes $r_k\gets|\widehat{\rho}^{\,Q}_k|$ for $k=1,\ldots,K$

\If{$T_m=\mathrm{static}$}
    \State Set $\tau\gets Z/\sqrt{N}$
\ElsIf{$T_m=\mathrm{percentile}$}
    \State Set $\tau\gets \mathrm{Percentile}(\{r_k\}_{k=1}^{K},\xi)$
\ElsIf{$T_m=\mathrm{std}$}
    \State Set $\tau\gets \mathrm{mean}(\{r_k\}_{k=1}^{K})+c_{\sigma}\,\mathrm{std}(\{r_k\}_{k=1}^{K})$
\Else
    \State Raise threshold-mode error
\EndIf

\State Define fallback threshold $\tau_f\gets\beta\tau$

\State Identify QPACF-screened AR lags
$\mathcal{L}_s\gets\{k:|\widehat{\rho}^{\,Q}_k|\geq\tau\}$

\State Identify fallback AR lags
$\mathcal{L}_f\gets\{k:\tau_f\leq|\widehat{\rho}^{\,Q}_k|<\tau\}$

\If{$f=\mathrm{True}$}
    \State Set $\mathcal{L}_{\mathrm{final}}\gets\mathcal{L}_s\cup\mathcal{L}_f$
\Else
    \State Set $\mathcal{L}_{\mathrm{final}}\gets\mathcal{L}_s$
\EndIf

\State \Return $\texttt{QPACF}$, $\mathcal{L}_{\mathrm{final}}$, $\tau$
\end{algorithmic}
\end{algorithm}
\FloatBarrier

\begin{table}[H]
\centering
\caption{Hyperparameters for QPACF Partial-Lag Screening}
\label{tab:pacf-hyperparams}
\begin{tabularx}{\linewidth}{@{} l l X @{}}
\toprule
\textbf{Symbol} & \textbf{Name} & \textbf{Description} \\
\midrule
$K$ & max lag & Maximum lag evaluated by the QPACF screening routine. \\
$S$ & shots & Number of compact swap-test measurements used for each aligned lag pair. \\
$T_m$ & threshold mode & Thresholding strategy used to select candidate AR lags: static, percentile, or std. \\
$Z$ & static numerator & Reference numerator for the static screening band, typically $1.96$ for a PACF-inspired $95\%$ reference band. \\
$\xi$ & percentile level & Percentile level used for dynamic thresholding of QPACF magnitudes. \\
$c_{\sigma}$ & std multiplier & Multiplier applied to the standard deviation of QPACF magnitudes in std mode. \\
$\beta$ & fallback ratio & Fraction of the primary threshold used to retain moderately informative fallback lags, where $0<\beta<1$. \\
$f$ & enable fallback & Boolean flag controlling whether fallback lags are included in the final candidate AR-lag set. \\
$\widehat{s}^{Q}_{t,k}$ & pair score & Compact-swap-test state-similarity score for aligned pair $(y_t,y_{t-k})$. \\
$\widehat{\rho}^{\,Q}_k$ & QPACF score & Average quantum partial-lag relevance score at lag $k$. \\
\bottomrule
\end{tabularx}
\end{table}
\FloatBarrier

\FloatBarrier
\subsection{AutoRegressive Order $p$}

In ARIMA modeling, the autoregressive (AR) order $p$ governs how many past values of a time series $y_t$ are used to predict its current value. Classical AR models rely on linear regression, using dot-product formulations over lagged vectors. In our quantum-inspired extension, we replace the classical regression-based fitting with a loss function that combines classical and quantum elements: phase-corrected cosine similarity from compact swap tests, entropy-based uncertainty, and misalignment penalties. The process is formulated as follows. For a lag vector $\mathbf{x}_t$ and coefficient vector $\mathbf{b}$ at time $t$, the classical prediction is:

\begin{eqnarray}
\hat{y}_t^{(\text{dot})} = \mathbf{x}_t^\top \mathbf{b}
\end{eqnarray}

We compute the classical cosine similarity as:
\begin{eqnarray}
\cos\theta_{\text{dot}} = \frac{\mathbf{x}_t^\top \hat{\mathbf{b}}}{\|\mathbf{x}_t\|}, \quad \text{where } \hat{\mathbf{b}} = \frac{\mathbf{b}}{\|\mathbf{b}\|}
\end{eqnarray}

The quantum-inspired variant estimates the cosine angle using a simulated swap test:
\begin{eqnarray}
\cos\theta_{\text{swap}} = \text{SWAP}(\mathbf{x}_t, \mathbf{b})
\end{eqnarray}

This estimate is phase-corrected based on the discrepancy between $\theta_{\text{dot}} = \arccos(\cos\theta_{\text{dot}})$ and $\theta_{\text{swap}} = \arccos(\cos\theta_{\text{swap}})$, resulting in the corrected prediction:

\begin{eqnarray}
\hat{y}_t^{(\text{quantum})} = \|\mathbf{x}_t\| \cdot \|\mathbf{b}\| \cdot \cos\left(\theta_{\text{swap}} + \lambda_{\text{phase}} (\theta_{\text{dot}} - \theta_{\text{swap}}) \right)
\end{eqnarray}

The total AR objective integrates prediction error, cosine-alignment regularization, and entropy regularization. The entropy term is not computed from the raw compact swap-test measurement probability $p_0^{\mathrm{meas}}$. Instead, the compact swap test first provides the state-alignment score $\widehat{s}_{\mathrm{swap},t}$, from which the swap-test angle $\theta_{\mathrm{swap},t}$ is obtained. The entropy regularizer then uses a separate overlap-derived uncertainty probability:
\begin{eqnarray}
\pi_t^{\mathrm{AR}} &=& 1-\cos^2\left(\theta_{\mathrm{swap},t}\right).
\end{eqnarray}

The resulting AR objective is:
\begin{eqnarray}
\label{ARloss}
\mathcal{L}_{\mathrm{AR}} &=& \sum_t\left(y_t-\widehat{y}_t^{(\mathrm{quantum})}\right)^2
\nonumber\\
&&+ \lambda_{\mathrm{cos}}\sum_t\left(\cos\theta_{\mathrm{dot},t} - \cos\theta_{\mathrm{swap},t}\right)^2 \nonumber\\
&&+ \lambda_{\mathrm{ent}} \sum_t H\left(\pi_t^{\mathrm{AR}}\right).
\end{eqnarray}

The binary entropy term is computed as:
\begin{eqnarray}
\label{Binary_Entropy}
H\left(\pi_t^{\mathrm{AR}}\right) &=& -\pi_t^{\mathrm{AR}} \log_2\left(\pi_t^{\mathrm{AR}}\right) \nonumber\\
&&- \left(1-\pi_t^{\mathrm{AR}}\right)\log_2\left(1-\pi_t^{\mathrm{AR}}\right).
\end{eqnarray}

The following algorithms compute the total loss $\mathcal{L}_{\text{AR}}$ for a given $p$ and coefficient vector $\mathbf{b}$:

\begin{algorithm}[H]
\caption{Quantum-Inspired AR Loss Estimation}
\label{alg:quantum-ar-loss}
\begin{algorithmic}[1]
\Require Lagged features $\{\mathbf{x}_t\}$, targets $\{y_t\}$, candidate AR coefficients $\mathbf{b}$, compact swap-test function, hyperparameters $\lambda_{\mathrm{cos}}, \lambda_{\mathrm{ent}}, \lambda_{\mathrm{phase}}$
\Ensure Total quantum-inspired AR loss $\mathcal{L}_{\mathrm{AR}}$

\State Initialize $\mathcal{L}_{\mathrm{AR}}\gets 0$
\State Normalize $\mathbf{b}$ to obtain unit vector $\mathbf{b}_{\mathrm{unit}}$

\For{each time step $t$}
    \State Pad $\mathbf{b}_{\mathrm{unit}}$ and $\mathbf{x}_t$ to power-of-two length
    \State Compute classical dot-product prediction:
    $\widehat{y}_t^{(\mathrm{dot})}\gets \mathbf{x}_t^\top \mathbf{b}$
    \State Compute classical cosine:
    $\cos\theta_{\mathrm{dot},t}\gets \frac{\mathbf{x}_t^\top \mathbf{b}_{\mathrm{unit}}}{\|\mathbf{x}_t\|}$
    \State Compute classical angle:
    $\theta_{\mathrm{dot},t}\gets \arccos\left(\cos\theta_{\mathrm{dot},t}\right)$

    \State Estimate raw swap-test measurement probability $p_{0,t}^{\mathrm{meas}}$ using the compact swap test
    \State Compute swap-test state-alignment score:
    $\widehat{s}_{\mathrm{swap},t}\gets \sqrt{2p_{0,t}^{\mathrm{meas}}-1}$
    \State Compute swap-test angle:
    $\theta_{\mathrm{swap},t}\gets \arccos\left(\widehat{s}_{\mathrm{swap},t}\right)$

    \State Compute angular correction:
    $\Delta\theta_t\gets \theta_{\mathrm{dot},t}-\theta_{\mathrm{swap},t}$
    \State Corrected angle:
    $\theta_{\mathrm{corr},t}\gets \theta_{\mathrm{swap},t}+\lambda_{\mathrm{phase}}\Delta\theta_t$
    \State Quantum prediction:
    $\widehat{y}_t^{(\mathrm{quantum})}\gets \|\mathbf{x}_t\|\,\|\mathbf{b}\|\cos\left(\theta_{\mathrm{corr},t}\right)$

    \State Compute AR entropy probability:
    $\pi_t^{\mathrm{AR}}\gets 1-\cos^2\left(\theta_{\mathrm{swap},t}\right)$
    \State Compute entropy:
    $H_t\gets -\pi_t^{\mathrm{AR}}\log_2\left(\pi_t^{\mathrm{AR}}\right)-\left(1-\pi_t^{\mathrm{AR}}\right)\log_2\left(1-\pi_t^{\mathrm{AR}}\right)$

    \State Accumulate loss:
    \Statex \(\begin{aligned}
    \mathcal{L}_{\mathrm{AR}} \gets{}& \mathcal{L}_{\mathrm{AR}}
    + \left(y_t-\widehat{y}_t^{(\mathrm{quantum})}\right)^2 \\
    &+ \lambda_{\mathrm{cos}}\left(\cos\theta_{\mathrm{dot},t}
    -\widehat{s}_{\mathrm{swap},t}\right)^2
    + \lambda_{\mathrm{ent}}H_t
    \end{aligned}\)
\EndFor

\State \Return $\mathcal{L}_{\mathrm{AR}}$
\end{algorithmic}
\end{algorithm}
\FloatBarrier

\begin{table}[H]
\centering
\caption{Hyperparameters for Quantum-Inspired AR Loss Estimation}
\label{tab:quantum-ar-loss-hyperparams}
\begin{tabularx}{\linewidth}{@{} l l X @{}}
\toprule
\textbf{Symbol} & \textbf{Name} & \textbf{Description} \\
\midrule
$S$                     & shots                    & Number of swap test measurement shots used for cosine similarity estimation. \\
$\lambda_{\text{cos}}$  & lambda\_cosine\_align    & Penalty weight applied to alignment difference between classical dot and swap-based cosine. \\
$\lambda_{\text{ent}}$  & lambda\_entropy          & Penalty weight for quantum entropy derived from swap test probability. \\
$\omega$                & phase\_weight            & Scaling factor to correct the phase discrepancy between classical and quantum cosine. \\
$\sigma$                & noise\_std               & Standard deviation of added Gaussian noise to simulate quantum fluctuations. \\
$P$                     & padded\_dim              & Size of padded vector, adjusted to nearest power-of-two for swap test register. \\
$f$                     & save\_csv                & Boolean flag to save log metrics and loss values for each AR trial. \\
\bottomrule
\end{tabularx}
\end{table}
\FloatBarrier

\FloatBarrier
\subsubsection{QPACF-Guided AR Order Evaluation via VQC Refinement}

In classical time-series analysis, the autoregressive order ($p$) is typically selected using information criteria such as AIC or BIC, or through grid search guided by PACF diagnostics~\cite{box2015time}. In QARIMA, the role of PACF-guided autoregressive lag discovery is preserved through the QPACF module. QPACF first identifies a candidate set of autoregressive orders ($\mathcal{P}$), and VQC-based refinement is then used to estimate and evaluate AR coefficients for each candidate ($p \in \mathcal{P}$). Thus, the VQC does not act as a standalone order-discovery mechanism; rather, it refines the AR coefficient vector within a QPACF-screened AR search space.

For each ($p \in \mathcal{P}$), the procedure is as follows:
\begin{enumerate}
\item Difference the series using the identified order ($d^\star$);
\item Build the AR design matrix ($\mathbf{X}^{(p)}$) from lagged values;
\item Initialize the coefficient vector ($\mathbf{b}^{(0)}$) using ordinary least squares (OLS);
\item Refine ($\mathbf{b}$) by minimizing the VQC-based AR state-alignment loss ($\mathcal{L}_{\mathrm{AR}}^{\mathrm{VQC}}$) in Eq.~\eqref{ARlossVQC}, using a depth-controlled VQC ansatz and a classical derivative-free optimizer such as COBYLA~\cite{cerezo2021variational,powell1994direct}.
\end{enumerate}

The VQC provides a shallow, hardware-compatible parameterization for refining AR coefficients while retaining the interpretability of QPACF-selected lag candidates. OLS initialization is used as a warm start to anchor the optimization and reduce susceptibility to barren plateaus and poor local minima~\cite{Grant_initialization_2019}. Let ($\mathbf{y}^{(p)} \in \mathbb{R}^{T-p}$) denote the target vector after differencing and lag alignment. For each ($p \in \mathcal{P}$), we form the AR design matrix ($\mathbf{X}^{(p)} \in \mathbb{R}^{(T-p) \times p}$) and compute the OLS initialization:
\begin{eqnarray}
\mathbf{b}^{(0)}
&=&
\arg\min_{\mathbf{b}}
\left|
\mathbf{y}^{(p)} - \mathbf{X}^{(p)}\mathbf{b}
\right|_2^2 .
\end{eqnarray}
\paragraph{VQC-to-AR coefficient map.}
Let $\boldsymbol{\beta}\in\mathbb{R}^{p}$ denote the trainable AR VQC parameter vector for a candidate autoregressive order $p$. The OLS initialization $\mathbf{b}^{(0)}$ is used to initialize the corresponding rotation angles. In the implementation used in this study, the VQC does not act as a separate measurement-to-coefficient black box. Instead, it provides a hardware-compatible parameterization of the AR coefficient vector through trainable $R_y$ rotation angles. The coefficient map is defined as
\begin{eqnarray}
\mathbf{b}(\boldsymbol{\beta}) &=& \mathrm{Proj}_{\mathcal{B}}
\left(g_{\mathrm{AR}}(\boldsymbol{\beta};\mathbf{b}^{(0)})\right),
\end{eqnarray}
where $g_{\mathrm{AR}}(\cdot)$ denotes the AR coefficient readout induced by the VQC parameterization, and $\mathrm{Proj}_{\mathcal{B}}(\cdot)$ denotes the coefficient scaling or clipping operation used for numerical stability. In the direct-angle parameterization used here, the trainable VQC angles correspond componentwise to the AR coefficient entries after projection:
\begin{eqnarray}
g_{\mathrm{AR}}(\boldsymbol{\beta};\mathbf{b}^{(0)}) 
&=& \boldsymbol{\beta}, \qquad \boldsymbol{\beta}^{(0)}=\mathbf{b}^{(0)}.
\end{eqnarray}
Thus, the optimized AR coefficient vector used in the ARIMA forecasting equation is
\begin{eqnarray}
\mathbf{b}^{\star} &=& \mathbf{b}(\boldsymbol{\beta}^{\star}),
\qquad \boldsymbol{\beta}^{\star} 
= \arg\min_{\boldsymbol{\beta}}
\mathcal{L}^{\mathrm{VQC}}_{\mathrm{AR}}
\left(\mathbf{b}(\boldsymbol{\beta})\right).
\end{eqnarray}
The OLS coefficient vector $\mathbf{b}^{(0)}$ is used to initialize the trainable VQC rotation parameters through
\[
\boldsymbol{\beta}^{(0)}=\mathbf{b}^{(0)}.
\]
During optimization, the trainable VQC parameter vector $\boldsymbol{\beta}$ is mapped to the AR coefficient vector through
\[
\mathbf{b}(\boldsymbol{\beta})
=
\mathrm{Proj}_{\mathcal{B}}(\boldsymbol{\beta}),
\]
where $\mathrm{Proj}_{\mathcal{B}}(\cdot)$ denotes the coefficient projection (or clipping) operation used for numerical stability. Under the direct-angle parameterization adopted in this work, each trainable rotation angle $\beta_i$ corresponds componentwise to the associated AR coefficient after projection.

\begin{figure}[htbp]
\centering
\includegraphics[width=0.98\linewidth]{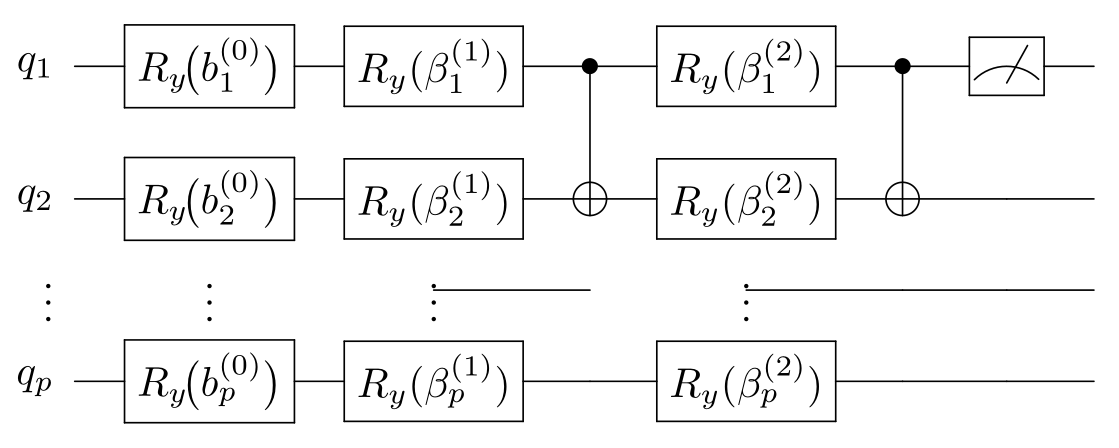}
\caption{VQC for AR order estimation with \(p\) qubits (one per lag). The first rotation layer uses OLS initialization \(R_y(b_i^{(0)})\); then \(r=2\) variational layers (shown) alternate trainable \(R_y(\beta_i^{(\ell)})\) rotations with a linear entangling pattern (CNOT ladder). The optimized VQC parameters are mapped to the AR coefficient vector through \(\mathbf{b}(\boldsymbol{\beta})=\mathrm{Proj}_{\mathcal{B}}(\boldsymbol{\beta})\). These coefficients are subsequently used in the norm-scaled phase-corrected projection to compute the AR prediction, while training minimizes \(\mathcal{L}_{\mathrm{AR}}\) consisting of prediction error, cosine-alignment, and entropy regularization.}
\label{fig:qcircuit-vqc-ar}
\end{figure}

The VQC architecture used for AR coefficient refinement is illustrated in Fig.~\ref{fig:qcircuit-vqc-ar}. It employs one qubit for each candidate AR lag, with OLS-initialized rotations followed by trainable rotation and entangling layers.

The VQC-based AR loss is defined as

\begin{eqnarray}
\label{ARlossVQC}
\mathcal{L}_{\mathrm{AR}}^{\mathrm{VQC}}(\boldsymbol{\beta})
&=&
\sum_t
\left(
y_t-
\hat{y}_t^{(\mathrm{quantum})}
\!\left(
\mathbf{b}(\boldsymbol{\beta})
\right)
\right)^2
\nonumber\\
&&+
\lambda_{\mathrm{cos}}
\sum_t
\left(
\cos\theta_{\mathrm{dot},t}
-
\cos\theta_{\mathrm{swap},t}
\right)^2
\nonumber\\
&&+
\lambda_{\mathrm{ent}}
\sum_t
H\!\left(
\pi_t^{\mathrm{AR}}
\right).
\end{eqnarray}

The overlap-derived probability used for entropy regularization is

\begin{eqnarray}
\pi_t^{\mathrm{AR}} &=& 1-\cos^2\!\left(\theta_{\mathrm{swap},t}\right).
\end{eqnarray}

The phase-corrected prediction is

\begin{equation}
\begin{aligned}
\hat{y}_t^{(\mathrm{quantum})}
\!\left(\mathbf{b}(\boldsymbol{\beta})\right)
={}&\|\mathbf{x}_t\|\,
\|\mathbf{b}(\boldsymbol{\beta})\| \\
&\times \cos\!\left[
\theta_{\mathrm{swap},t}
+\omega\left(
\theta_{\mathrm{dot},t}-\theta_{\mathrm{swap},t}
\right)\right].
\end{aligned}
\end{equation}

where $\mathbf{x}_t$ is the lag vector at time $t$, and $\mathbf{b}(\boldsymbol{\beta})$ is the VQC-refined AR coefficient vector. The normalized coefficient vector and the classical cosine component are given by

\begin{eqnarray}
\widehat{\mathbf{b}}(\boldsymbol{\beta})
&=& \frac{\mathbf{b}(\boldsymbol{\beta})}{\left\|\mathbf{b}(\boldsymbol{\beta})\right\|}
\nonumber\\
\cos\theta_{\mathrm{dot},t} &=& \frac{\mathbf{x}_t^\top\widehat{\mathbf{b}}(\boldsymbol{\beta})}{\left\|\mathbf{x}_t\right\|}.
\end{eqnarray}
The term ($\theta_{\mathrm{swap},t}$) denotes the angle implied by the compact-swap-test-derived state similarity between the encoded lag vector and the encoded coefficient state. Here, $\pi_t^{\mathrm{AR}}$ denotes the overlap-derived entropy-regularization probability, not the raw control-qubit measurement probability $p_{0,t}^{\mathrm{meas}}$. The entropy regularization term is defined as:
\begin{eqnarray}
H(p) &=& -p\log_2 p - (1-p)\log_2(1-p).
\end{eqnarray}

The optimal AR order is selected from the QPACF-screened candidate set after VQC refinement:
\begin{eqnarray}
p^\star &=& \arg\min_{p \in \mathcal{P}}
\mathcal{L}_{\mathrm{AR}}^{\mathrm{VQC}}
\left(\mathbf{b}^\star_{(p)}\right),
\end{eqnarray}
where ($\mathbf{b}^\star_{(p)}$) denotes the refined coefficient vector obtained for candidate order ($p$). This selection rule preserves the AR-order discovery role of QPACF while using VQC-based state-alignment to evaluate the coefficient quality associated with each candidate AR structure.

\begin{algorithm}[H]
\caption{VQC-Based AR Order Estimation}
\label{alg:vqc-p-estimation}
\begin{algorithmic}[1]
\Require Differenced series $\mathbf{y}$, best differencing order $d^\star$, candidate set $\mathcal{P}$, number of VQC layers $r$, maximum iterations $T_{\mathrm{max}}$
\Ensure Best AR order $p^\star$, optimized AR coefficients $\mathbf{b}^\star$

\State Initialize summary table $\mathcal{S}_{\mathrm{AR}}\gets[\,]$

\For{each $p \in \mathcal{P}$}
    \State Build delay matrix $\mathbf{X}^{(p)}$ from $\mathbf{y}$
    \State Construct aligned target vector $\mathbf{y}^{(p)}$
    \State Initialize AR coefficients by OLS:
    \[\mathbf{b}^{(0)} \gets \arg\min_{\mathbf{b}}
    \left\|\mathbf{y}^{(p)}-\mathbf{X}^{(p)}\mathbf{b}
    \right\|_2^2\]

    \State Initialize VQC trainable parameters using the OLS coefficients:
    \[\boldsymbol{\beta}^{(0)}\gets\mathbf{b}^{(0)}\]

    \State Define the VQC-to-coefficient map:
    \[\mathbf{b}(\boldsymbol{\beta}) \gets \mathrm{Proj}_{\mathcal{B}}(\boldsymbol{\beta})    \]
    where $\mathrm{Proj}_{\mathcal{B}}(\cdot)$ denotes the coefficient scaling or clipping operation used for numerical stability.

    \State Build a depth-$r$ VQC with trainable rotation parameters $\boldsymbol{\beta}$
    \State Optimize the VQC parameters:
    \[\boldsymbol{\beta}^{\star}_{(p)} \gets \arg\min_{\boldsymbol{\beta}}
    \mathcal{L}_{\mathrm{AR}}^{\mathrm{VQC}}
    \left(\mathbf{b}(\boldsymbol{\beta})\right)\]
    using COBYLA for at most $T_{\mathrm{max}}$ iterations

    \State Recover the refined AR coefficient vector:
    \[\mathbf{b}^{\star}_{(p)} \gets \mathbf{b}
    \left(\boldsymbol{\beta}^{\star}_{(p)}\right)\]

    \State Compute refined AR loss:
    \[\mathcal{L}^{\star}_{\mathrm{AR},p} \gets \mathcal{L}_{\mathrm{AR}}^{\mathrm{VQC}}
    \left(\mathbf{b}^{\star}_{(p)}\right)
    \]

    \State Append $\left(p,\mathbf{b}^{\star}_{(p)},\mathcal{L}^{\star}_{\mathrm{AR},p}\right)$ to $\mathcal{S}_{\mathrm{AR}}$
\EndFor

\State Select the best AR order:
\[p^\star \gets \arg\min_{p\in\mathcal{P}}\mathcal{L}^{\star}_{\mathrm{AR},p}\]

\State Set $\mathbf{b}^{\star}\gets\mathbf{b}^{\star}_{(p^\star)}$
\State \Return $p^\star$, $\mathbf{b}^{\star}$
\end{algorithmic}
\end{algorithm}
\FloatBarrier

\begin{table}[H]
\centering
\caption{Hyperparameters and Coefficient Map for VQC-Based AR Order Estimation}
\label{tab:vqc-p-hyperparams}
\scriptsize
\setlength{\tabcolsep}{3pt}
\renewcommand{\arraystretch}{1.12}
\begin{tabularx}{\linewidth}{@{} 
>{\raggedright\arraybackslash}p{0.18\linewidth}
>{\raggedright\arraybackslash}p{0.22\linewidth}
>{\raggedright\arraybackslash}X 
@{}}
\toprule
\textbf{Symbol} & \textbf{Name} & \textbf{Description} \\
\midrule
$\mathcal{P}$ & candidate orders & QPACF-screened candidate AR orders evaluated by the VQC refinement stage. \\
$d^\star$ & selected differencing order & Differencing order selected by quantum state-similarity differencing assessment. \\
$\mathbf{X}^{(p)}$ & AR design matrix & Lagged design matrix constructed for candidate order $p$. \\
$\mathbf{y}^{(p)}$ & aligned target vector & Target vector aligned with $\mathbf{X}^{(p)}$ after differencing and lag construction. \\
$\mathbf{b}^{(0)}$ & OLS initialization & Initial AR coefficient vector obtained by ordinary least squares for each candidate order $p$. \\
$\boldsymbol{\beta}$ & VQC parameters & Trainable VQC rotation-parameter vector used to refine AR coefficients. \\
$\boldsymbol{\beta}^{(0)}$ & VQC initialization & Initial VQC parameter vector, set as $\boldsymbol{\beta}^{(0)}=\mathbf{b}^{(0)}$. \\
$\mathbf{b}(\boldsymbol{\beta})$ & AR coefficient map & Coefficient vector obtained from VQC parameters as $\mathbf{b}(\boldsymbol{\beta})=\mathrm{Proj}_{\mathcal{B}}(\boldsymbol{\beta})$. \\
$\mathrm{Proj}_{\mathcal{B}}(\cdot)$ & coefficient projection & Scaling or clipping operation applied for numerical stability. \\
$r$ & VQC layers & Number of VQC entangling layers used in coefficient refinement. \\
$S$ & shots & Number of compact swap-test shots used inside AR loss evaluation. \\
$T_{\mathrm{max}}$ & max iterations & Maximum number of COBYLA iterations for VQC optimization. \\
$\lambda_{\mathrm{cos}}$ & cosine weight & Penalty weight for the cosine-alignment term in the AR loss. \\
$\lambda_{\mathrm{ent}}$ & entropy weight & Penalty weight for entropy regularization in the AR loss. \\
$\lambda_{\mathrm{phase}}$ & phase weight & Weight controlling interpolation between swap-test and cosine-projection angles. \\
$\mathcal{L}_{\mathrm{AR}}^{\mathrm{VQC}}$ & AR refinement loss & VQC-based AR state-alignment loss minimized for each $p\in\mathcal{P}$. \\
$\mathbf{b}^{\star}_{(p)}$ & refined AR coefficients & Optimized AR coefficient vector for candidate order $p$. \\
$p^\star$ & selected AR order & Candidate order with minimum refined VQC-based AR loss. \\
\bottomrule
\end{tabularx}
\normalsize
\end{table}
\FloatBarrier

\FloatBarrier
\subsubsection{Weak Lag Refinement for Extended AR Order (VQC-Based)}

After selecting the anchor AR order $p^\star$ from the QPACF-screened candidate set and refining its coefficients through the VQC-based AR estimation stage, some higher-order lags may still contain weak but non-negligible predictive signal. These lags are not intended to replace the anchor AR structure. Instead, they provide an optional extension around the selected anchor model. We therefore extend the anchor AR model by adding a small number of weak lag terms and refining only their coefficients while keeping the anchor coefficients fixed.

\paragraph{Selecting weak-lag initializations}
Let $\mathbf{b}^\star$ denote the refined AR coefficients for the selected anchor order $p^\star$, and let $\mathcal{S}$ denote the collection of VQC-refined coefficient vectors obtained across the tested AR orders in the summary table. From $\mathcal{S}$, we extract a pool of candidate coefficient magnitudes, remove magnitudes already represented in the anchor coefficient set ${|b_i^\star|}_{i=1}^{p^\star}$, and retain the $k$ smallest remaining nonzero magnitudes as weak-lag initializations:
\begin{eqnarray}
\mathbf{w} &=& {w_1,\ldots,w_k},
\qquad
p'=p^\star + k .
\end{eqnarray}
This produces an extended AR order $p'$, consisting of the fixed anchor component of length $p^\star$ and a weak-lag component of length $k$. The weak-lag terms are treated as sparse refinements rather than as a re-estimation of the full AR model.

\paragraph{Objective with adaptive penalties:}
Let $\mathbf{b}*{\mathrm{weak}}\in\mathbb{R}^{k}$ denote the weak-lag coefficient vector and define the full extended coefficient vector as
\begin{eqnarray}
\mathbf{b}_{\mathrm{full}}
&=&
[\mathbf{b}^\star;\mathbf{b}_{\mathrm{weak}}].
\end{eqnarray}
Using the quantum-inspired AR loss $\mathcal{L}_{\mathrm{AR}}(\cdot)$ from Eq.~\eqref{ARloss}, only $\mathbf{b}_{\mathrm{weak}}$ is optimized. The anchor vector $\mathbf{b}^\star$ remains fixed. The weak-lag objective is:

\begin{equation}
\label{eq:weak-objective}
\begin{aligned}
\mathcal{L}_{\mathrm{weak}}(\mathbf{b}_{\mathrm{weak}})
={}&\mathcal{L}_{\mathrm{AR}}\!\bigl(\mathbf{b}_{\mathrm{full}}\bigr) \\
&+\lambda_{\mathrm{dev}}
\bigl\|\mathbf{b}_{\mathrm{weak}}-\mathbf{w}\bigr\|_2^2 \\
&+\lambda_{\mathrm{mag}}
\bigl\|\mathbf{b}_{\mathrm{weak}}\bigr\|_1.
\end{aligned}
\end{equation}
The deviation penalty encourages the refined weak-lag coefficients to remain close to their weak-signal initializations $\mathbf{w}$, while the magnitude penalty promotes sparsity and discourages the weak block from overwhelming the anchor AR structure. In practice, the penalty weights are scaled relative to a baseline loss $L_{\mathrm{base}}$, taken from the best loss in the AR summary table:
\begin{eqnarray}
\lambda_{\mathrm{dev}}
&=&
10^{-3}L_{\mathrm{base}},
\qquad
\lambda_{\mathrm{mag}}
= 5\times 10^{-4}L_{\mathrm{base}} .
\end{eqnarray}

\paragraph{VQC parameterization:}
The weak-lag block is parameterized using a depth-$r$ VQC ansatz. The trainable rotation angles correspond to the entries of $\mathbf{b}_{\mathrm{weak}}$, while the anchor block $\mathbf{b}^\star$ remains fixed throughout optimization. A classical derivative-free optimizer such as COBYLA is used to refine $\mathbf{b}_{\mathrm{weak}}$ by minimizing $\mathcal{L}_{\mathrm{weak}}$.

\begin{figure}[htbp]
\centering
\includegraphics[width=0.98\linewidth]{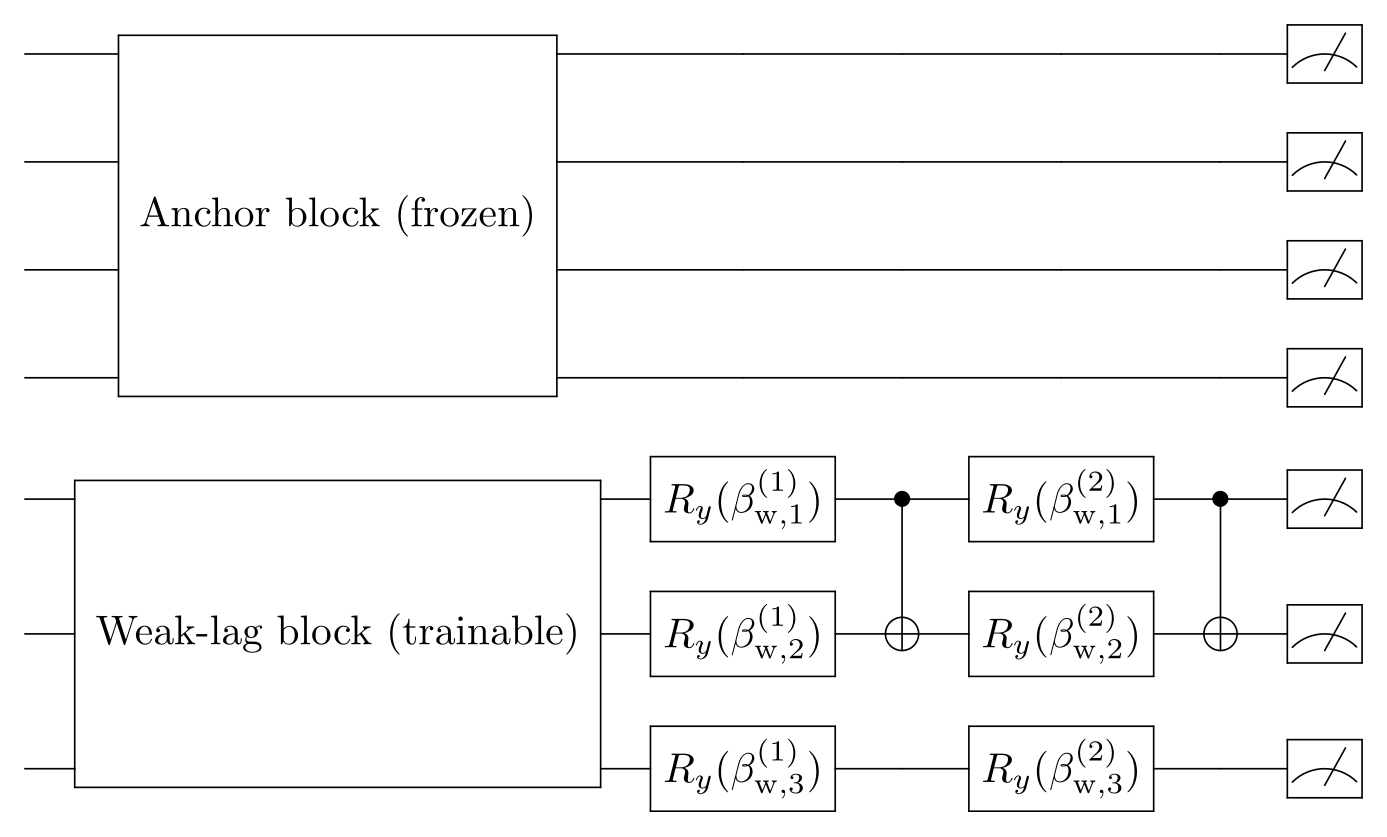}
\caption{VQC for weak-lag refinement with \(p^\star\) frozen anchors in the top block and \(k\) trainable weak-lag qubits in the bottom block. Anchors are initialized as \(R_y\big(b^{(p^\star)}_i\big)\) and kept fixed, shown as a grouped frozen block. Only weak-lag qubits carry trainable \(R_y(\beta^{(\ell)}_{\mathrm{w},j})\) rotations and are entangled through a CNOT ladder restricted to the weak block. Training minimizes \(\mathcal{L}_{\text{weak}}=\mathcal{L}_{\mathrm{AR}}+\lambda_{\text{dev}}\lVert\mathbf{b}_{\text{weak}}-\mathbf{w}\rVert_2^2+\lambda_{\text{mag}}\lVert\mathbf{b}_{\text{weak}}\rVert_1\).}
\label{fig:qcircuit-vqc-weak}
\end{figure}

The corresponding frozen-anchor and trainable weak-lag VQC architecture is illustrated in Fig.~\ref{fig:qcircuit-vqc-weak}. The upper circuit block represents the fixed coefficients of the QPACF-selected anchor model, whereas the lower block contains the trainable weak-lag rotations and entangling operations optimized during refinement.

This preserves the interpretability of the QPACF-selected anchor order $p^\star$, while allowing a controlled sparse extension to $p'$ when weak lag information improves the AR state-alignment objective.

\begin{algorithm}[H]
\caption{VQC-Based Weak Lag Extension and Refinement}
\label{alg:weak-lag-extension-vqc}
\begin{algorithmic}[1]
\Require Time series $y$, differencing order $d^\star$, best AR order $p^\star$, anchor coefficients $\mathbf{b}^\star$, loss summary $\mathsf{Summary}$, weak count $k$, VQC depth $r$, max iterations $T_{\max}$
\Ensure Extended order $p'$, refined coefficients $\mathbf{b}^{\star}_{(p')}$, refined loss $\mathcal{L}_{\mathrm{weak}}$

\State Extract candidate magnitudes from $\mathsf{Summary}$
\State Remove magnitudes already represented in ${|b_i^\star|}_{i=1}^{p^\star}$
\State Keep the $k$ smallest remaining nonzero magnitudes as $\mathbf{w}=\{w_1,\dots,w_k\}$
\State Set $p' \gets p^\star+k$
\State Difference $y$ using $d^\star$ and build the extended AR design matrix $\mathbf{X}^{(p')}$
\State Initialize $\mathbf{b}_{\mathrm{weak}}^{(0)} \gets \mathbf{w}$
\State Form $\mathbf{b}_{\mathrm{full}}^{(0)} \gets [\mathbf{b}^\star;\mathbf{b}_{\mathrm{weak}}^{(0)}]$
\State Build a depth-$r$ VQC with $k$ trainable angles for $\mathbf{b}_{\mathrm{weak}}$
\State Define $\mathcal{L}_{\mathrm{weak}}$ using Eq.~\eqref{eq:weak-objective}
\State Keep $\mathbf{b}^\star$ fixed and optimize only $\mathbf{b}_{\mathrm{weak}}$
\State Minimize $\mathcal{L}_{\mathrm{weak}}$ using COBYLA with budget $T_{\max}$
\State Obtain refined weak-lag coefficients $\mathbf{b}_{\mathrm{weak}}^\star$
\State Form $\mathbf{b}^{\star}_{(p')} \gets [\mathbf{b}^\star;\mathbf{b}_{\mathrm{weak}}^\star]$
\State Evaluate the final weak-lag loss $\mathcal{L}_{\mathrm{weak}}$
\State \textbf{return} $p'$, $\mathbf{b}^{\star}_{(p')}$, and $\mathcal{L}_{\mathrm{weak}}$
\end{algorithmic}
\end{algorithm}
\FloatBarrier

\begin{table}[H]
\centering
\caption{Hyperparameters for VQC-Based Weak Lag Refinement}
\label{tab:weak-lag-hyperparams}
\begin{tabularx}{\linewidth}{@{} l l X @{}}
\toprule
\textbf{Symbol} & \textbf{Name} & \textbf{Description} \\
\midrule
$p^\star$ & best\_p & AR order chosen by PACF\,+\,VQC refinement. \\
$\mathbf{b}^\star$ & anchor\_coeffs & Fixed anchor coefficients for AR($p^\star$). \\
$k$ & weak\_count & Number of weak lags added to form $p' = p^\star + k$. \\
$\mathbf{w}$ & weak\_inits & Weak-lag initialization magnitudes (from summary, excluding $|\mathbf{b}^\star|$). \\
$r$ & vqc\_depth & Number of entangling layers in the weak-lag VQC block. \\
$T_{\max}$ & max\_iter & Max COBYLA iterations for minimizing $\mathcal{L}_{\mathrm{weak}}$. \\
$\lambda_{\mathrm{dev}}$ & deviation\_penalty & Weight on $\|\mathbf{b}_{\mathrm{weak}}-\mathbf{w}\|_2^2$ (scaled by $L_{\mathrm{base}}$). \\
$\lambda_{\mathrm{mag}}$ & magnitude\_penalty & Weight on $\|\mathbf{b}_{\mathrm{weak}}\|_1$ (scaled by $L_{\mathrm{base}}$). \\
\bottomrule
\end{tabularx}
\end{table}
\FloatBarrier

\FloatBarrier
\subsubsection{Progressive Weak Lag Refinement via VQC}

After identifying a viable extended lag order $p'$, we further decompose its structure through a \textit{progressive refinement} strategy. Starting from the best anchor coefficients $\mathbf{b}^{(p^\star)}$ obtained from QPACF-guided VQC refinement, we incrementally add one weak lag at a time and re-optimize the extended coefficient vector. This stepwise inclusion facilitates interpretability of higher-order autoregressive models while minimizing the risk of overfitting by allowing early termination if no improvement is observed. Let $\mathbf{w} = \{w_1, w_2, \dots, w_k\}$ denote the initialization values for the $k$ weak lags, ranked by their absolute magnitude. For each $p = p^\star + 1$ to $p^\star + k$, we extend the model to include the first $j = p - p^\star$ weak lags, forming an initial coefficient vector
\begin{eqnarray}
\mathbf{b}^{(p)}_{\mathrm{init}} = \left[ \mathbf{b}^{(p^\star)}, w_1, \dots, w_j \right].
\end{eqnarray}
We employ a VQC-based refinement over the weak-lag block, keeping the anchor coefficients fixed. The weak-lag optimization problem is formulated as:
\begin{equation}
\label{eq:progressive-weak-vqc-loss}
\begin{aligned}
\mathcal{L}_{\mathrm{weak}}^{(p)}
={}&\mathcal{L}_{\mathrm{AR}}\left(\mathbf{b}^{(p)}\right) \\
&+\lambda_{\mathrm{dev}}
\left\|\mathbf{b}_{\mathrm{weak}}^{(p)}-\mathbf{w}_{1:j}\right\|_2^2 \\
&+\lambda_{\mathrm{mag}}
\left\|\mathbf{b}_{\mathrm{weak}}^{(p)}\right\|_1.
\end{aligned}
\end{equation}
where $\mathcal{L}_{\mathrm{AR}}$ is the quantum-inspired AR loss defined in Eq.~\eqref{ARloss}, $\mathbf{b}_{\mathrm{weak}}^{(p)}$ denotes the coefficients of the weak-lag block, $\lambda_{\mathrm{dev}}$ penalizes deviations from the weak-lag initializations, and $\lambda_{\mathrm{mag}}$ encourages sparsity in the weak-lag coefficients. The penalties are adaptively scaled by the baseline loss $\mathcal{L}_{\mathrm{base}} = \mathcal{L}_{\mathrm{AR}}(\mathbf{b}^{(p^\star)})$:
\begin{eqnarray}
\lambda_{\mathrm{dev}} = 10^{-3} \,\mathcal{L}_{\mathrm{base}}, \quad
\lambda_{\mathrm{mag}} = 5 \times 10^{-4} \,\mathcal{L}_{\mathrm{base}}.
\end{eqnarray}

For each refinement step, the weak-lag parameters are embedded into a $j$-qubit VQC with $R_y$ rotations and $r$ entangling layers, and optimized using COBYLA. The process continues progressively until the loss improvement

\begin{eqnarray}
\Delta^{(p)} = \mathcal{L}_{\mathrm{base}} - \mathcal{L}_{\mathrm{weak}}^{(p)} .
\end{eqnarray}

falls below a predefined threshold, at which point refinement halts.

\begin{algorithm}[H]
\caption{Progressive Weak Lag Refinement via VQC}
\label{alg:progressive-weak-vqc}
\begin{algorithmic}[1]
\Require Differenced series $\mathbf{y}$; best differencing order $d^\star$; best AR order $p^\star$; anchor coefficients $\mathbf{b}^{(p^\star)}$; weak-lag inits $\mathbf{w}$; VQC depth $r$; max iterations $T_{\max}$; threshold $\tau$
\Ensure Sequence $\{(p, \mathbf{b}^{(p)}, \mathcal{L}^{(p)})\}$
\State $\mathcal{L}_{\mathrm{base}} \gets \mathcal{L}_{\mathrm{AR}}(\mathbf{b}^{(p^\star)})$
\For{$j = 1$ \textbf{to} $|\mathbf{w}|$}
    \State $p \gets p^\star + j$
    \State Initialize $\mathbf{b}^{(p)}_{\mathrm{init}} \gets [\mathbf{b}^{(p^\star)}, w_1, \dots, w_j]$
    \State Build a $j$-qubit VQC (depth $r$) for the weak-lag block
    \State Minimize $\mathcal{L}^{(p)}$ (Eq.~\eqref{eq:progressive-weak-vqc-loss}) via COBYLA for $T_{\max}$ iterations
    \State Record $(p, \mathbf{b}^{(p)}, \mathcal{L}_{\mathrm{weak}}^{(p)})$
    \State Compute $\Delta^{(p)} \gets \mathcal{L}_{\mathrm{base}}-\mathcal{L}_{\mathrm{weak}}^{(p)}$    
    \If{$\Delta^{(p)} < \tau$}
        \State \textbf{break}
    \EndIf
\EndFor
\end{algorithmic}
\end{algorithm}
\FloatBarrier

\begin{table}[H]
\centering
\caption{Hyperparameters for Progressive VQC-Based Weak Lag Refinement}
\label{tab:progressive-weak-hyperparams}
\begin{tabularx}{\linewidth}{@{} l l X @{}}
\toprule
\textbf{Symbol} & \textbf{Name} & \textbf{Description} \\
\midrule
$\mathbf{w}$ & weak\_lags & Ordered list of weak-lag initialization values. \\
$r$ & vqc\_depth & Number of entangling layers in weak-lag VQC. \\
$T_{\max}$ & max\_iter & Maximum COBYLA iterations per refinement step. \\
$\lambda_{\mathrm{dev}}, \lambda_{\mathrm{mag}}$ & penalty\_weights & Adaptive regularization weights based on $\mathcal{L}_{\mathrm{base}}$. \\
\bottomrule
\end{tabularx}
\end{table}
\FloatBarrier

\FloatBarrier
\subsection{Residual Evaluation for Quantum-Inspired AR Models}

Once autoregressive candidate orders have been screened through QPACF and their corresponding coefficient vectors have been refined through the VQC-based AR state-alignment loss, the next step is to compute residuals. These residuals represent the remaining error between the observed differenced series and the AR prediction generated by the quantum-refined coefficient vector. They serve two purposes in the QARIMA pipeline: first, they evaluate the adequacy of the AR component, and second, they provide the residual sequence used by the QACF module for subsequent moving-average order selection.

For each VQC-refined AR candidate order $p$, a delay matrix $\mathbf{X}^{(p)}$ is constructed from $p$ prior values of the differenced series $\Delta^{d^\star}y_t$. Let $\mathbf{x}_t$ denote the row vector of lagged values used to predict the current target value $\Delta^{d^\star}y_t$, and let $\boldsymbol{b}$ denote the VQC-refined AR coefficient vector for that candidate order. The AR prediction is computed through the norm-scaled cosine projection:
\begin{eqnarray}
\hat{y}_t &=& \left\|\boldsymbol{b}\right\|\left\|\mathbf{x}_t\right\|\cos(\theta_t),\
\cos(\theta_t) = \frac{\mathbf{x}_t^\top\widehat{\boldsymbol{b}}}{\left\|\mathbf{x}_t\right\|},\
\widehat{\boldsymbol{b}} = \frac{\boldsymbol{b}}{\left\|\boldsymbol{b}\right\|}.
\end{eqnarray}

The residual at time $t$ is then computed as:
\begin{eqnarray}
\varepsilon_t &=& \Delta^{d^\star}y_t - \hat{y}_t .
\end{eqnarray}

Although the residuals are computed in the standard observed-minus-predicted form, the prediction itself is induced by the QPACF-guided and VQC-refined AR state-alignment mechanism. Therefore, the residual sequence captures the temporal dependence that remains after the quantum-refined AR component has been fitted. From the residual sequence ${\varepsilon_t}$, we compute the residual mean $\mu_\varepsilon$, residual standard deviation $\sigma_\varepsilon$, and preserve the full residual vector for each candidate AR order $p$. The retained residual sequence is then passed to QACF, where residual dependence is screened to propose candidate MA orders $q$. The procedure is summarized in Algorithm~\ref{alg:residual-estimation}.

\begin{algorithm}[H]
\caption{Residual Evaluation for Quantum-Inspired AR Models}
\label{alg:residual-estimation}
\begin{algorithmic}[1]
\Require Differenced time series $\Delta^{d^\star}y$, selected differencing order $d^\star$, AR candidate orders ${p_i}$, VQC-refined coefficients ${\boldsymbol{b}_i}$
\Ensure Residual statistics for each AR model

\State Initialize residual log $\mathcal{R} \gets [;]$

\For{each row $i$ in AR results}
\State Set $p \gets p_i$ and $\boldsymbol{b} \gets \boldsymbol{b}_i$
\State Construct delay matrix $\mathbf{X}^{(p)}$ using $p$ lags from $\Delta^{d^\star}y$
\State Set target vector $\mathbf{y}_{\mathrm{target}} \gets {\Delta^{d^\star}y_t}_{t=p+1}^{N}$
\State Normalize $\boldsymbol{b}$ to obtain $\widehat{\boldsymbol{b}}$

\For{each paired row $\mathbf{x}_t$ and target $y_t^{\mathrm{target}}$}
    \State Compute norm $\left\|\mathbf{x}_t\right\|$
    \State Compute cosine projection $\cos(\theta_t) \gets \frac{\mathbf{x}_t^\top\widehat{\boldsymbol{b}}}{\left\|\mathbf{x}_t\right\|}$
    \State Compute prediction $\hat{y}_t \gets \left\|\boldsymbol{b}\right\|\left\|\mathbf{x}_t\right\|\cos(\theta_t)$
    \State Compute residual $\varepsilon_t \gets y_t^{\mathrm{target}}-\hat{y}_t$
\EndFor

\State Compute residual mean $\mu_\varepsilon$ and residual standard deviation $\sigma_\varepsilon$
\State Append $(p,\mu_\varepsilon,\sigma_\varepsilon,\{\varepsilon_t\})$ to $\mathcal{R}$

\EndFor

\State \textbf{return} residual summary table $\mathcal{R}$
\end{algorithmic}
\end{algorithm}

\FloatBarrier
\subsection{Quantum-Inspired Moving Average Modeling}

In classical time-series modeling, the moving-average component models serial dependence in residual errors, capturing short-term structure not explained by autoregressive terms. In QARIMA, the MA stage plays the same modelling role, but the residual-dependence and coefficient-estimation steps are reconstructed through quantum state-similarity mechanisms. After the QPACF-guided AR stage has produced VQC-refined AR coefficients, the residual sequence ${\varepsilon_t}$ is computed and passed to the QACF module. QACF screens the residual sequence for remaining lagged dependence and proposes candidate MA orders $q$.

Given the AR residuals ${\varepsilon_t}$, the MA component is modelled as an MA$(q)$ process with a trainable coefficient vector $\boldsymbol{\theta}\in\mathbb{R}^{q}$. The objective is to estimate $\boldsymbol{\theta}$ for each candidate $q$ using a quantum-inspired state-alignment loss based on compact swap-test similarity, phase-corrected projection, and entropy-aware regularization. Thus, the MA stage is not treated as an independent forecasting engine; it is a residual-correction module that extends the QPACF-guided AR model into an ARMA-style representation.

The overall MA modeling framework consists of the following steps:

\begin{itemize}
\item {Residual dependence screening}: QACF is applied to the AR residual sequence ${\varepsilon_t}$ to identify candidate MA orders $q$.
\item {Loss computation}: For a given $\boldsymbol{\theta}$ and candidate $q$, compact swap-test similarity and phase-corrected projection are used to evaluate the residual-prediction error.
\item {MA coefficient estimation}: A constrained classical optimizer refines $\boldsymbol{\theta}$ by minimizing the quantum-regularized MA loss.
\item {ARMA extension}: The refined MA component is combined with the selected AR structure to produce the final ARMA model summary.
\end{itemize}

Each component is described in the following subsections.

\FloatBarrier
\subsubsection{Quantum MA Loss Estimation}

To estimate the quality of MA coefficients $\boldsymbol{\theta}$, we define a quantum-regularized loss function $\mathcal{L}_{\mathrm{MA}}$ that combines residual prediction with compact-swap-test-derived state similarity, phase-corrected projection, and entropy-aware regularization~\cite{schuld2021machine, havlivcek2019supervised}. In the QARIMA pipeline, the MA stage operates on the AR residual sequence ${\varepsilon_t}$ produced after QPACF-guided AR modelling. Thus, the objective is not to refit the original series directly, but to model the remaining residual dependence that was not explained by the AR component.

For a candidate MA order $q$, define the delayed residual vector at time $t$ as:
\begin{eqnarray}
\boldsymbol{\varepsilon}_t &=& [\varepsilon_{t-1},\varepsilon_{t-2},\ldots,\varepsilon_{t-q}] .
\end{eqnarray}
Let $\boldsymbol{\theta}\in\mathbb{R}^{q}$ denote the trainable MA coefficient vector. The predicted residual contribution is computed using a norm-scaled phase-corrected projection:
\begin{eqnarray}
\widehat{\varepsilon}_t &=& \left\|\boldsymbol{\theta}\right\|
\left\|\boldsymbol{\varepsilon}_t\right\|
\cos\left(\phi_t^{\mathrm{corrected}}\right).
\end{eqnarray}

Here, $\phi_t^{\mathrm{corrected}}$ is a phase-adjusted angle between the normalized coefficient vector and the delayed residual vector. It combines the classical dot-product angle with the compact-swap-test-derived angle:
\begin{eqnarray}
\cos\left(\phi_t^{\mathrm{corrected}}\right) &=& \cos\left(\theta_{\mathrm{swap},t} 
+
\omega \left(\theta_{\mathrm{dot},t}-\theta_{\mathrm{swap},t}\right)\right),
\end{eqnarray}
where
\begin{eqnarray}
\theta_{\mathrm{dot},t} &=& \arccos\left(\left\langle\boldsymbol{\theta}_{\mathrm{unit}},
\boldsymbol{\varepsilon}_{t,\mathrm{unit}}\right\rangle\right).
\end{eqnarray}
The term $\theta_{\mathrm{swap},t}$ denotes the angle inferred from the compact swap-test similarity between the encoded MA coefficient state and the encoded delayed residual state, while $\omega$ controls the contribution of the phase correction.

The total quantum MA loss is defined as:

\begin{eqnarray}
\label{MA_loss}
\mathcal{L}_{\mathrm{MA}}
&=& \sum_t \left(\varepsilon_t-\widehat{\varepsilon}_t\right)^2 \nonumber\\
&&+ \lambda_{\mathrm{cos}}\sum_t\left[\cos\left(\theta_{\mathrm{dot},t}\right)
- \cos\left(\phi_t^{\mathrm{corrected}}\right)\right]^2 \nonumber\\
&&+ \lambda_{\mathrm{ent}} \sum_t\mathcal{H}
\left(\pi_t^{\mathrm{MA}}\right) 
+
\lambda_{\mathrm{L2}}\left\|\boldsymbol{\theta}\right\|_2^2.
\end{eqnarray}

The MA entropy term is computed from a separate overlap-derived uncertainty
probability:
\begin{eqnarray}
\pi_t^{\mathrm{MA}} &=& 1-\cos^2\left(\theta_{\mathrm{swap},t}\right), \nonumber\\
\mathcal{H}_t &=& -\pi_t^{\mathrm{MA}}\log_2\left(\pi_t^{\mathrm{MA}}\right) \nonumber\\
&&- \left(1-\pi_t^{\mathrm{MA}}\right)\log_2\left(1-\pi_t^{\mathrm{MA}}\right).
\end{eqnarray}

\vspace{1em}

\begin{algorithm}[H]
\caption{Quantum-Inspired MA Loss Estimation}
\label{alg:quantum-ma-loss}
\begin{algorithmic}[1]
\Require MA coefficients $\boldsymbol{\theta}$, residual matrix $\mathbf{E}$,
target vector $\boldsymbol{\varepsilon}_{\mathrm{target}}$,
weights $\lambda_{\mathrm{cos}}$, $\lambda_{\mathrm{ent}}$,
$\lambda_{\mathrm{L2}}$, phase weight $\omega$, shots $S$
\Ensure Loss $\mathcal{L}_{\mathrm{MA}}$

\State Normalize $\boldsymbol{\theta}$ and initialize
$\mathcal{L}_{\mathrm{MA}}\gets 0$

\For{each residual window $\boldsymbol{\varepsilon}_t$ in $\mathbf{E}$}
    \State Normalize and pad $\boldsymbol{\varepsilon}_t$ and
    $\boldsymbol{\theta}$ for swap-test encoding
    \State Compute $\theta_{\mathrm{dot},t}$ and estimate
    $\theta_{\mathrm{swap},t}$ using $S$ shots
    \State Set 
    $\phi_t^{\mathrm{corrected}} \gets \theta_{\mathrm{swap},t}
    + \omega(\theta_{\mathrm{dot},t}-\theta_{\mathrm{swap},t})$
    \State Compute $\widehat{\varepsilon}_t \gets \|\boldsymbol{\theta}\|    \|\boldsymbol{\varepsilon}_t\|\cos(\phi_t^{\mathrm{corrected}})$
    \State Set $\pi_t^{\mathrm{MA}} \gets 1-\cos^2(\theta_{\mathrm{swap},t})$
    \State Accumulate
    \Statex \(\begin{aligned}
    \mathcal{L}_{\mathrm{MA}} \gets{}& \mathcal{L}_{\mathrm{MA}}
    +(\varepsilon_{\mathrm{target},t}-\widehat{\varepsilon}_t)^2 \\
    &+\lambda_{\mathrm{cos}}\left[
    \cos(\theta_{\mathrm{dot},t})
    -\cos(\phi_t^{\mathrm{corrected}})\right]^2 \\
    &+\lambda_{\mathrm{ent}}\mathcal{H}(\pi_t^{\mathrm{MA}})
    \end{aligned}\)
\EndFor

\State Add
$\lambda_{\mathrm{L2}}\|\boldsymbol{\theta}\|_2^2$
to $\mathcal{L}_{\mathrm{MA}}$
\State \Return $\mathcal{L}_{\mathrm{MA}}$
\end{algorithmic}
\end{algorithm}
\FloatBarrier

\vspace{1em}

\begin{table}[H]
\centering
\caption{Hyperparameters for Quantum MA Loss}
\label{tab:ma-loss-hyperparams}
\begin{tabularx}{\linewidth}{@{} l l X @{}}
\toprule
\textbf{Symbol} & \textbf{Name} & \textbf{Description} \\
\midrule
$\lambda_{\mathrm{cos}}$ & cosine alignment weight & Penalty for misalignment between classical dot-product projection and phase-corrected swap-test projection. \\
$\lambda_{\mathrm{ent}}$ & entropy weight & Weight for binary entropy regularization based on the overlap-derived MA uncertainty probability $\pi_t^{\mathrm{MA}}$. \\
$\omega$ & phase weight & Phase-correction factor for angular adjustment. \\
$S$ & shots & Number of measurements used in the swap-test simulation. \\
$\lambda_{\mathrm{L2}}$ & L2 weight & $\ell_2$ regularization weight on the magnitude of $\boldsymbol{\theta}$. \\
\bottomrule
\end{tabularx}
\end{table}
\FloatBarrier

\FloatBarrier
\subsubsection{MA Order Estimation via Quantum Residual Modeling}

To determine the MA order $q$, the quantum-inspired MA loss  $\mathcal{L}_{\mathrm{MA}}$ is evaluated across a candidate set of residual lag orders. QACF first produces the screened set
\begin{eqnarray}
\mathcal{Q}_{\mathrm{QACF}}
&=& \left\{k\in\{1,\ldots,K\}:
\widehat{\rho}^{\,\mathrm{Q}}_k \geq \tau_{\mathrm{QACF}}\right\}.
\end{eqnarray}

When this set is nonempty, it defines the candidate MA orders. Otherwise,
the procedure reverts to the predefined bounded search set
\begin{eqnarray}
\mathcal{Q}_{0} &=& \left\{ q_{\min},q_{\min}+1,\ldots,q_{\max}\right\}.
\end{eqnarray}

The final candidate set is therefore
\begin{eqnarray}
\mathcal{Q}_{\mathrm{cand}} &=&
\begin{cases}
\mathcal{Q}_{\mathrm{QACF}},
&
\mathcal{Q}_{\mathrm{QACF}}\neq\varnothing,
\\
\mathcal{Q}_{0},
&
\mathcal{Q}_{\mathrm{QACF}}=\varnothing.
\end{cases}
\end{eqnarray}

For each candidate $q\in\mathcal{Q}_{\mathrm{cand}}$, a delayed residual
matrix $\mathbf{E}^{(q)}$ is constructed from the AR residual sequence
$\{\varepsilon_t\}$. The corresponding aligned residual target vector is
denoted by $\boldsymbol{\varepsilon}_{\mathrm{target}}^{(q)}$. The MA
coefficient vector
$\boldsymbol{\theta}^{(q)}\in\mathbb{R}^{q}$ is then initialized and refined
by minimizing the quantum MA loss defined in Eq.~\eqref{MA_loss}. The
optimized coefficient vector for candidate order $q$ is
\begin{eqnarray}
\boldsymbol{\theta}^{(q)\star}
&=&
\arg\min_{\boldsymbol{\theta}^{(q)}}
\mathcal{L}_{\mathrm{MA}}^{(q)}
\left(
\boldsymbol{\theta}^{(q)};
\mathbf{E}^{(q)},
\boldsymbol{\varepsilon}_{\mathrm{target}}^{(q)}
\right).
\end{eqnarray}

The optimal MA order is selected by comparing the minimized losses across
the candidate set:
\begin{eqnarray}
q^\star &=& \arg\min_{q\in\mathcal{Q}_{\mathrm{cand}}}
\mathcal{L}_{\mathrm{MA}}^{(q)}\left(\boldsymbol{\theta}^{(q)\star}\right).
\end{eqnarray}

Each candidate order therefore generates a separate delayed residual matrix,
aligned target vector, and optimized MA coefficient vector. During broad
candidate-order screening, the initial MA coefficient vector is sampled from
a standard Gaussian distribution:
\begin{eqnarray}
\boldsymbol{\theta}^{(q,0)}
&\sim&
\mathcal{N}
\left(
\mathbf{0},
\mathbf{I}_q
\right).
\end{eqnarray}

A fixed random seed and the same optimization budget are used across all
candidate orders so that their minimized losses remain procedurally
comparable.

The loss function $\mathcal{L}_{\mathrm{MA}}$ incorporates compact-swap-test
state similarity~\cite{buhrman2001quantum}, phase correction combining
cosine-projection and swap-test-derived angular information
~\cite{schuld2021machine}, and entropy regularization based on the
overlap-derived MA uncertainty probability
~\cite{havlivcek2019supervised}. Optimization of each
$\boldsymbol{\theta}^{(q)}$ is performed using a classical derivative-free
optimizer such as COBYLA~\cite{powell1994direct}, consistent with hybrid
variational quantum optimization settings~\cite{cerezo2021variational}.
\vspace{1em}

\begin{algorithm}[H]
\caption{Quantum MA Order Estimation}
\label{alg:ma-q-estimation}
\begin{algorithmic}[1]
\Require Residuals $\{\varepsilon_t\}$, candidate orders $\mathcal{Q}_{\mathrm{cand}}$, loss $\mathcal{L}_{\mathrm{MA}}$,shots $S$, random seed $s$, maximum iterations $T_{\max}$
\Ensure Optimal order $q^\star$, coefficients $\boldsymbol{\theta}^{(q^\star)\star}$

\State Initialize $\mathcal{S}\gets[\ ]$

\For{each $q\in\mathcal{Q}_{\mathrm{cand}}$}
    \State Construct $\mathbf{E}^{(q)}$ and $\boldsymbol{\varepsilon}_{\mathrm{target}}^{(q)}$
    \State Initialize the random generator using the fixed seed $s$
    \State Sample
     $\boldsymbol{\theta}^{(q,0)}
     \sim
     \mathcal{N}(\mathbf{0},\mathbf{I}_q)$
    \State Optimize $\boldsymbol{\theta}^{(q)}$ using COBYLA for at most $T_{\max}$ iterations
    \State Store $\left(q,\boldsymbol{\theta}^{(q)\star},
    \mathcal{L}_{\mathrm{MA}}^{(q)\star}\right)$ in $\mathcal{S}$
\EndFor

\State Select
\[
q^\star
\gets
\arg\min_{q\in\mathcal{Q}_{\mathrm{cand}}}
\mathcal{L}_{\mathrm{MA}}^{(q)\star}
\]
\State \Return $q^\star$, $\boldsymbol{\theta}^{(q^\star)\star}$
\end{algorithmic}
\end{algorithm}
\FloatBarrier

\vspace{1em}

\begin{table}[H]
\centering
\caption{Hyperparameters for MA Order Estimation}
\label{tab:ma-q-estimation-hyperparams}
\begin{tabularx}{\linewidth}{@{} l l X @{}}
\toprule
\textbf{Symbol} & \textbf{Name} & \textbf{Description} \\
\midrule
$q_{\min},q_{\max}$ & q range & Bounds of the fallback MA-order search. \\
$\mathcal{Q}_{\mathrm{QACF}}$ & screened orders & Orders satisfying $\widehat{\rho}^{\,\mathrm{Q}}_k\geq\tau_{\mathrm{QACF}}$. \\
$\mathcal{Q}_{0}$ & fallback set & Bounded order set used when QACF selects no lag. \\
$\mathcal{Q}_{\mathrm{cand}}$ & candidate orders & Final MA-order set passed to loss evaluation. \\
$\boldsymbol{\theta}^{(q,0)}$ & theta init & Standard Gaussian initialization
$\mathcal{N}(\mathbf{0},\mathbf{I}_q)$ used during candidate-order screening. \\
$\mathcal{L}_{\mathrm{MA}}^{(q)}$ & MA loss & Quantum MA loss for candidate order $q$. \\
$S$ & shots & Swap-test measurements per loss evaluation. \\
$s$ & random seed & Fixed seed used for reproducible initialization across candidate orders. \\
$T_{\max}$ & max iterations & Common maximum COBYLA budget applied to every candidate order. \\
\bottomrule
\end{tabularx}
\end{table}
\FloatBarrier

\FloatBarrier
\subsubsection{MA Coefficient Optimization via VQC Refinement}

After selecting the optimal MA order $q^\star$ through QACF-gated residual screening and loss-based comparison, we perform final MA coefficient estimation by minimizing the quantum-regularized MA loss $\mathcal{L}_{\mathrm{MA}}(\boldsymbol{\theta})$ with respect to the coefficient vector $\boldsymbol{\theta}\in\mathbb{R}^{q^\star}$. The loss retains the structure defined in Eq.~\eqref{MA_loss}, comprising:
\begin{enumerate}
\item Residual prediction error computed using phase-corrected cosine similarity,
\item Cosine alignment penalty between dot-product and swap-test-derived projections,
\item Entropy penalty on the swap-test-derived distribution,
\item $\ell_2$ Regularization for coefficient stability.
\end{enumerate}

\paragraph{VQC-to-MA coefficient map}
Let $\boldsymbol{\eta}\in\mathbb{R}^{q^\star}$ denote the trainable MA VQC parameter vector for the selected MA order $q^\star$. The MA coefficient vector is obtained through the map
\begin{eqnarray}
\boldsymbol{\theta}(\boldsymbol{\eta})
&=& \mathrm{Proj}_{\Theta}
\left(g_{\mathrm{MA}}(\boldsymbol{\eta};\boldsymbol{\theta}^{(0)})\right),
\end{eqnarray}
where $g_{\mathrm{MA}}(\cdot)$ denotes the MA coefficient readout induced by the VQC parameterization, and $\mathrm{Proj}_{\Theta}(\cdot)$ denotes the elementwise projection used to enforce the coefficient bounds. In the direct-angle parameterization used in this work,
\begin{eqnarray}
g_{\mathrm{MA}}(\boldsymbol{\eta};\boldsymbol{\theta}^{(0)})
&=& \boldsymbol{\eta}, \qquad \boldsymbol{\eta}^{(0)}=\boldsymbol{\theta}^{(0)}.
\end{eqnarray}
Therefore, the optimized MA coefficient vector is
\begin{eqnarray}
\boldsymbol{\theta}^{\star}
&=& \boldsymbol{\theta}(\boldsymbol{\eta}^{\star}), \qquad
\boldsymbol{\eta}^{\star} = \arg\min_{\boldsymbol{\eta}}
\mathcal{L}_{\mathrm{MA}}
\left(\boldsymbol{\theta}(\boldsymbol{\eta})\right).\label{eq:vqc-ma-opt}
\end{eqnarray}

\textbf{VQC-based parameterization.}
Rather than optimizing the MA coefficient vector directly as an unconstrained Euclidean parameter, QARIMA introduces a trainable VQC parameter vector $\boldsymbol{\eta}\in\mathbb{R}^{q^\star}$. The entries of $\boldsymbol{\eta}$ are represented as trainable $R_y(\eta_j)$ rotation angles in a shallow VQC denoted \texttt{build\_vqc\_ma}$(q^\star,r)$, where $r$ is the number of entangling repetitions. The MA coefficient vector used in the residual-correction equation is then obtained through the coefficient map $\boldsymbol{\theta}(\boldsymbol{\eta})=\mathrm{Proj}_{\Theta}(\boldsymbol{\eta})$. This provides a hardware-compatible parameterization while preserving a direct correspondence between VQC rotation parameters and MA coefficients. In the hybrid optimization loop, compact swap-test projections are used inside the MA loss evaluation, and a classical derivative-free optimizer such as COBYLA~\cite{powell1994direct} updates $\boldsymbol{\eta}$ to minimize $\mathcal{L}_{\mathrm{MA}}(\boldsymbol{\theta}(\boldsymbol{\eta}))$.

\begin{figure}[htbp]
\centering
\includegraphics[width=0.86\linewidth]{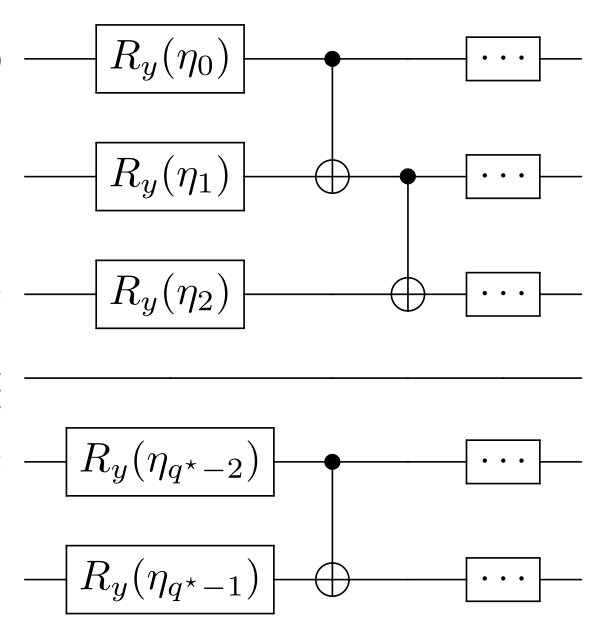}
\caption{Schematic VQC ansatz for MA coefficient refinement using \(\texttt{build\_vqc\_ma}\) with arguments \(q^\star\) and \(r\). Each qubit \(q_i\) receives a trainable single-qubit \(R_y(\eta_i)\) rotation, followed by an entangling CNOT chain. The block is repeated \(r\) times, indicated by \(\cdots\). The trainable VQC parameter vector \(\boldsymbol{\eta}\) is mapped to the MA coefficient vector through \(\boldsymbol{\theta}(\boldsymbol{\eta})=\mathrm{Proj}_{\Theta}(\boldsymbol{\eta})\). Compact swap-test projections are then used inside the MA loss evaluation for \(\mathcal{L}_{\mathrm{MA}}\).}
\label{fig:vqc-ma-ansatz}
\end{figure}

The corresponding VQC ansatz for MA coefficient refinement is illustrated in Fig.~\ref{fig:vqc-ma-ansatz}. Each of the $q^\star$ qubits represents one trainable MA parameter through an $R_y(\eta_j)$ rotation. A nearest-neighbour CNOT chain introduces inter-parameter coupling, and the rotation--entanglement block is repeated $r$ times before the optimized VQC parameters are mapped to the MA coefficient vector $\boldsymbol{\theta}(\boldsymbol{\eta})$.

\textbf{Initialization.}
Candidate MA orders are screened using standard Gaussian initialization with a fixed random seed and an identical optimization budget. After selecting $q^\star$, the final VQC-based refinement uses conditional least-squares initialization whenever the residual design matrix is well-conditioned. From the AR residual stream $\varepsilon_t$, we form the delayed residual matrix
\begin{eqnarray}
\mathbf{E}^{(q^\star)} &=&
\left[\varepsilon_{t-1} \quad \varepsilon_{t-2} \quad \cdots \quad \varepsilon_{t-q^\star}
\right],
\end{eqnarray}
aligned to the MA residual target vector $\boldsymbol{\varepsilon}_{\mathrm{target}}$. The initial coefficient vector is obtained as:
\begin{eqnarray}
\boldsymbol{\theta}^{(0)} &=& \arg\min_{\boldsymbol{\theta}} \left\| \boldsymbol{\varepsilon}_{\mathrm{target}} - \mathbf{E}^{(q^\star)}\boldsymbol{\theta}
\right\|_2^2 .
\end{eqnarray}
The resulting entries are clipped elementwise to $[-1,1]$ for numerical stability. The VQC parameter vector is then initialized from the coefficient initialization:
\begin{eqnarray}
\boldsymbol{\eta}^{(0)} &=& \boldsymbol{\theta}^{(0)}.
\end{eqnarray}
If conditional least-squares initialization fails due to an ill-conditioned residual design or insufficient data, we fall back to:
\begin{eqnarray}
\boldsymbol{\eta}^{(0)} &\sim& \mathcal{U}(-1,1).
\end{eqnarray}

\textbf{Post-training evaluation.}
After convergence, the optimized VQC parameter vector $\boldsymbol{\eta}^{\star}$ is mapped to the final MA coefficient vector
\begin{eqnarray}
\boldsymbol{\theta}^{\star} &=& \boldsymbol{\theta}(\boldsymbol{\eta}^{\star}).
\end{eqnarray}
The optimized MA coefficients $\boldsymbol{\theta}^{\star}$ are then used to compute the predicted residual contribution $\widehat{\varepsilon}_t$. The remaining MA-stage error is:
\begin{eqnarray}
\zeta_t &=& \varepsilon_t-\widehat{\varepsilon}_t.
\end{eqnarray}
We then compute the empirical standard deviation of the remaining residual error:
\begin{eqnarray}
\sigma_{\mathrm{MA}} &=& \sqrt{\frac{1}{T_q}\sum_t\zeta_t^2}.
\end{eqnarray}
where $T_q$ denotes the number of aligned residual observations available after forming the $q^\star$-lag residual matrix. This quantity, together with the final convergence loss, serves as a post-training quality indicator for the MA residual-correction component.

\begin{algorithm}[H]
\caption{VQC-Refined MA Training for $q^\star$}
\label{alg:vqc-ma-train}
\begin{algorithmic}[1]
\Require AR residual sequence $\{\varepsilon_t\}$, optimal MA order $q^\star$, VQC depth $r$, maximum iterations $T_{\max}$, shots $S$, loss weights $\lambda_{\mathrm{cos}}$, $\lambda_{\mathrm{ent}}$, $\lambda_{\mathrm{L2}}$, phase weight $\omega$
\Ensure Optimized MA coefficients $\boldsymbol{\theta}^\star$, MA residual standard deviation $\sigma_{\mathrm{MA}}$

\State Construct delayed residual matrix $\mathbf{E}^{(q^\star)}$ from $\{\varepsilon_t\}$
\State Construct aligned residual target vector $\boldsymbol{\varepsilon}_{\mathrm{target}}$
\State Initialize $\boldsymbol{\theta}^{(0)}$ using conditional least squares; fallback to $\mathcal{U}(-1,1)$ if initialization fails
\State Clip $\boldsymbol{\theta}^{(0)}$ elementwise to $[-1,1]$
\State Initialize VQC parameters $\boldsymbol{\eta}^{(0)}\gets\boldsymbol{\theta}^{(0)}$
\State Define the MA coefficient map $\boldsymbol{\theta}(\boldsymbol{\eta})\gets\mathrm{Proj}_{\Theta}(\boldsymbol{\eta})$, where $\mathrm{Proj}_{\Theta}$ clips coefficients to $[-1,1]$
\State Build $\texttt{build\_vqc\_ma}(q^\star,r)$ with trainable rotation parameters $\boldsymbol{\eta}$
\State Optimize
\[\boldsymbol{\eta}^{\star} \gets \arg\min_{\boldsymbol{\eta}}
\mathcal{L}_{\mathrm{MA}} \left( \boldsymbol{\theta}(\boldsymbol{\eta})\right)\]
using COBYLA for at most $T_{\max}$ iterations
\State Recover optimized MA coefficients $\boldsymbol{\theta}^{\star}\gets\boldsymbol{\theta}(\boldsymbol{\eta}^{\star})$
\State Compute predicted residual contribution $\widehat{\varepsilon}_t$ using $\boldsymbol{\theta}^{\star}$
\State Compute remaining MA-stage error $\zeta_t\gets\varepsilon_t-\widehat{\varepsilon}_t$
\State Compute $\sigma_{\mathrm{MA}}\gets\sqrt{\frac{1}{T_q}\sum_t\zeta_t^2}$
\State \Return $\boldsymbol{\theta}^{\star}$, $\sigma_{\mathrm{MA}}$
\end{algorithmic}
\end{algorithm}
\FloatBarrier

\begin{table}[H]
\centering
\caption{Hyperparameters and Coefficient Map for VQC-Refined MA Training}
\label{tab:vqc-ma-hyperparams}
\scriptsize
\setlength{\tabcolsep}{3pt}
\renewcommand{\arraystretch}{1.12}
\begin{tabularx}{\linewidth}{@{} 
>{\raggedright\arraybackslash}p{0.20\linewidth}
>{\raggedright\arraybackslash}p{0.22\linewidth}
>{\raggedright\arraybackslash}X 
@{}}
\toprule
\textbf{Symbol} & \textbf{Name} & \textbf{Description} \\
\midrule
$\boldsymbol{\theta}^{(0)}$ & MA coefficient initialization & Initial MA coefficient vector obtained from conditional least squares, clipped to $[-1,1]$, with fallback to $\mathcal{U}(-1,1)$. \\
$\boldsymbol{\eta}$ & VQC parameters & Trainable VQC rotation-parameter vector used to refine the MA coefficients. \\
$\boldsymbol{\eta}^{(0)}$ & VQC initialization & Initial VQC parameter vector, set from the MA coefficient initialization as $\boldsymbol{\eta}^{(0)}=\boldsymbol{\theta}^{(0)}$. \\
$\boldsymbol{\theta}(\boldsymbol{\eta})$ & MA coefficient map & MA coefficient vector obtained from VQC parameters through $\boldsymbol{\theta}(\boldsymbol{\eta})=\mathrm{Proj}_{\Theta}(\boldsymbol{\eta})$. \\
$\mathrm{Proj}_{\Theta}(\cdot)$ & coefficient projection & Elementwise projection or clipping operation enforcing the coefficient bounds $[-1,1]$. \\
$[-1,1]$ & coefficient bounds & Elementwise bounds imposed on each MA coefficient $\theta_j$ for numerical stability. \\
$r$ & VQC repetitions & Number of entangling repetitions in the VQC ansatz. \\
$T_{\max}$ & max iterations & Maximum number of COBYLA iterations used for VQC parameter optimization. \\
$S$ & shots & Number of compact swap-test shots used during MA loss evaluation. \\
$\lambda_{\mathrm{cos}}$ & cosine weight & Penalty weight for the cosine-alignment term in $\mathcal{L}_{\mathrm{MA}}$. \\
$\lambda_{\mathrm{ent}}$ & entropy weight & Penalty weight for entropy regularization based on the overlap-derived probability. \\
$\lambda_{\mathrm{L2}}$ & L2 weight & Weight for $\ell_2$ regularization on the MA coefficient vector. \\
$\omega$ & phase weight & Phase-correction factor used to combine swap-test and cosine-projection angular information. \\
$\boldsymbol{\theta}^{\star}$ & optimized MA coefficients & Final MA coefficient vector recovered from the optimized VQC parameters as $\boldsymbol{\theta}^{\star}=\boldsymbol{\theta}(\boldsymbol{\eta}^{\star})$. \\
$\zeta_t$ & remaining MA error & Remaining MA-stage error after residual correction, defined as $\zeta_t=\varepsilon_t-\widehat{\varepsilon}_t$. \\
$\sigma_{\mathrm{MA}}$ & MA residual std & Standard deviation of the remaining MA-stage residual error after fitting $\boldsymbol{\theta}^{\star}$. \\
\bottomrule
\end{tabularx}
\normalsize
\end{table}
\FloatBarrier

\FloatBarrier

\FloatBarrier
\subsubsection{ARMA Finalization Across All AR Models}

Finally, the full ARMA summary is constructed by applying the MA estimation and VQC-refined MA training procedure to each retained AR model. Each AR candidate order $p$ has its own residual sequence ${\varepsilon_t^{(p)}}$; therefore, the selected MA order is conditional on the AR residual structure and is denoted $q_p^\star$. For each AR model, the residual sequence is passed through QACF-gated MA-order screening, followed by quantum-regularized MA coefficient optimization. The resulting summary records the final $(p,d^\star,q_p^\star)$ configuration together with the corresponding AR coefficients, MA coefficients, residual statistics, and convergence loss. This final table enables direct comparison of candidate QARIMA/ARMA configurations before selecting the reported model.

\begin{algorithm}[H]
\caption{Run MA Estimation for All AR Models}
\label{alg:run-ma-all}
\begin{algorithmic}[1]
\Require Differenced series $\Delta^{d^\star}y$, selected differencing order $d^\star$, AR model summary $\mathcal{A}={(p,\boldsymbol{b}^{(p)},{\varepsilon_t^{(p)}},\sigma_{\mathrm{AR}}^{(p)})}$, MA loss $\mathcal{L}_{\mathrm{MA}}$
\Ensure Final summary table of ARMA $(p,d^\star,q_p^\star)$ models

\State Initialize final summary table $\mathcal{F}\gets[;]$

\For{each AR model $(p,\boldsymbol{b}^{(p)},{\varepsilon_t^{(p)}},\sigma_{\mathrm{AR}}^{(p)})$ in $\mathcal{A}$}
\State Retrieve AR residual sequence ${\varepsilon_t^{(p)}}$
\State Apply QACF-gated MA order estimation to obtain candidate MA orders
\State Select $q_p^\star$ through MA loss comparison on ${\varepsilon_t^{(p)}}$
\State Train VQC-refined MA coefficients $\boldsymbol{\theta}^{(q_p^\star)\star}$
\State Compute MA residual standard deviation $\sigma_{\mathrm{MA}}^{(p,q_p^\star)}$
\State Store $(p,d^\star,q_p^\star,\boldsymbol{b}^{(p)},\boldsymbol{\theta}^{(q_p^\star)\star},\sigma_{\mathrm{AR}}^{(p)},\sigma_{\mathrm{MA}}^{(p,q_p^\star)})$ in $\mathcal{F}$
\EndFor

\State \Return final ARMA summary table $\mathcal{F}$
\end{algorithmic}
\end{algorithm}
\FloatBarrier

\begin{table}[H]
\centering
\caption{Output Structure for Final ARMA Summary}
\label{tab}
\begin{tabularx}{\linewidth}{@{} l l X @{}}
\toprule
\textbf{Column} & \textbf{Type} & \textbf{Description} \\
\midrule
$p$ & Integer & AR order selected or retained after QPACF-guided VQC refinement. \\
$d^\star$ & Integer & Differencing order selected by the quantum state-similarity differencing assessment. \\
$q_p^\star$ & Integer & MA order selected for the residual sequence associated with AR order $p$. \\
$\boldsymbol{b}^{(p)}$ & Vector & VQC-refined AR coefficient vector for order $p$. \\
$\boldsymbol{\theta}^{(q_p^\star)\star}$ & Vector & VQC-refined MA coefficient vector for the selected residual MA order. \\
$\sigma_{\mathrm{AR}}^{(p)}$ & Float & Standard deviation of the AR residuals before MA correction. \\
$\sigma_{\mathrm{MA}}^{(p,q_p^\star)}$ & Float & Standard deviation of the remaining residual error after MA correction. \\
$\mathcal{L}_{\mathrm{MA}}$ & Float & Final minimized quantum-regularized MA loss for the selected $q_p^\star$. \\
\bottomrule
\end{tabularx}
\end{table}
\FloatBarrier

\FloatBarrier
\subsection{Computational Complexity and Practical Implementation}

The proposed QARIMA framework introduces additional computational overhead
relative to classical ARIMA because state-similarity estimates require repeated
quantum measurements and coefficient estimation uses iterative hybrid
optimization. Let $N$ denote the training-series length, $K$ the maximum lag,
and $S$ the number of measurement shots.

The implemented QACF performs one compact swap-test comparison between two
complete aligned residual vectors for each lag. Its measurement-execution cost
is therefore approximately
\begin{equation}
\mathcal{O}(SK),
\end{equation}
with an additional classical vector-construction cost of approximately
\begin{equation}
\mathcal{O}(NK).
\end{equation}

QPACF evaluates compact-swap-test similarities across aligned observations for
each candidate lag, resulting in an approximate measurement-execution cost of
\begin{equation}
\mathcal{O}(SNK).
\end{equation}
Consequently, the combined lag-discovery stage is generally dominated by
QPACF. By comparison, classical ACF/PACF evaluation requires approximately
\begin{equation}
\mathcal{O}(NK),
\end{equation}
excluding implementation-specific optimizations.

The AR and MA coefficient-estimation stages introduce additional circuit
evaluations through classical optimization. If $E$ optimizer iterations are
performed and each iteration requires $C$ circuit evaluations, the corresponding
optimization cost is approximately
\begin{equation}
\mathcal{O}(EC)
\end{equation}
for each candidate order. These expressions describe measurement and
optimization costs and exclude hardware-dependent state-preparation,
communication, and simulator-execution overheads.

The present work does not claim a computational quantum speedup over
classical ARIMA. Its objective is to investigate whether quantum-compatible
state-similarity and variational parameter-estimation mechanisms can produce
useful forecasting behaviour while preserving the statistical structure and
interpretability of the classical Box--Jenkins methodology.

\FloatBarrier
\section{Results: Evaluation Protocol Across Datasets}
\label{sec:results-protocol}
We evaluated Quantum-ARIMA (QARIMA) against a classical \texttt{pmdarima} baseline on five public time series:
\emph{Sunspots}, \emph{Mauna Loa CO\textsubscript{2}}, \emph{Australian Woollen Yarn Production}, \emph{Australian Beer Production}, and \emph{Sydney 2024 Weather}. For each dataset, we followed the Methods pipeline: (i) identical preprocessing and differencing, (ii) generation of candidate QARIMA$(p,d,q)$ models via the manuscript procedure, and (iii) training of a matched classical ARIMA comparator on the same training data. Model quality was assessed \emph{out-of-sample} (OOS) on a held-out segment defined per dataset.
We report error metrics (MSE, MAPE) for every candidate model and then conduct Diebold-Mariano (DM) tests comparing each QARIMA against the classical baseline under two loss functions (MSE and MAE).

\FloatBarrier
\subsection{Error-Based Evaluation}
\label{subsec:results-arima-metrics}
For every dataset we report out-of-sample (OOS) accuracy using two standard ARIMA metrics: Mean Squared Error (MSE) and Mean Absolute Percentage Error (MAPE). MSE reflects point-forecast fidelity and penalizes large deviations, while MAPE provides a scale-free view that is comparable across series with different magnitudes. For each QARIMA candidate we list its OOS value alongside the matched classical baseline and, when useful, the absolute improvement
\begin{equation}
\begin{aligned}
\Delta\mathrm{MSE}
&=\mathrm{MSE}_{\mathrm{classical}}-\mathrm{MSE}_{\mathrm{quantum}},\\
\Delta\mathrm{MAPE}
&=\mathrm{MAPE}_{\mathrm{classical}}-\mathrm{MAPE}_{\mathrm{quantum}}.
\end{aligned}
\end{equation}
Individual dataset subsections interpret these numbers in context (trend-like vs.\ seasonal series, short vs.\ long OOS windows) and highlight cases where both metrics move in the same favourable direction.

\FloatBarrier
\subsection{Statistical Comparison}
\label{subsec:results-dm}

To determine whether the observed error gaps are more than numerical fluctuations, we run Diebold-Mariano (DM) tests \cite{diebold1995comparing} for each quantum–classical pair on the same OOS segment. We evaluate DM under two loss differentials: squared-error (MSE-style) and absolute-error (MAE-style). For every comparison we report the DM statistic, its $p$-value, and mark results that satisfy $p \le \alpha$ (typically $\alpha = 0.05$). Alongside significance, we also show the corresponding classical and quantum mean losses and their difference
\begin{eqnarray}
\Delta = \text{loss}_{\text{classical}} - \text{loss}_{\text{quantum}},
\end{eqnarray}
which acts as an effect-size indicator. The per-dataset result sections that follow use this common procedure to comment on when a QARIMA variant is not only better in value (MSE/MAPE) but also statistically supported by DM.

\paragraph{Interpretation rule.}
We consider a QARIMA model to \emph{reliably outperform} the classical baseline on a dataset when it (i) improves OOS MSE/MAPE and (ii) achieves DM significance ($p \le 0.05$) under at least one loss (preferably both).
When multiple QARIMA specifications are significant, we prioritize those with the largest positive mean-loss deltas and consistent gains across both error metrics. In subsequent subsections, we will present the results achieved by each dataset and our interpretation of QARIMA  as per the results.

\FloatBarrier
\subsection{Sunspots dataset}
We use the classic annual mean sunspot counts provided in \texttt{statsmodels.datasets.sunspots} (column \texttt{SUNACTIVITY}), spanning 1700-2008. These counts proxy solar magnetic activity and exhibit the well-known $\sim$11-year Schwabe cycle, but with quasi-periodic behavior: cycle length and amplitude drift over time, peaks are asymmetric, and multi-decadal envelopes (amplitude/phase modulation) are common. Statistically, the series is nonstationary in level and nonlinear in its dynamics; simple linear AR models often underfit long-range dependence and changing cycle shape, motivating differencing and richer lag structure. Sunspots have total of 308 annual data points  out of which we have use 181 data points for training and 128 data points for Out of Sample test. The generated QARIMA models are evaluated against the classical pmdarima(2,0,0).  The OOS results of MSE , MAPE and DM tests are presented in Table \ref{tab:Sunspots_MSE_MAPE}.

\begin{table}[H]
\centering
\caption{Sunspots Classical Vs Quantum MSE MAPE}
\label{tab:Sunspots_MSE_MAPE}
\scriptsize
\setlength{\tabcolsep}{2pt}
\begin{tabular}{|l|l|l|l|} 
\hline
\multicolumn{4}{|l|}{\textbf{~Sunspots Classical Vs Quantum OOS~ MSE MAPE}}                                                                           \\ 
\hline
\textbf{Model}                               & \textbf{N} & $MSE$                                        & \textbf{MAPE}                              \\ 
\hline
Classical (pmdarima) (2, 0, 0) seasonal=None & 128        & {\cellcolor[rgb]{1,0.922,0.518}}2181.589     & {\cellcolor[rgb]{0.984,0.808,0.82}}1.661   \\ 
\hline
Quantum (pdq=(3, 1, 1))                      & 128        & {\cellcolor[rgb]{0.388,0.745,0.482}}2146.926 & {\cellcolor[rgb]{0.973,0.412,0.42}}1.790   \\ 
\hline
Quantum (pdq=(10, 1, 1))                     & 128        & {\cellcolor[rgb]{0.635,0.816,0.494}}2160.998 & {\cellcolor[rgb]{0.863,0.898,0.953}}1.534  \\ 
\hline
Quantum (pdq=(7, 1, 3))                      & 128        & {\cellcolor[rgb]{0.875,0.882,0.51}}2174.622  & {\cellcolor[rgb]{0.988,0.906,0.918}}1.629  \\ 
\hline
Quantum (pdq=(4, 1, 1))                      & 128        & {\cellcolor[rgb]{0.961,0.91,0.514}}2179.405  & {\cellcolor[rgb]{0.98,0.675,0.682}}1.705   \\ 
\hline
Quantum (pdq=(9, 1, 3))                      & 128        & {\cellcolor[rgb]{1,0.914,0.518}}2182.978     & {\cellcolor[rgb]{0.988,0.988,1}}1.601      \\ 
\hline
Quantum (pdq=(8, 1, 3))                      & 128        & {\cellcolor[rgb]{1,0.906,0.518}}2184.006     & {\cellcolor[rgb]{0.945,0.957,0.984}}1.580  \\ 
\hline
Quantum (pdq=(6, 1, 3))                      & 128        & {\cellcolor[rgb]{0.992,0.729,0.482}}2210.434 & {\cellcolor[rgb]{0.525,0.663,0.835}}1.349  \\ 
\hline
Quantum (pdq=(5, 1, 3))                      & 128        & {\cellcolor[rgb]{0.973,0.412,0.42}}2256.788  & {\cellcolor[rgb]{0.353,0.541,0.776}}1.254  \\
\hline
\end{tabular}
\end{table}
\FloatBarrier

\paragraph{Sunspots OOS performance assessment.}

\paragraph{MSE / MAPE comparison}

\begin{figure}[H]
  \centering
  \includegraphics[width=.49\linewidth]{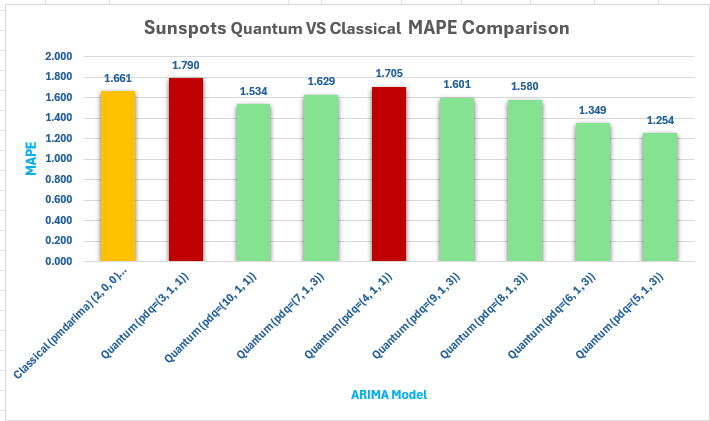}\hfill
  \includegraphics[width=.49\linewidth]{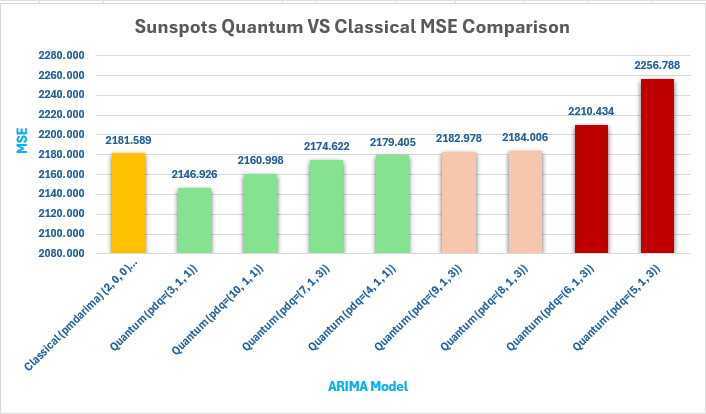}
  \caption{Sunspots OOS  MSE, MAPE. QARIMA vs.\ classical.}
  \label{fig:sunspots-mse-mape}
\end{figure}
\FloatBarrier

All QARIMA specifications were benchmarked against the classical non-seasonal pmdarima ARIMA(2,0,0) baseline on the 128-step OOS window. The results in Table~\ref{tab:Sunspots_MSE_MAPE} and Fig.~\ref{fig:sunspots-mse-mape} show that QARIMA identifies several competitive and improved configurations for the quasi-periodic Sunspots series. The most balanced model is Q(10,1,1), which improves both MSE and MAPE relative to the classical baseline. Q(3,1,1), Q(7,1,3), and Q(4,1,1) also reduce MSE, with Q(7,1,3) simultaneously improving MAPE. In addition, the MA-enriched higher-order models Q(6,1,3) and Q(5,1,3) achieve the strongest MAPE reductions, indicating that the quantum-derived lag and residual structure improves relative-error tracking over the long OOS horizon. Thus, the Sunspots experiment does not show a single isolated improvement; it shows that QARIMA generates a family of viable ARIMA configurations with different error trade-offs, including models that improve squared-error behaviour, relative-error behaviour, or both.

\paragraph{DM analysis}

\begin{figure}[H]
  \centering
  \includegraphics[width=.49\linewidth]{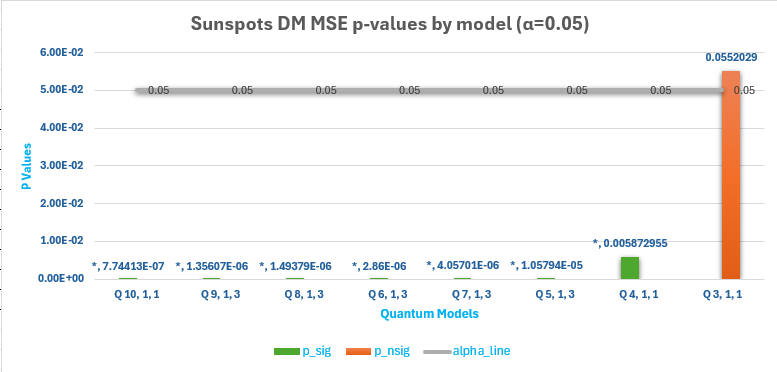}\hfill
  \includegraphics[width=.49\linewidth]{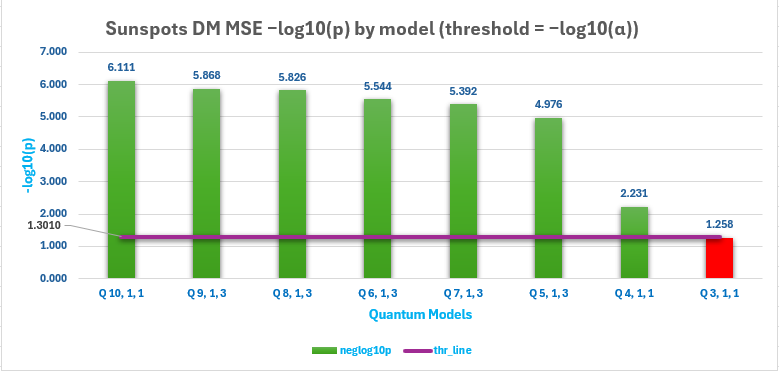}\hfill
  \caption{Sunspots DM (MSE) p-values with $\alpha$ reference line; significant QARIMA bars appear in green.}
  \label{fig:sunspots-dm-mse-p}
\end{figure}
\FloatBarrier

To assess whether the observed forecast-loss differences are statistically meaningful, we ran Diebold-Mariano tests between each QARIMA forecast and the classical ARIMA$(2,0,0)$ forecast. The MSE-based DM results in Fig.~\ref{fig:sunspots-dm-mse-p} show strong support for the higher-order QARIMA configurations. Seven of the eight quantum specifications, Q(10,1,1), Q(9,1,3), Q(8,1,3), Q(7,1,3), Q(6,1,3), Q(5,1,3), and Q(4,1,1), fall below the $\alpha=0.05$ threshold. In the corresponding DM summary table, these models also show positive classical-minus-quantum mean-loss differences, indicating that the significant loss differences favour QARIMA under the aligned MSE comparison. Q(3,1,1) is the only MSE case that does not cross the threshold, remaining marginally above significance; this suggests that its aggregate OOS MSE improvement is less stable under the aligned DM loss blocks than the higher-order QARIMA group.

\begin{figure}[H]
  \centering
  \includegraphics[width=.49\linewidth]{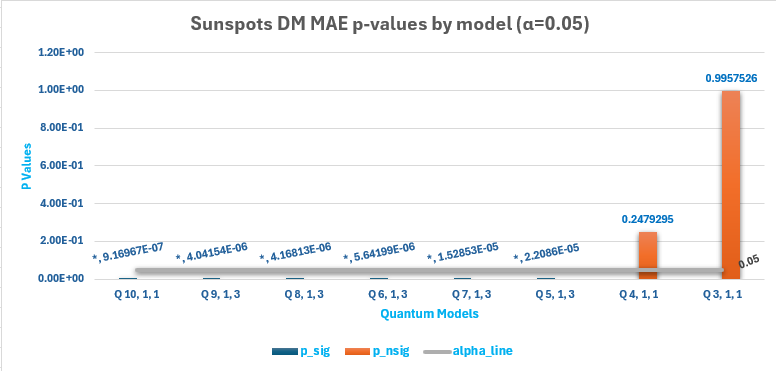}\hfill
  \includegraphics[width=.49\linewidth]{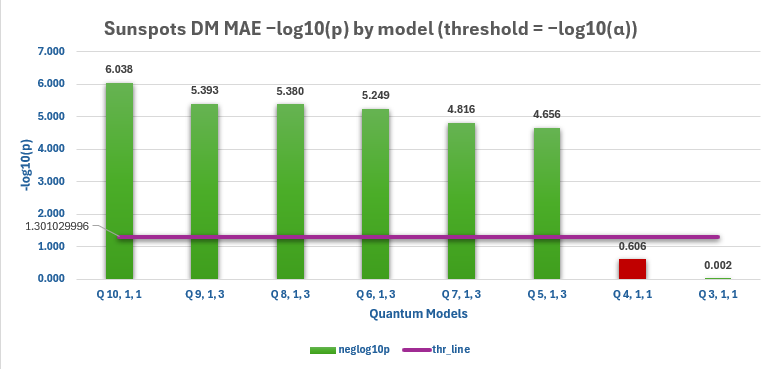}\hfill
  \caption{Sunspots DM (MAE) $-\log_{10}(p)$ with threshold at $-\log_{10}(\alpha)$.}
  \label{fig:sunspots-dm-mae-p}
\end{figure}
\FloatBarrier

The MAE-based DM results in Fig.~\ref{fig:sunspots-dm-mae-p} provide a similar but slightly more selective pattern. Six QARIMA models, led by Q(10,1,1) and followed by the MA-enriched higher-order variants, remain significant at $\alpha=0.05$. Q(3,1,1) and Q(4,1,1) do not show statistically significant MAE differences, indicating that their advantage is mainly expressed through squared-error behaviour rather than absolute-error stability. Overall, the DM analysis supports the main Sunspots conclusion: QARIMA does not merely produce visually comparable forecasts, but generates statistically distinguishable forecast-loss sequences for most higher-order quantum configurations, with Q(10,1,1) providing the most balanced evidence across both OOS error metrics and aligned DM comparisons. Full DM statistics, including DM\_stat, $p$-value, classical and quantum mean losses, and $\Delta$, are provided in Appendix Tables~\ref{Tab:Sunspots_DM_MSE}-\ref{tab:sunspots_dm_mae} for reproducibility.

\paragraph{Interpretation of Sunspots results}
The Sunspots series provides a useful structural validation case for QARIMA because it combines long-memory behaviour, quasi-periodicity, and drifting amplitude/phase associated with the $\sim$11-year Schwabe cycle. Within the proposed pipeline, quantum state-similarity differencing, QPACF/QACF-guided lag discovery, and shallow VQC-based coefficient refinement preserve the Box--Jenkins structure while allowing richer lag interactions to be screened and estimated through state-alignment losses. As shown in Table~\ref{tab:Sunspots_MSE_MAPE}, QARIMA identifies meaningful higher-order configurations rather than merely reproducing the classical ARIMA$(2,0,0)$ baseline. In particular, Q(10,1,1) provides the most balanced evidence by improving both OOS MSE and MAPE, while the MA-enriched higher-order models provide strong aligned-loss support in the DM panels in Figs.~\ref{fig:sunspots-dm-mse-p}--\ref{fig:sunspots-dm-mae-p}. Thus, Sunspots supports the central claim that quantum state-similarity modules can discover useful lag and residual structures for quasi-periodic dynamics while retaining the interpretability of ARIMA.

\FloatBarrier
\subsection{Mauna Loa CO2 dataset}
\begin{figure}[H]
  \centering
  \includegraphics[width=.49\linewidth]{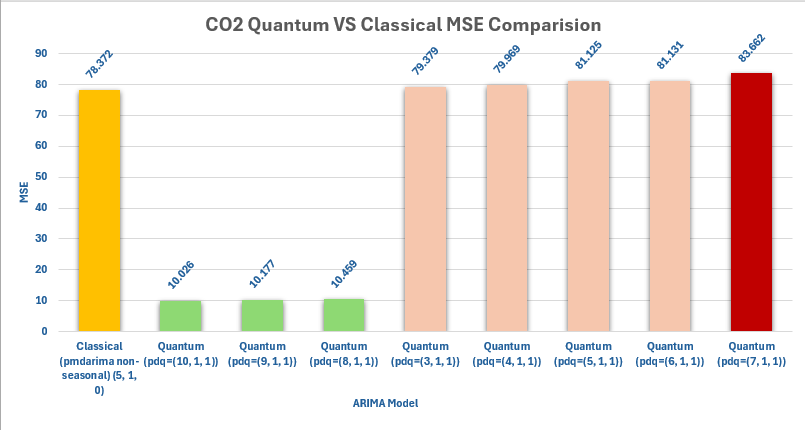}\hfill
  \includegraphics[width=.49\linewidth]{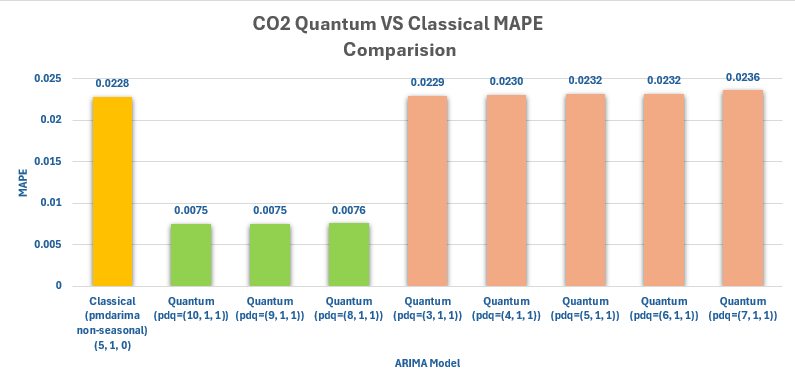}
  \caption{CO2 OOS  MSE, MAPE. QARIMA vs.\ classical.}
  \label{fig:CO2-mse-mape}
\end{figure}

\begin{figure}[H]
  \centering
  \includegraphics[width=.49\linewidth]{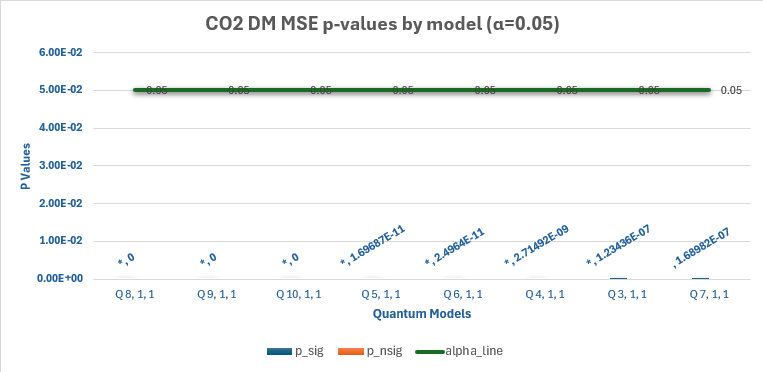}\hfill
  \includegraphics[width=.49\linewidth]{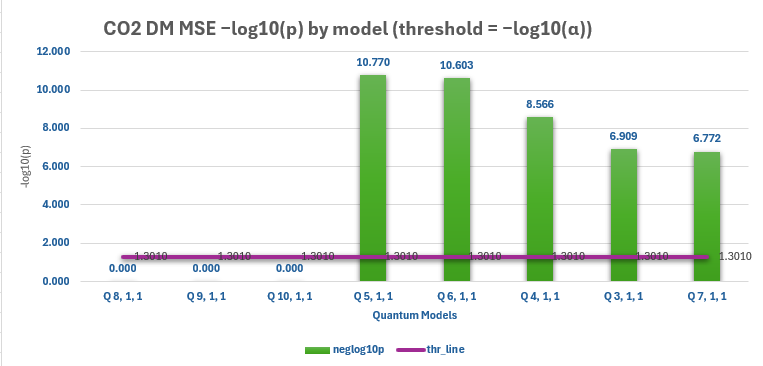}\hfill
  \caption{CO2 DM (MSE) p-values with $\alpha$ reference line; significant QARIMA bars appear in green.}
  \label{fig:CO2-dm-mse-p}
\end{figure}

\begin{figure}[H]
  \centering
  \includegraphics[width=.49\linewidth]{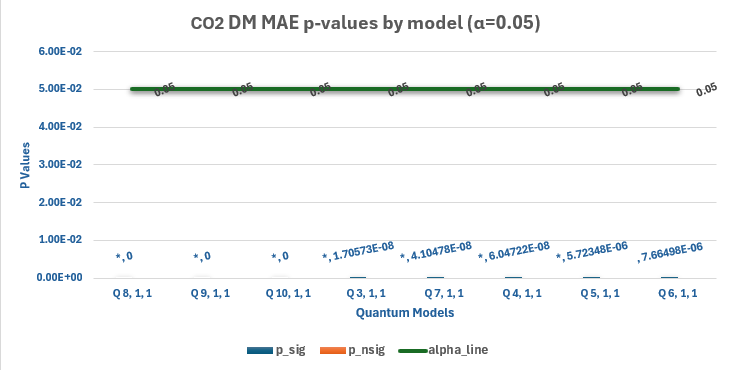}\hfill
  \includegraphics[width=.49\linewidth]{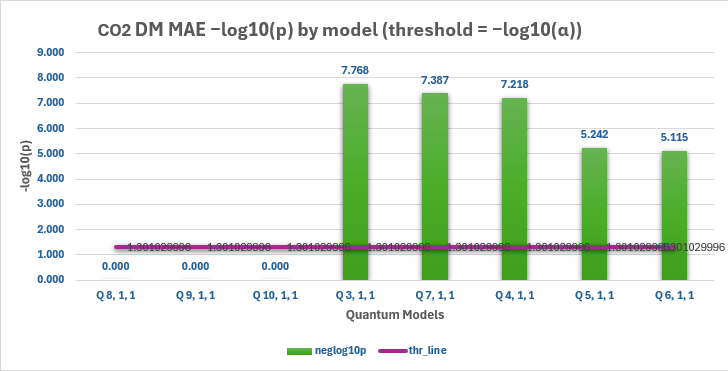}\hfill
  \caption{CO2 DM (MAE) $-\log_{10}(p)$ with threshold at $-\log_{10}(\alpha)$.}
  \label{fig:CO2-dm-mae-p}
\end{figure}

We analyze the canonical Mauna Loa atmospheric CO\textsubscript{2} record (monthly ppm). The series features a strong \emph{upward trend} (Keeling curve) and a pronounced \emph{annual cycle} whose amplitude drifts slowly over time. Hence it is nonstationary in level with deterministic seasonality and gradual modulation, stressing purely linear, low–order ARIMA baselines. The Mauna Loa CO\textsubscript{2} series comprises monthly observations and after removing initial missing values it contains $468$ observations in total. We reserve the last $120$ months for out-of-sample (OOS) testing and use the preceding $348$ months for training. The generated QARIMA models are evaluated against a classical non-seasonal baseline \texttt{pmdarima} ARIMA(5,1,0). OOS results for MSE and MAPE, together with DM tests (MSE/MAE), are reported in Table~\ref{tab:co2-mse-mape} and Figs.~\ref{fig:CO2-mse-mape}–\ref{fig:CO2-dm-mae-p}.

\begin{table}[H]
\centering
\caption{CO2 Classical Vs Quantum OOS}
\scriptsize
\setlength{\tabcolsep}{2pt}
\begin{tabular}{|>{\hspace{0pt}}m{0.592\linewidth}|>{\hspace{0pt}}m{0.063\linewidth}|>{\hspace{0pt}}m{0.138\linewidth}|>{\hspace{0pt}}m{0.138\linewidth}|} 
\hline
\multicolumn{4}{|>{\hspace{0pt}}m{0.931\linewidth}|}{\textbf{CO2 Classical Vs Quantum OOS}} \\ 
\hline
\textbf{Model} & \textbf{N} & \textbf{MSE} & \textbf{MAPE} \\ 
\hline
Classical (pmdarima non-seasonal) (5, 1, 0) & 120 & {\cellcolor[rgb]{0.988,0.918,0.514}}78.37204 & {\cellcolor[rgb]{0.98,0.98,0.996}}0.022805 \\ 
\hline
Quantum (pdq=(10, 1, 1)) & 120 & {\cellcolor[rgb]{0.388,0.745,0.482}}10.02575 & {\cellcolor[rgb]{0.353,0.541,0.776}}0.007472 \\ 
\hline
Quantum (pdq=(9, 1, 1)) & 120 & {\cellcolor[rgb]{0.388,0.745,0.482}}10.17655 & {\cellcolor[rgb]{0.353,0.541,0.776}}0.007496 \\ 
\hline
Quantum (pdq=(8, 1, 1)) & 120 & {\cellcolor[rgb]{0.388,0.745,0.482}}10.45898 & {\cellcolor[rgb]{0.357,0.541,0.776}}0.007573 \\ 
\hline
Quantum (pdq=(3, 1, 1)) & 120 & {\cellcolor[rgb]{1,0.922,0.518}}79.3785 & {\cellcolor[rgb]{0.988,0.988,1}}0.022943 \\ 
\hline
Quantum (pdq=(4, 1, 1)) & 120 & {\cellcolor[rgb]{1,0.855,0.506}}79.96909 & {\cellcolor[rgb]{0.988,0.906,0.918}}0.023035 \\ 
\hline
Quantum (pdq=(5, 1, 1)) & 120 & {\cellcolor[rgb]{0.992,0.714,0.478}}81.12483 & {\cellcolor[rgb]{0.984,0.749,0.757}}0.023214 \\ 
\hline
Quantum (pdq=(6, 1, 1)) & 120 & {\cellcolor[rgb]{0.992,0.714,0.478}}81.13141 & {\cellcolor[rgb]{0.984,0.745,0.757}}0.023215 \\ 
\hline
Quantum (pdq=(7, 1, 1)) & 120 & {\cellcolor[rgb]{0.973,0.412,0.42}}83.66155 & {\cellcolor[rgb]{0.973,0.412,0.42}}0.023586 \\
\hline
\end{tabular}\label{tab:co2-mse-mape}
\end{table}
\FloatBarrier

\paragraph{Mauna Loa CO2 Performance assessment (OOS).}
\paragraph{MSE and MAPE comparison}

\FloatBarrier

Following the same protocol as Sunspots, we compare multiple QARIMA$(p,d,q)$ models against a non–seasonal classical comparator ARIMA(5,1,0) on an OOS window of $N{=}120$ months.
Table~\ref{tab:co2-mse-mape} and Fig.~\ref{fig:CO2-mse-mape} show that \emph{high–$p$} quantum models with a small MA term Q(10,1,1), Q(9,1,1), Q(8,1,1), achieve \textbf{large} gains over the classical baseline:
MSE drops from $\approx 78.37$ to $\approx 10$ and MAPE from $\approx 2.28\%$ to $\approx 0.75\%$.
Lower–order quantum variants Q(3–7,1,1) do not improve upon the baseline (MSE $\approx 79$–$86$, MAPE $\approx 2.29$–$2.36\%$), indicating that the CO\textsubscript{2} series benefits from richer autoregressive memory with modest $q$ when using a non–seasonal specification.

\paragraph{Diebold–Mariano (DM) tests}

\FloatBarrier

\FloatBarrier

DM results (Figs.~\ref{fig:CO2-dm-mse-p}–\ref{fig:CO2-dm-mae-p}) corroborate the error patterns. Where available, models with lower mean loss than the classical baseline (e.g., the strong Q(10/9/8,1,1) group) show \emph{significant} differences under MSE/MAE DM (bars below $\alpha$ or above the $-\log_{10}(\alpha)$ threshold).Conversely, quantum models with higher mean loss (Q(3–7,1,1)) are also significantly different, but the direction (worse than classical) matches their positive classical–minus–quantum loss differentials. Complete per–model DM tables and mean–loss deltas are provided in the Appendix.

\paragraph{Interpretation of CO2 results.}
In line with our pipeline \emph{quantum-informed differencing} (Q-$d$), quantum-assisted lag discovery for $(p,d,q)$, and \emph{VQC-only} coefficient estimation under a fixed optimizer budget with a \emph{shallow} variational embedding (\texttt{reps}=1) QARIMA keeps the Box–Jenkins structure but replaces linear estimation by a unitary, learned feature map over lags. For CO\textsubscript{2}, this embedding enhances representation of \emph{long memory} (via larger $p$) and introduces mild \emph{nonlinear mixing} that effectively tracks the annual cycle’s amplitude drift \emph{without} an explicit seasonal operator. As a result, high–$p$ QARIMA models (Q(10/9/8,1,1)) deliver large, statistically significant OOS error reductions, while lower–order quantum variants (Q(3–7,1,1)) confirm via DM that the classical baseline is preferable when autoregressive memory is too limited.

\FloatBarrier
\subsection{Australian Beer Production Dataset}
The quarterly \texttt{ausbeer} series contains $211$ observations. We follow the same protocol as for the other datasets and keep the last $8$ quarters for out-of-sample (OOS) evaluation, using the first $203$ quarters for training:
\begin{eqnarray}
N_{\text{total}} = 211,\quad N_{\text{train}} = 203,\quad N_{\text{OOS}} = 8.
\end{eqnarray}
All quantum ARIMA (QARIMA) models are compared against a simple classical non-seasonal baseline \texttt{pmdarima} ARIMA$(0,1,1)$.

\begin{table}[H]
\centering
\caption{AusBeer classical vs.\ quantum OOS (last 8 quarters).}
\label{tab:ausbeer-mse-mape}
\scriptsize
\setlength{\tabcolsep}{2pt}
\begin{tabular}{|l|c|r|r|}
\hline
\rowcolor[RGB]{0,55,99}
\textcolor{white}{\textbf{Model}} &
\textcolor{white}{\textbf{N}} &
\textcolor{white}{\textbf{MSE}} &
\textcolor{white}{\textbf{MAPE}} \\ \hline

Classical (pmdarima non-seasonal) (0, 1, 1) & 8 &
\cellcolor[HTML]{F4CCCC}1491.762063 &
\cellcolor[HTML]{F4CCCC}0.080529 \\ \hline

Quantum (pdq=(7, 1, 1)) & 8 &
\cellcolor[HTML]{C6EFCE}59.792337 &
\cellcolor[HTML]{C6EFCE}0.016178 \\ \hline

Quantum (pdq=(10, 1, 1)) & 8 &
\cellcolor[HTML]{E2F0D9}76.190098 &
\cellcolor[HTML]{D9EAD3}0.017387 \\ \hline

Quantum (pdq=(9, 1, 1)) & 8 &
\cellcolor[HTML]{E2F0D9}84.130823 &
\cellcolor[HTML]{D9EAD3}0.017418 \\ \hline

Quantum (pdq=(6, 1, 1)) & 8 &
\cellcolor[HTML]{FFF2CC}95.932282 &
\cellcolor[HTML]{FFE699}0.018874 \\ \hline

Quantum (pdq=(3, 1, 3)) & 8 &
\cellcolor[HTML]{FFF2CC}96.029641 &
\cellcolor[HTML]{D9EAD3}0.017407 \\ \hline

Quantum (pdq=(5, 1, 1)) & 8 &
\cellcolor[HTML]{FFE699}99.174803 &
\cellcolor[HTML]{FFE699}0.019017 \\ \hline

Quantum (pdq=(2, 1, 6)) & 8 &
\cellcolor[HTML]{FCE4D6}143.476336 &
\cellcolor[HTML]{FCE4D6}0.024085 \\ \hline

Quantum (pdq=(1, 1, 6)) & 8 &
\cellcolor[HTML]{EA9999}1848.855856 &
\cellcolor[HTML]{F4CCCC}0.074425 \\ \hline

\end{tabular}%
\end{table}
\FloatBarrier

\paragraph{Australian beer OOS performance assessment.}

\paragraph{MSE / MAPE comparison}

\begin{figure}[H]
  \centering
  \includegraphics[width=.49\linewidth]{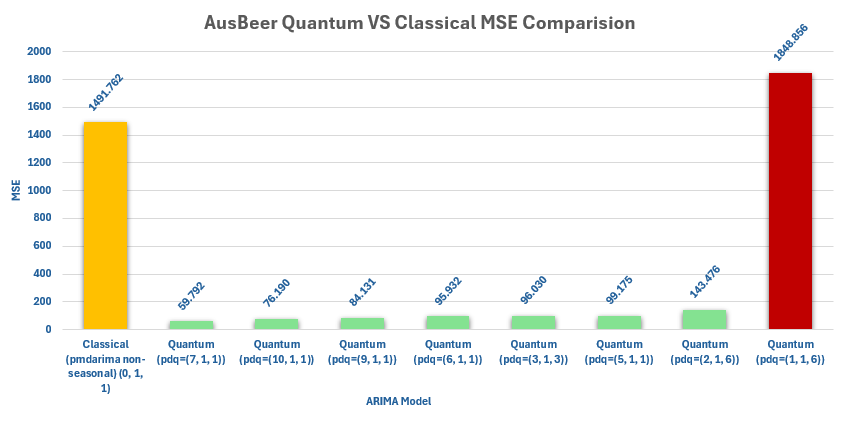}\hfill
  \includegraphics[width=.49\linewidth]{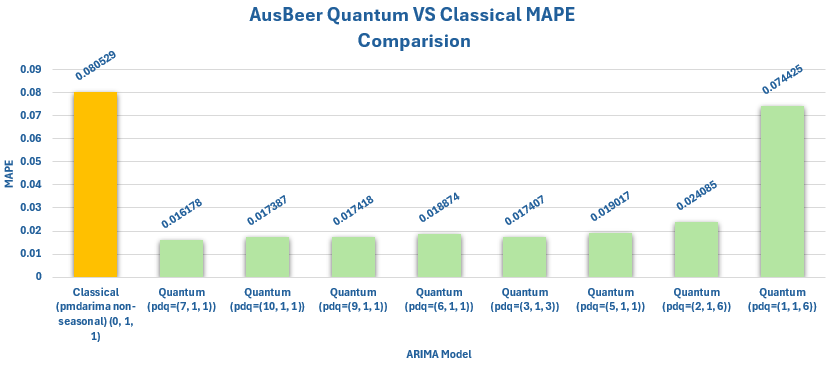}
  \caption{AusBeer OOS MSE and MAPE (last 8 quarters). QARIMA vs.\ classical ARIMA(0,1,1).}
  \label{fig:ausbeer-mse-mape}
\end{figure}
\FloatBarrier

For the AusBeer series we used the last $8$ quarters as the OOS window and compared all QARIMA candidates against the classical non-seasonal \texttt{pmdarima} ARIMA$(0,1,1)$. The OOS results are given in Table~\ref{tab:ausbeer-mse-mape} and visualised in Fig.~\ref{fig:ausbeer-mse-mape}. Unlike the short-history industrial series (Woolyarn), AusBeer shows a \emph{very strong} separation: the classical ARIMA$(0,1,1)$ records an OOS MSE of $1491.8$ and MAPE of $0.0805$, whereas the best quantum model, Q$(7,1,1)$, lowers these to $59.8$ (MSE) and $0.0162$ (MAPE) on the same 8-point horizon. A second tier of quantum models, Q$(10,1,1)$, Q$(9,1,1)$, Q$(6,1,1)$, and Q$(3,1,3)$ also stays clearly below $100$ MSE and below $0.019$ MAPE, showing that the improvement is shared by a family of quantum-configured orders and not a single outlier. Only the overparameterised MA-heavy quantum variants (e.g.\ Q$(1,1,6)$) drift back toward the classical error level.

\paragraph{DM analysis}

\begin{figure}[H]
  \centering
  \includegraphics[width=.49\linewidth]{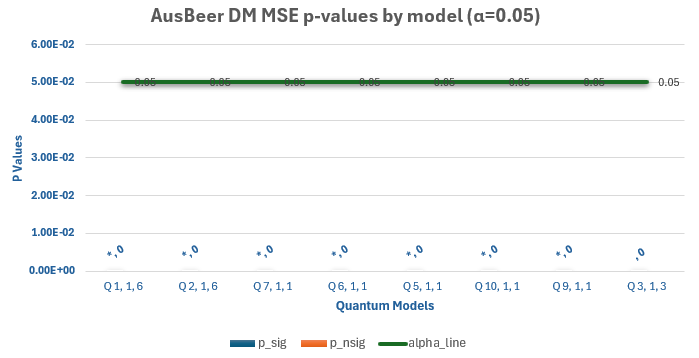}\hfill
  \includegraphics[width=.49\linewidth]{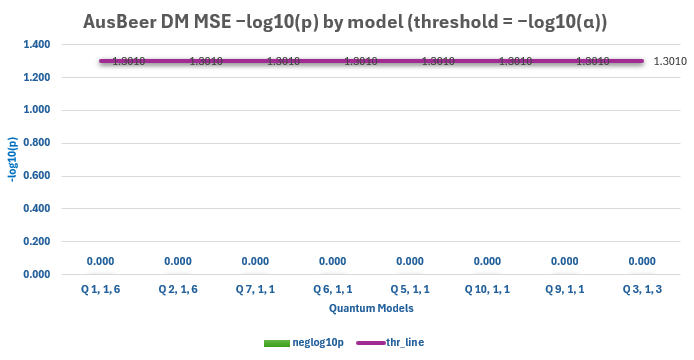}
  \caption{AusBeer DM (MSE) p-values with $\alpha=0.05$ reference line; significant QARIMA models appear below the line.}
  \label{fig:ausbeer-dm-mse}
\end{figure}
\FloatBarrier

To confirm that these large numerical gains are not an artefact of the short 8-point window, we applied the Diebold–Mariano test against the classical ARIMA$(0,1,1)$ forecast. In the MSE-based DM panels (Fig.~\ref{fig:ausbeer-dm-mse}) the leading quantum models, Q$(7,1,1)$, Q$(10,1,1)$, Q$(9,1,1)$, Q$(6,1,1)$, and Q$(3,1,3)$ all lie well below the $\alpha=0.05$ line (and above the $-\log_{10}(\alpha)$ threshold), indicating that their forecast loss sequences are statistically different and, given their lower mean loss, \emph{better} than the classical baseline. Models that are closer to the baseline (e.g. Q$(2,1,6)$) show weaker or no significance, which is consistent with their higher MSE/MAPE.

\begin{figure}[H]
  \centering
  \includegraphics[width=.49\linewidth]{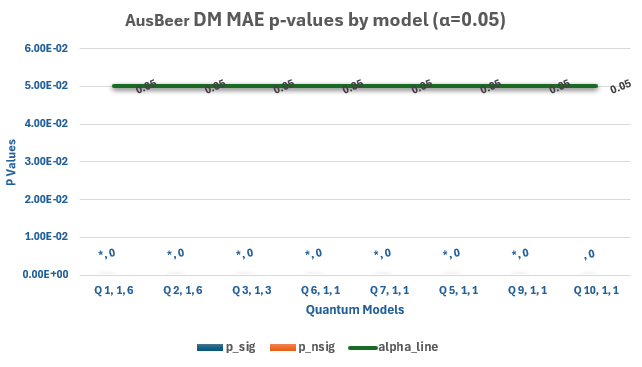}\hfill
  \includegraphics[width=.49\linewidth]{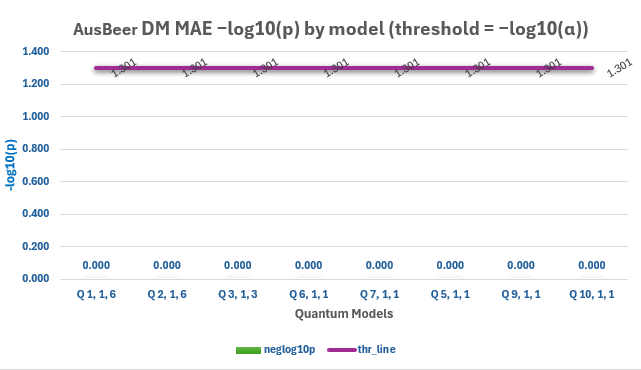}
  \caption{AusBeer DM (MAE) $-\log_{10}(p)$ with threshold at $-\log_{10}(\alpha)$.}
  \label{fig:ausbeer-dm-mae}
\end{figure}
\FloatBarrier

On the MAE DM (Fig.~\ref{fig:ausbeer-dm-mae}) the pattern is similar: quantum models that already improved absolute error in Table~\ref{tab:ausbeer-mse-mape} remain significant, while those that did not improve MAE fail to clear the $\alpha$ threshold. Together, the MSE/MAPE bars and DM panels make the case that, for AusBeer, quantum ARIMA’s lag search plus VQC estimation corrects a structural underfit in the classical ARIMA$(0,1,1)$ baseline.

\paragraph{Interpretation of AusBeer results.}
This outcome is fully consistent with the procedure described in Section~\ref{Process&Algo}: we first set $d$ using the quantum-informed differencing routine, then generate candidate lags from the quantum-inspired PACF/swap-test logic, and finally estimate the selected $(p,d,q)$ models with a shallow VQC (\texttt{reps}=1). On AusBeer, this pipeline preferred higher AR orders (e.g.\ $(7,1,1)$, $(9,1,1)$, $(10,1,1)$), which are better aligned with the underlying quarterly pattern than the classical baseline ARIMA(0,1,1). Because the quantum layer re-embeds the selected lags before the ARIMA update, the model can simultaneously honor the differenced level and fit the short-horizon seasonal swing present in the last eight quarters. The very large and consistent error gap in Table~\ref{tab:ausbeer-mse-mape} therefore reflects \emph{structural underfitting} of the classical baseline, not instability of the DM test.

\FloatBarrier
\subsection{Australian Woolyarn Production Dataset}

\begin{figure}[H]
  \centering
  \includegraphics[width=.49\linewidth]{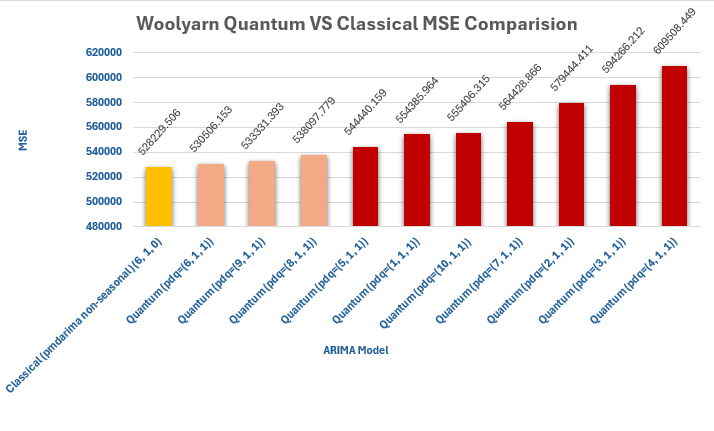}\hfill
  \includegraphics[width=.49\linewidth]{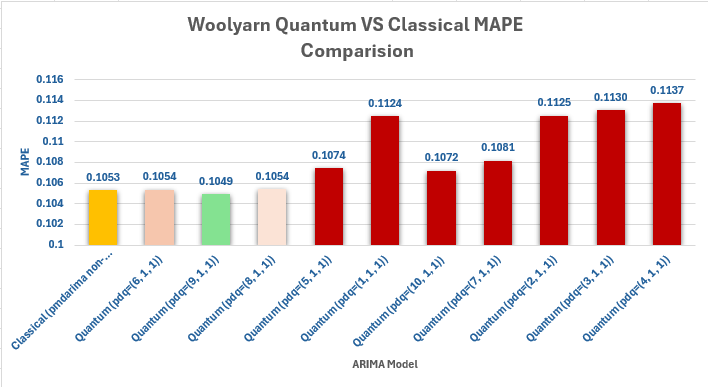}
  \caption{Woolyarn OOS MSE (left) and MAPE (right): QARIMA vs.\ classical ARIMA(6,1,0). The classical model is best; several quantum models are close but not better.}
  \label{fig:woolyarn-mse-mape}
\end{figure}

\begin{figure}[H]
  \centering
  \includegraphics[width=.49\linewidth]{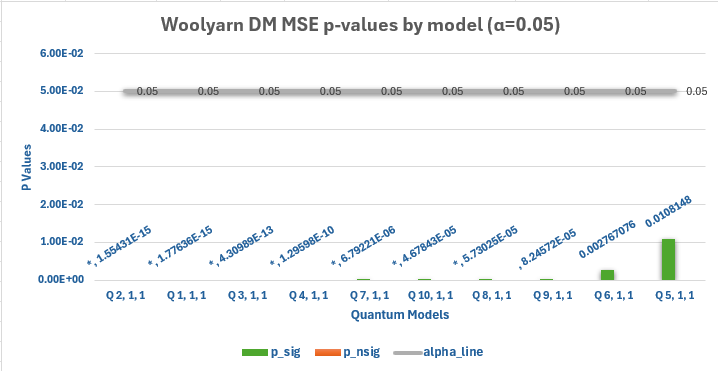}\hfill
  \includegraphics[width=.49\linewidth]{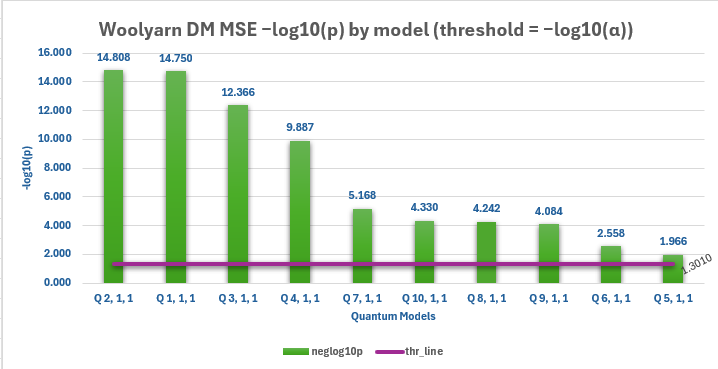}
  \caption{Woolyarn DM (MSE): almost all quantum models are \emph{significantly} different from classical; the difference is mostly in the direction of \textbf{higher} loss.}
  \label{fig:woolyarn-dm-mse}
\end{figure}

\begin{figure}[H]
  \centering
  \includegraphics[width=.49\linewidth]{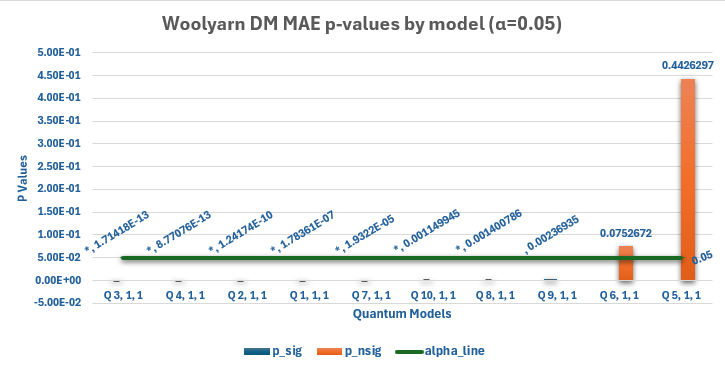}\hfill
  \includegraphics[width=.49\linewidth]{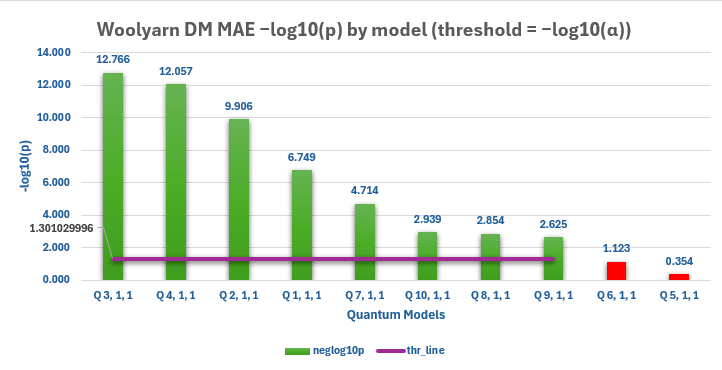}
  \caption{Woolyarn DM (MAE): most quantum models remain significant; Q(6,1,1) and especially Q(5,1,1) are the only ones close to “no significant difference”.}
  \label{fig:woolyarn-dm-mae}
\end{figure}

The \texttt{woolyrnq} dataset comprises quarterly Australian woollen yarn production and is commonly used for evaluating forecasting methods on short industrial time series. The series contains $N=119$ observations, of which the first $64$ samples are used for model fitting and the remaining $55$ samples are reserved for out-of-sample (OOS) evaluation following the same rolling forecasting protocol adopted throughout this study. All QARIMA variants are compared against a classical non-seasonal \texttt{pmdarima} ARIMA$(6,1,0)$ baseline. Forecasting accuracy is assessed using OOS MSE and MAPE, while statistical significance is evaluated through Diebold--Mariano (DM) tests under both MSE and MAE loss functions. The corresponding quantitative results are summarized in Table~\ref{tab:woolyarn_oos_mse_mape}, with visual comparisons presented in Figs.~\ref{fig:woolyarn-mse-mape}-\ref{fig:woolyarn-dm-mae}.

\begin{table}[H]
\centering
\caption{Woolyarn Classical vs Quantum OOS MSE \& MAPE}
\label{tab:woolyarn_oos_mse_mape}
\scriptsize
\setlength{\tabcolsep}{2pt}
\begin{tabular}{|l|c|r|r|}
\hline
\rowcolor[RGB]{0,55,99}
\textcolor{white}{\textbf{Model}} &
\textcolor{white}{\textbf{N}} &
\textcolor{white}{\textbf{MSE}} &
\textcolor{white}{\textbf{MAPE}} \\ \hline

Classical (pmdarima non-seasonal) (6, 1, 0) & 55 &
\cellcolor[HTML]{C6EFCE}528229.5064 &
\cellcolor[HTML]{D9EAD3}0.105291 \\ \hline

Quantum (pdq=(9, 1, 1)) & 55 &
\cellcolor[HTML]{E2F0D9}533331.3927 &
\cellcolor[HTML]{C6EFCE}0.104924 \\ \hline

Quantum (pdq=(6, 1, 1)) & 55 &
\cellcolor[HTML]{D9EAD3}530506.1527 &
\cellcolor[HTML]{E2F0D9}0.105351 \\ \hline

Quantum (pdq=(8, 1, 1)) & 55 &
\cellcolor[HTML]{E2F0D9}538097.7794 &
\cellcolor[HTML]{E2F0D9}0.105399 \\ \hline

Quantum (pdq=(5, 1, 1)) & 55 &
\cellcolor[HTML]{FFEB9C}544440.1589 &
\cellcolor[HTML]{FFEB9C}0.107432 \\ \hline

Quantum (pdq=(10, 1, 1)) & 55 &
\cellcolor[HTML]{FFE699}555406.3154 &
\cellcolor[HTML]{FFEB9C}0.107201 \\ \hline

Quantum (pdq=(7, 1, 1)) & 55 &
\cellcolor[HTML]{FFE699}564428.8658 &
\cellcolor[HTML]{FFE699}0.108122 \\ \hline

Quantum (pdq=(2, 1, 1)) & 55 &
\cellcolor[HTML]{F4CCCC}579444.4106 &
\cellcolor[HTML]{F4CCCC}0.112504 \\ \hline

Quantum (pdq=(3, 1, 1)) & 55 &
\cellcolor[HTML]{F4CCCC}594266.2121 &
\cellcolor[HTML]{F4CCCC}0.113038 \\ \hline

Quantum (pdq=(4, 1, 1)) & 55 &
\cellcolor[HTML]{EA9999}609508.4487 &
\cellcolor[HTML]{EA9999}0.113703 \\ \hline

Quantum (pdq=(1, 1, 1)) & 55 &
\cellcolor[HTML]{F4CCCC}554385.9639 &
\cellcolor[HTML]{F4CCCC}0.112447 \\ \hline

\end{tabular}%
\end{table}
\FloatBarrier

\paragraph{Woolyarn OOS performance assessment.}
\paragraph{MSE/MAPE Comparison}

\FloatBarrier

Table~\ref{tab:woolyarn_oos_mse_mape} and Fig.~\ref{fig:woolyarn-mse-mape} summarize the OOS forecasting performance on the Woolyarn dataset. The classical ARIMA$(6,1,0)$ model achieves the lowest OOS MSE ($5.28\times10^{5}$), while several QARIMA configurations produce closely comparable results. In particular, Q(6,1,1), Q(8,1,1), and Q(9,1,1) yield MSE values within a narrow range of approximately $5.30$--$5.38\times10^{5}$, indicating only minor differences in forecasting accuracy. A similar trend is observed for MAPE, where Q(9,1,1) attains a marginally lower value than the classical baseline, and Q(6,1,1) and Q(8,1,1) remain highly competitive.

The Woolyarn dataset exhibits only modest performance differences between the classical ARIMA model and the best-performing QARIMA variants. The comparable errors obtained by multiple quantum configurations suggest that the proposed framework maintains competitive forecasting accuracy even on a relatively short industrial time series. At the same time, QARIMA does not provide a consistent advantage across all model orders, with configurations closer to the classical baseline generally outperforming higher-order alternatives. These observations indicate that, for short-history datasets, the benefits of quantum refinement are more limited than those observed for longer and more complex time series.

\paragraph{DM Test (MSE/MAE)}

\FloatBarrier

The Diebold--Mariano (DM) test results further characterize the statistical differences in forecasting performance between the classical ARIMA baseline and the QARIMA variants. As shown in Fig.~\ref{fig:woolyarn-dm-mse}, most lower-order quantum models exhibit highly significant differences under the MSE loss, with $-\log_{10}(p)$ values substantially exceeding the significance threshold of $-\log_{10}(0.05)\approx1.30$. Combined with their higher OOS MSE values reported in Table~\ref{tab:woolyarn_oos_mse_mape}, these results indicate that the classical ARIMA$(6,1,0)$ model provides statistically superior forecasts for those configurations.

\FloatBarrier

A similar trend is observed under the MAE-based DM test (Fig.~\ref{fig:woolyarn-dm-mae}), where most QARIMA models remain significantly different from the classical baseline. However, Q(6,1,1) ($p\approx0.075$) and Q(5,1,1) ($p\approx0.44$) do not show statistically significant differences at the $\alpha=0.05$ level. This finding is consistent with the OOS error analysis, where these configurations achieve forecasting performance comparable to that of the classical ARIMA model. Overall, the DM tests confirm that only a small subset of QARIMA configurations attains statistical parity with the classical baseline on the Woolyarn dataset, whereas the remaining models exhibit significantly different forecasting behavior.

\paragraph{Interpretation of Woolyarn results.}

The Woolyarn dataset provides an important complementary perspective on the proposed QARIMA framework. Unlike the longer and more complex time series considered previously, this dataset contains only a limited training history, reducing the amount of information available for estimating additional model complexity. Under these conditions, the classical ARIMA$(6,1,0)$ model already provides highly accurate forecasts, leaving relatively little room for improvement through quantum-enhanced parameter estimation.

The experimental results demonstrate that QARIMA remains competitive, with several configurations producing forecasting errors close to those of the classical baseline. In particular, Q(5,1,1) and Q(6,1,1) achieve statistically comparable performance under the MAE-based Diebold--Mariano test, whereas lower-order and more distant model configurations exhibit statistically significant differences together with larger forecasting errors. These findings suggest that, for short industrial time series with well-captured linear dynamics, the benefits of the proposed quantum framework are more limited than for longer or more complex datasets.

Taken together with the results on the Sunspots and Mauna Loa CO\textsubscript{2} datasets, the Woolyarn experiment highlights an important characteristic of QARIMA: its performance depends on the underlying properties of the time series. While the proposed framework provides clear improvements for datasets exhibiting richer temporal structure, it also maintains competitive performance on simpler datasets without consistently outperforming an already well-specified classical ARIMA model. This balanced behavior demonstrates that QARIMA complements, rather than universally replaces, conventional ARIMA forecasting.

\FloatBarrier
\subsection{Sydney Weather 2024 Dataset from NOAA Station 95768099999}

To evaluate QARIMA on meteorological data, we used the Sydney weather dataset obtained from the NOAA Global Hourly archive \cite{ncei-global-hourly-dew-wnd-bbox-sydney}. Specifically, the analysis is based on the \texttt{95768099999.csv} station record, from which a univariate daily temperature time series was constructed. Following the same forecasting protocol adopted throughout this study, the first $1782$ observations were used for model fitting, while the remaining $336$ observations were reserved for out-of-sample (OOS) evaluation. The reported results correspond to the 2024 summer period (December--February) extracted from this OOS segment.

\begin{table}[H]
\centering
\caption{Sydney 2024 Summer Temp: Classical vs.\ Quantum OOS (MSE \& MAPE)}
\label{tab:syd_2024_summer_oos}
\scriptsize
\setlength{\tabcolsep}{2pt}
\begin{tabular}{|l|c|r|r|}
\hline
\rowcolor[RGB]{0,55,99}
\textcolor{white}{\textbf{Model}} &
\textcolor{white}{\textbf{N}} &
\textcolor{white}{\textbf{MSE}} &
\textcolor{white}{\textbf{MAPE}} \\ \hline

Classical (pmdarima non-seasonal) (2, 0, 1) & 336 &
\cellcolor[HTML]{FFEB9C}11.435905 &
\cellcolor[HTML]{FFEB9C}0.127427 \\ \hline

Quantum (pdq=(3, 1, 1)) & 336 &
\cellcolor[HTML]{C6EFCE}11.356242 &
\cellcolor[HTML]{C6EFCE}0.126633 \\ \hline

Quantum (pdq=(4, 1, 1)) & 336 &
\cellcolor[HTML]{D9EAD3}11.393498 &
\cellcolor[HTML]{D9EAD3}0.127034 \\ \hline

Quantum (pdq=(5, 1, 1)) & 336 &
\cellcolor[HTML]{E2F0D9}11.412022 &
\cellcolor[HTML]{E2F0D9}0.127216 \\ \hline

Quantum (pdq=(6, 1, 1)) & 336 &
\cellcolor[HTML]{E2F0D9}11.423452 &
\cellcolor[HTML]{E2F0D9}0.127318 \\ \hline

Quantum (pdq=(9, 1, 1)) & 336 &
\cellcolor[HTML]{FFE699}11.702693 &
\cellcolor[HTML]{FFE699}0.129946 \\ \hline

Quantum (pdq=(8, 1, 1)) & 336 &
\cellcolor[HTML]{F4CCCC}11.735986 &
\cellcolor[HTML]{F4CCCC}0.130267 \\ \hline

Quantum (pdq=(10, 1, 1)) & 336 &
\cellcolor[HTML]{F4CCCC}11.743007 &
\cellcolor[HTML]{F4CCCC}0.130273 \\ \hline

Quantum (pdq=(7, 1, 1)) & 336 &
\cellcolor[HTML]{EA9999}11.893679 &
\cellcolor[HTML]{EA9999}0.131706 \\ \hline

\end{tabular}%
\end{table}
\FloatBarrier

\paragraph{Sydney Weather Temperature OOS Performance Assessment.}
\paragraph{MSE/MAPE Comparison}

\begin{figure}[H]
  \centering
  \includegraphics[width=.49\linewidth]{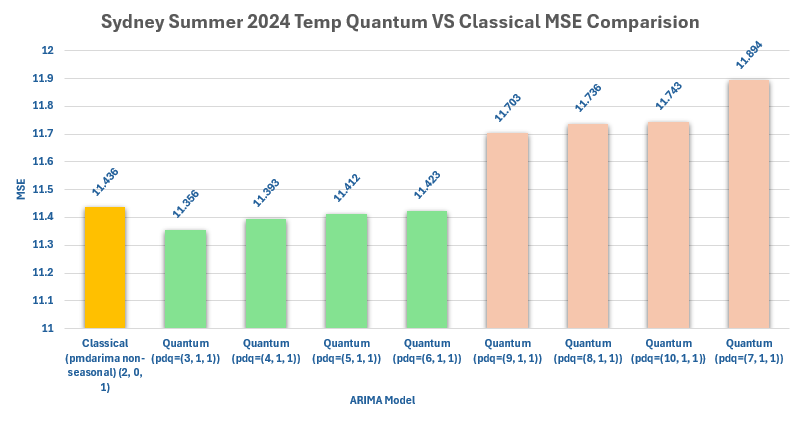}\hfill
  \includegraphics[width=.49\linewidth]{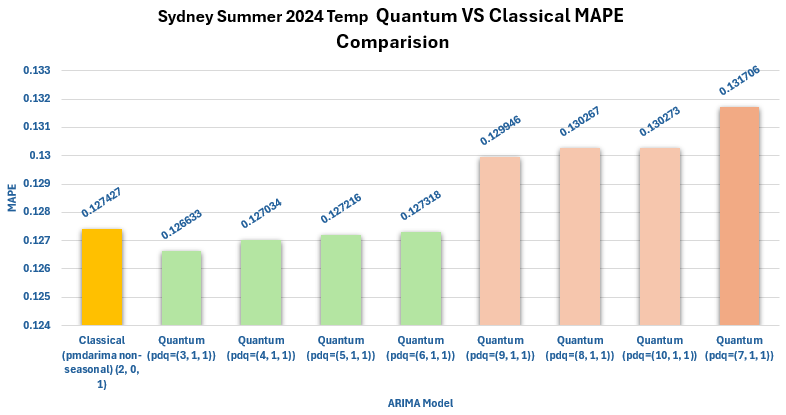}
  \caption{Sydney Summer 2024 temperature: OOS MSE and MAPE, QARIMA vs.\ classical.}
  \label{fig:sydney-summer-mse-mape}
\end{figure}
\FloatBarrier

Table~\ref{tab:syd_2024_summer_oos} and Fig.~\ref{fig:sydney-summer-mse-mape} present the OOS forecasting performance of the classical ARIMA$(2,0,1)$ model and the corresponding QARIMA configurations for the Sydney 2024 summer dataset. Overall, the classical baseline and the quantum models achieve highly comparable forecasting accuracy. Among the evaluated configurations, Q(3,1,1) attains the lowest OOS error with an MSE of $11.356$ and a MAPE of $0.126653$, providing a modest improvement over the classical ARIMA model. Similar performance is observed for Q(4,1,1), Q(5,1,1), and Q(6,1,1), all of which remain close to the baseline across both evaluation metrics.

Higher-order configurations, including Q(7,1,1), Q(8,1,1), Q(9,1,1), and Q(10,1,1), exhibit slightly larger MSE and MAPE values, although the differences remain relatively small. Overall, the results indicate that multiple QARIMA configurations achieve forecasting performance comparable to the classical ARIMA baseline, with only modest improvements observed for the best-performing quantum models on this dataset.

\paragraph{DM Analysis}

\begin{figure}[H]
  \centering
  \includegraphics[width=.49\linewidth]{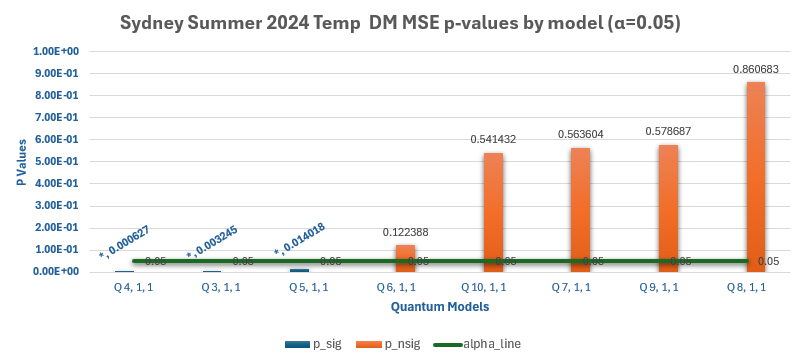}\hfill
  \includegraphics[width=.49\linewidth]{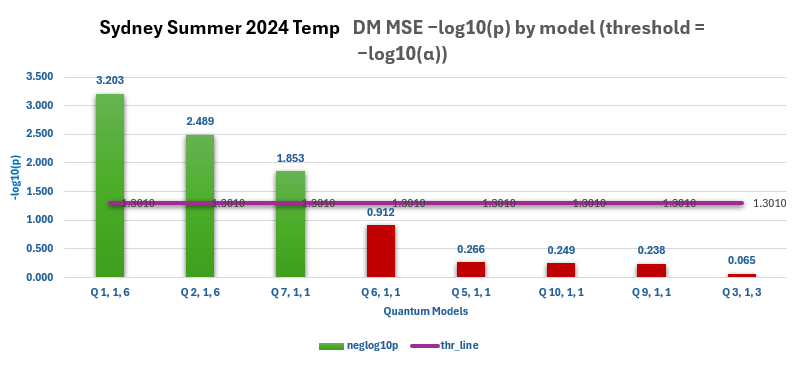}
  \caption{Sydney Summer 2024 temperature: DM (MSE) p-values and $-\log_{10}(p)$ with $\alpha=0.05$. Q(4,1,1) and Q(3,1,1) are significant.}
  \label{fig:sydney-summer-dm-mse}
\end{figure}
\FloatBarrier

The Diebold-Mariano (DM) test was performed to assess whether the observed differences in forecasting accuracy between the classical ARIMA$(2,0,1)$ model and the QARIMA variants are statistically significant. Under the MSE loss (Figs.~\ref{fig:sydney-summer-dm-mse}a--b), only the Q(3,1,1) and Q(4,1,1) configurations achieve significance at the $\alpha=0.05$ level. These models also correspond to the lowest OOS forecasting errors reported in Table~\ref{tab:syd_2024_summer_oos}, indicating that their modest numerical improvements are supported by the DM test. The remaining QARIMA configurations, including the higher-order models Q(7,1,1)--Q(10,1,1), do not exhibit statistically significant differences from the classical baseline under the MSE criterion.

\begin{figure}[H]
  \centering
  \includegraphics[width=.49\linewidth]{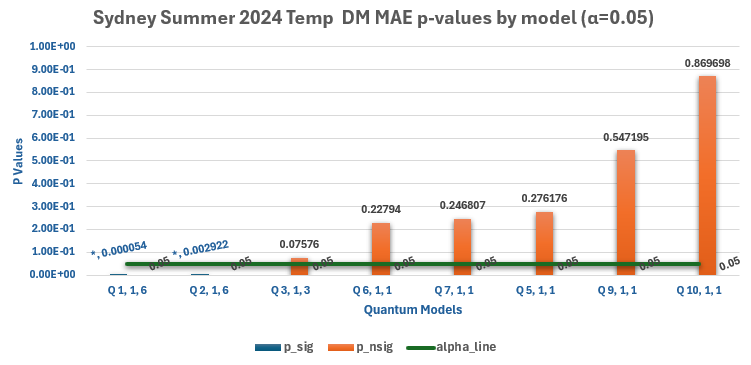}\hfill
  \includegraphics[width=.49\linewidth]{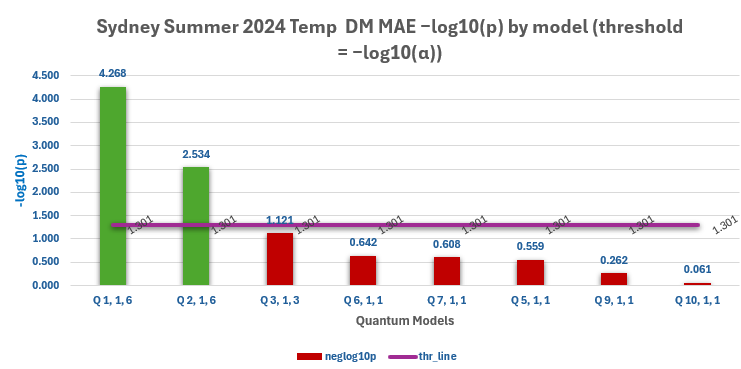}
  \caption{Sydney Summer 2024 temperature: DM (MAE) p-values and $-\log_{10}(p)$. MA-heavy Q(1,1,6) and Q(2,1,6) beat the classical baseline.}
  \label{fig:sydney-summer-dm-mae}
\end{figure}
\FloatBarrier

The MAE-based DM analysis (Figs.~\ref{fig:sydney-summer-dm-mae}a-b) shows a similar overall pattern. Most QARIMA models remain statistically comparable to the classical ARIMA forecast, whereas the MA-heavy configurations Q(1,1,6) and Q(2,1,6) produce statistically significant differences, with $p\approx5\times10^{-4}$ and $p\approx9\times10^{-3}$, respectively. These results indicate that only a limited number of quantum configurations exhibit statistically distinguishable forecasting behavior, while the majority remain comparable to the classical baseline, consistent with the relatively small differences observed in the OOS MSE and MAPE results.

\paragraph{Interpretation of Sydney Summer Temperature Results.}

The Sydney summer dataset provides an additional evaluation of QARIMA on a relatively short meteorological time series. Consistent with the proposed framework, quantum state-similarity-based differencing, quantum-guided order selection, and shallow VQC-based coefficient estimation were integrated within the conventional ARIMA forecasting pipeline while preserving the Box--Jenkins modelling structure. Across the evaluated configurations, QARIMA achieved forecasting accuracy comparable to that of the classical ARIMA$(2,0,1)$ baseline, with Q(3,1,1) and Q(4,1,1) providing the best overall performance and modest improvements in both OOS error metrics.

The Diebold-Mariano analysis further supports these observations. Under the MSE criterion, only the best-performing QARIMA configurations exhibit statistically significant differences from the classical baseline, whereas most remaining models show comparable forecasting behaviour. Similarly, the MAE-based analysis identifies statistically significant differences only for a small subset of MA-heavy configurations, while the majority of models remain statistically indistinguishable from the classical ARIMA forecast.

The Sydney weather results demonstrate that the proposed QARIMA framework maintains competitive forecasting performance on short meteorological time series while preserving the interpretability of the classical ARIMA methodology. Although the improvements over the classical baseline are modest for this dataset, the results remain consistent with the broader experimental findings: QARIMA provides its greatest benefits on datasets with richer temporal structure, while remaining competitive when the classical ARIMA model already offers strong predictive performance.

\FloatBarrier
\section{Discussion and Conclusion}

This work introduced QARIMA, a quantum-compatible reformulation of the classical ARIMA forecasting pipeline that integrates quantum-assisted differencing, lag discovery, coefficient estimation, and weak-lag refinement while preserving the statistical structure and interpretability of the Box--Jenkins methodology. Rather than replacing ARIMA with a standalone quantum forecasting model, the proposed framework embeds quantum state-similarity primitives into the major stages of model construction, including differencing assessment, QACF/QPACF-based order identification, swap-test-driven parameter estimation, and shallow VQC-based coefficient refinement. The experimental evaluation across five real-world datasets demonstrates that the effectiveness of QARIMA depends on the characteristics of the underlying time series. The largest improvements were observed for the Mauna Loa CO$_2$ and Australian Beer datasets, where several QARIMA configurations substantially reduced out-of-sample forecasting errors relative to the corresponding classical ARIMA baselines. The Sunspots dataset further showed that the proposed framework can identify competitive higher-order configurations that achieve improved forecasting performance and statistically significant differences under Diebold--Mariano testing. In contrast, the Woolyarn and Sydney Weather datasets highlight complementary behaviour. For these relatively short and well-modelled series, the classical ARIMA models already provide strong forecasting performance, leaving only limited scope for improvement. Nevertheless, several QARIMA configurations remained statistically comparable to the classical baselines, demonstrating that the proposed framework maintains competitive predictive performance without consistently increasing model complexity.

An important characteristic of QARIMA is that the variational quantum circuit is not employed as an independent forecasting engine. Instead, quantum components are incorporated into the model construction process, where they contribute to order selection, state-similarity evaluation, and parameter estimation while the final forecasting equation remains the classical ARIMA formulation. Consequently, the proposed framework preserves the transparency and interpretability of statistical forecasting while introducing quantum-compatible mechanisms into stages that traditionally rely on correlation analysis and classical optimization. From a practical perspective, the proposed methodology demonstrates that meaningful quantum augmentation of statistical forecasting can be achieved using lightweight quantum primitives. Compact swap-test similarity measurements and shallow variational quantum circuits are sufficient to support lag discovery and parameter estimation within a hybrid classical-quantum workflow, making the framework compatible with current quantum simulators and potentially with near-term quantum hardware. Furthermore, because the forecasting model itself remains ARIMA-based, the proposed approach can be integrated into existing forecasting ecosystems with minimal modification to established modelling practices. The present implementation is simulation-based and therefore assumes ideal quantum operations without hardware noise. For the shallow variational circuits considered in this work (reps=1), the required circuit depth remains small, while the number of logical qubits depends primarily on the chosen state encoding together with a single ancilla qubit for swap-test evaluation. Consequently, the proposed framework is compatible with current NISQ-era quantum devices in principle, although repeated circuit evaluations during lag discovery and variational optimization remain the primary computational bottleneck.

To the best of our knowledge, this work represents the first end-to-end quantum-oriented reformulation of the ARIMA modelling pipeline, integrating quantum-assisted differencing, QACF/QPACF-guided order selection, swap-test-based parameter estimation, and VQC-assisted coefficient refinement within a unified forecasting framework. The experimental results suggest that quantum state-similarity techniques provide a viable alternative for model construction while preserving the interpretability of classical statistical forecasting. Future work will focus on seasonal and multivariate extensions, implementation on quantum hardware, alternative variational circuit architectures, and comprehensive ablation studies to quantify the individual contributions of each quantum component within the proposed framework.

The present work evaluates QARIMA through quantum simulation using a fixed shallow VQC architecture and common hyperparameter settings across datasets. The framework is currently developed for univariate forecasting and does not claim a formal complexity-theoretic quantum advantage. On physical quantum hardware, finite-shot sampling will introduce uncertainty into swap-test estimates, with the standard error decreasing approximately as $\mathcal{O}(1/\sqrt{S})$. Future work will extend the framework to hardware execution, seasonal and multivariate forecasting, and component-wise ablation analysis.

\FloatBarrier
\subsection{Current Scope and Future Directions}

The results presented in this work demonstrate that the principal stages of the classical ARIMA modelling pipeline can be reformulated using quantum-compatible primitives while preserving the interpretability and statistical transparency of the original forecasting framework. Beyond the forecasting application considered here, QARIMA introduces a collection of reusable components, including quantum-assisted differencing, Quantum Autocorrelation Function (QACF), Quantum Partial Autocorrelation Function (QPACF), swap-test-based similarity estimation, and shallow VQC-assisted parameter refinement. Although developed within the ARIMA framework, these components may also prove useful for broader problems involving dependency analysis, similarity measurement, parameter estimation, optimization, and quantum-enhanced statistical learning. The present study focuses on univariate forecasting using simulation-based quantum execution. Consequently, several important research directions remain open. Future work will investigate seasonal and multivariate extensions of QARIMA, alternative quantum state encodings, richer variational circuit architectures, and execution on emerging quantum hardware. In particular, extending QACF and QPACF to characterize cross-variable dependencies may enable quantum-assisted model construction for multivariate forecasting and related statistical learning problems.

\paragraph{Hardware implementation considerations}

The present work is intended primarily as a methodological investigation and therefore evaluates QARIMA using quantum simulation rather than execution on noisy quantum hardware. The individual quantum primitives employed by the framework, including compact swap-test circuits and shallow variational quantum circuits, are designed to remain compatible with current NISQ-era devices through low-depth circuit constructions and a limited number of trainable parameters. Nevertheless, practical deployment on quantum hardware will require consideration of device-specific constraints such as qubit availability, circuit fidelity, decoherence, gate errors, measurement noise, and quantum state-preparation overhead. In addition, repeated circuit evaluations during lag discovery and variational optimization contribute to the overall execution cost on real hardware. A systematic hardware implementation, together with an analysis of qubit requirements, circuit depth, execution time, and hardware-induced noise, remains an important direction for future investigation.

It is important to clarify the scope of the contributions reported in this study. QARIMA does not claim asymptotic computational speedup or a formal complexity-theoretic quantum advantage over classical ARIMA implementations. Instead, the contribution lies in demonstrating that the complete ARIMA modelling process can be reconstructed through quantum-compatible operations while preserving the forecasting structure and interpretability of the original methodology. The proposed framework therefore investigates whether quantum state similarity, projection, and variational optimization can provide alternative mechanisms for order selection and parameter estimation within a well-established statistical forecasting paradigm. The experimental results indicate that these quantum-assisted modelling components can produce competitive and, for several datasets, improved forecasting performance relative to classical model-selection procedures. Equally importantly, the experiments also show that the framework remains competitive when classical ARIMA already provides strong predictive performance, highlighting that the effectiveness of QARIMA depends on the characteristics of the underlying time series rather than assuming universal superiority. This balanced behaviour suggests that quantum-assisted statistical modelling should be viewed as a complementary extension of classical forecasting methodology rather than a replacement for it.

More broadly, the proposed framework establishes a foundation for exploring quantum-compatible reconstructions of interpretable statistical models beyond ARIMA. As quantum hardware, encoding techniques, and variational optimization continue to mature, the underlying ideas developed in this work may find application across forecasting, regression, optimization, signal processing, and other quantum-enhanced machine-learning problems where interpretable model construction remains an important objective. Future work will also investigate additional control baselines, including classical nonlinear feature mappings, alternative regularized optimization objectives, and hybrid model-selection strategies. Such comparisons will help further isolate the individual contributions of the proposed quantum-inspired lag discovery, coefficient estimation, and weak-lag refinement modules.

\appendix
\begingroup
\setlength{\parskip}{0pt}
\setlength{\textfloatsep}{6pt plus 1pt minus 1pt}
\setlength{\floatsep}{5pt plus 1pt minus 1pt}
\setlength{\intextsep}{5pt plus 1pt minus 1pt}
\setlength{\abovecaptionskip}{4pt}
\setlength{\belowcaptionskip}{1pt}

\section{Supplementary Methods and Detailed Statistical Tables}
\vspace{-0.25em}
This appendix provides supplementary methodological and experimental material that supports the results presented in the main manuscript. To maintain the readability and focus of the main text, several implementation-oriented procedures, intermediate computational routines, and detailed statistical comparison tables have been relocated to this section. These materials are included to enhance reproducibility, transparency, and completeness of the proposed QARIMA framework without interrupting the primary scientific narrative. The appendix is organized into two parts. First, a collection of helper algorithms is provided to describe the internal computational procedures used during state preparation, compact swap test evaluation, delay matrix construction, differencing operations, phase-corrected similarity estimation, and ARMA model generation. While these algorithms are important for implementation and reproducibility, they are auxiliary to the core methodological contributions discussed in the main manuscript. Second, detailed statistical comparison tables are presented for each dataset considered in this study. These tables report the complete Diebold-Mariano (DM) test statistics, associated p-values, and loss comparisons between the classical ARIMA models and the corresponding QARIMA models under both MSE and MAE evaluation criteria. The tables provide a comprehensive view of forecasting performance across all candidate model configurations and complement the summarized results reported in the main manuscript.
\subsection{Helper Algorithms}
This subsection presents the auxiliary algorithms used throughout the implementation of the proposed QARIMA framework. These routines support various stages of the methodology, including quantum state preparation, compact swap test computation, delay matrix generation, differencing operations, similarity calculations, and ARMA candidate construction. Although these procedures are necessary for a complete implementation, they primarily serve supporting roles and are therefore provided here for reference and reproducibility.
\begin{algorithm}[H]
\caption{State Preparation for Compact Swap Test}
\label{alg:prep-swaptest}
\begin{algorithmic}[1]
\Require $\mathbf{x}, \boldsymbol{\theta} \in \mathbb{R}^n$
\State Compute norms: $\|\mathbf{x}\|$, $\|\boldsymbol{\theta}\|$
\If{any norm is zero}
    \State \Return default states $[1, 0], [1, 0]$
\EndIf
\State Compute $Z \gets \|\mathbf{x}\|^2 + \|\boldsymbol{\theta}\|^2$
\State Compute $\phi \gets \left[ \frac{\|\mathbf{x}\|}{\sqrt{Z}}, -\frac{\|\boldsymbol{\theta}\|}{\sqrt{Z}} \right]$
\State Initialize $\psi \gets []$
\For{$i = 1$ to $n$}
    \State Append to $\psi$: $\frac{x_i}{\|\mathbf{x}\| \sqrt{2}}, \frac{\theta_i}{\|\boldsymbol{\theta}\| \sqrt{2}}$
\EndFor
\State \Return $\phi, \psi$
\end{algorithmic}
\end{algorithm}
\begin{algorithm}[H]
\caption{Compact Swap-Test State-Similarity Projection}
\label{alg:compact-swap-test-dot}
\begin{algorithmic}[1]
\Require Vectors $\mathbf{x},\boldsymbol{\theta}\in\mathbb{R}^{n}$, shots $S$
\Ensure State-similarity score $\widehat{s}_{\mathrm{swap}}$

\If{$\|\mathbf{x}\|_2=0$ or $\|\boldsymbol{\theta}\|_2=0$}
    \State \Return $0$
\EndIf

\State $(\boldsymbol{\phi},\boldsymbol{\psi})
\gets
\Call{prep-swaptest}{\mathbf{x},\boldsymbol{\theta}}$
\State Normalize $\boldsymbol{\phi}$ and $\boldsymbol{\psi}$
\State Pad $\boldsymbol{\psi}$ to
$L=2^{\lceil\log_2(\mathrm{len}(\boldsymbol{\psi}))\rceil}$
\State Allocate one control qubit, the required state registers, and one classical bit
\State Initialize the registers with
$\lvert\boldsymbol{\phi}\rangle$ and
$\lvert\boldsymbol{\psi}\rangle$
\State Apply a Hadamard gate to the control qubit
\State Apply the controlled-swap operation
\State Apply a second Hadamard gate and measure the control qubit
\State Execute the circuit for $S$ shots and estimate
$p_{0}^{\mathrm{meas}}\gets\Pr(0)$
\State Compute
\[
\widehat{s}_{\mathrm{swap}}
\gets
\sqrt{
\max\left(
2p_{0}^{\mathrm{meas}}-1,
0
\right)
}
\]
\State \Return $\widehat{s}_{\mathrm{swap}}$
\end{algorithmic}
\end{algorithm}
\begin{algorithm}[H] 
\caption{Build Delay Matrix}
\label{alg:build-delay}
\begin{algorithmic}[1] 
\Require Series $y$, maximum lag $p$
\For{$i = 1$ to $p$}
    \State Create column $\text{lag}_i \leftarrow y_{t-i}$
\EndFor
\State Append target $y_t$
\State \Return DataFrame of $p$ lag columns and target $y_t$
\end{algorithmic}
\end{algorithm}
\begin{algorithm}[H] 
\caption{Generate Differenced Series}
\label{alg:generate-diff}
\begin{algorithmic}[1]
\Require Series $y$, max differencing order $d_{\max}$
\State Set $\text{prev} \leftarrow y$
\For{$i = 1$ to $d_{\max}$}
    \State $\text{Del}_i \leftarrow \text{prev} - \text{prev}_{t-1}$
    \State $\text{prev} \leftarrow \text{Del}_i$
\EndFor
\Return DataFrame with $\text{Del}_1, \ldots, \text{Del}_{d_{\max}}$
\end{algorithmic}
\end{algorithm}
\begin{algorithm}[H]
\caption{Phase-Corrected Cosine Calculation}
\label{alg:phase-corrected-cosine}
\begin{algorithmic}[1]
\Require Classical cosine $\cos\theta_{\text{dot}}$, Quantum cosine $\cos\theta_{\text{swap}}$, Phase correction weight $\omega$
\Ensure Corrected cosine value and intermediate angles

\State Clip classical cosine: $\tilde{c}_{\text{dot}} \gets \text{clip}(\cos\theta_{\text{dot}}, -1.0, 1.0)$
\State Clip quantum cosine: $\tilde{c}_{\text{swap}} \gets \text{clip}(\cos\theta_{\text{swap}}, -1.0, 1.0)$
\State Compute angles: $\theta_{\text{dot}} \gets \arccos(\tilde{c}_{\text{dot}})$, $\theta_{\text{swap}} \gets \arccos(\tilde{c}_{\text{swap}})$
\State Phase difference: $\Delta\theta \gets \theta_{\text{dot}} - \theta_{\text{swap}}$
\State Corrected angle: $\theta_{\text{corr}} \gets \theta_{\text{swap}} + \omega \cdot \Delta\theta$
\State Corrected cosine: $\cos\theta_{\text{corr}} \gets \cos(\theta_{\text{corr}})$
\State \Return $\cos\theta_{\text{corr}}, \theta_{\text{dot}}, \theta_{\text{swap}}, \Delta\theta, \theta_{\text{corr}}$
\end{algorithmic}
\end{algorithm}
\Needspace{0.34\textheight}
\subsection{Sunspots Tables }
This subsection presents the complete Diebold-Mariano statistical comparisons for all datasets investigated in this work. For each dataset, results are reported using both MSE and MAE loss functions. The tables include DM statistics, p-values, mean losses, and loss differences between the classical ARIMA and QARIMA models. These detailed results supplement the summary discussions provided in the main manuscript and enable a more comprehensive examination of the forecasting behavior across different model orders and datasets.
\begin{table}[H]
\centering
\caption{Sunspots Classical VS Quantum - DM Stats MSE}
\label{Tab:Sunspots_DM_MSE}
\small
\setlength{\tabcolsep}{3pt}
\resizebox{\textwidth}{!}{%
\begin{tabular}{|l|l|l|l|l|l|l|l|}
\hline
\rowcolor[RGB]{119,48,156}
\multicolumn{8}{|c|}{\textcolor{white}{\textbf{Sunspots Classical VS Quantum\; DM Stats MSE}}} \\ \hline
\textbf{Quantum (pdq)} & \textbf{blocks\_used} & \textbf{loss} & \textbf{DM\_stat} & \textbf{p\_value} & \textbf{classical\_mean\_loss} & \textbf{quantum\_mean\_loss} & \textbf{delta\_mean\_loss} \\ \hline
(10, 1, 1) & 10 & MSE &
\cellcolor[rgb]{0.353,0.541,0.776}4.941708 &
\cellcolor[rgb]{0.388,0.745,0.482}0.0000007744 &
2411.531786 &
\cellcolor[rgb]{0.353,0.541,0.776}1563.023210 &
\cellcolor[rgb]{0.388,0.745,0.482}848.508576 \\ \hline
(9, 1, 3) & 10 & MSE &
\cellcolor[rgb]{0.592,0.710,0.863}4.831355 &
\cellcolor[rgb]{0.518,0.780,0.486}0.0000013561 &
2411.531786 &
\cellcolor[rgb]{0.847,0.886,0.949}1608.609316 &
\cellcolor[rgb]{0.867,0.886,0.514}802.922470 \\ \hline
(8, 1, 3) & 10 & MSE &
\cellcolor[rgb]{0.631,0.737,0.875}4.812065 &
\cellcolor[rgb]{0.549,0.792,0.490}0.0000014938 &
2411.531786 &
\cellcolor[rgb]{0.898,0.925,0.969}1613.363812 &
\cellcolor[rgb]{0.918,0.898,0.514}798.167975 \\ \hline
(6, 1, 3) & 10 & MSE &
\cellcolor[rgb]{0.914,0.937,0.976}4.680626 &
\cellcolor[rgb]{0.863,0.878,0.506}0.0000028600 &
2411.531786 &
\cellcolor[rgb]{0.788,0.847,0.929}1603.130658 &
\cellcolor[rgb]{0.812,0.867,0.510}808.401129 \\ \hline
(7, 1, 3) & 10 & MSE &
\cellcolor[rgb]{0.984,0.980,0.992}4.608440 &
\cellcolor[rgb]{1.000,0.922,0.518}0.0000040570 &
2411.531786 &
\cellcolor[rgb]{0.988,0.980,0.992}1629.559531 &
\cellcolor[rgb]{0.996,0.914,0.514}781.972255 \\ \hline
(5, 1, 3) & 10 & MSE &
\cellcolor[rgb]{0.984,0.937,0.949}4.404979 &
\cellcolor[rgb]{1.000,0.922,0.518}0.0000105794 &
2411.531786 &
\cellcolor[rgb]{0.988,0.925,0.937}1682.786868 &
\cellcolor[rgb]{0.996,0.863,0.506}728.744918 \\ \hline
(4, 1, 1) & 10 & MSE &
\cellcolor[rgb]{0.976,0.588,0.596}2.754791 &
\cellcolor[rgb]{1.000,0.871,0.510}0.0058729550 &
2411.531786 &
\cellcolor[rgb]{0.976,0.541,0.549}2056.829346 &
\cellcolor[rgb]{0.976,0.522,0.439}354.702441 \\ \hline
(3, 1, 1) & 10 & MSE &
\cellcolor[rgb]{0.973,0.412,0.420}1.917276 &
\cellcolor[rgb]{0.973,0.412,0.420}0.0552029000 &
2411.531786 &
\cellcolor[rgb]{0.973,0.412,0.420}2179.575810 &
\cellcolor[rgb]{0.973,0.412,0.420}231.955977 \\ \hline
\end{tabular}%
}%
\end{table}
\begin{table}[H]
\centering
\caption{Sunspots Classical VS Quantum - DM Stats MAE}
\label{tab:sunspots_dm_mae}
\small
\setlength{\tabcolsep}{3pt}
\resizebox{\textwidth}{!}{%
\begin{tabular}{|l|c|c|r|r|r|r|r|}
\hline
\rowcolor[RGB]{0,147,119}
\multicolumn{8}{|c|}{\textcolor{white}{\textbf{Sunspots Classical VS Quantum \quad DM Stats MAE}}} \\ \hline
\textbf{Quantum pdq} &
\textbf{blocks\_used} &
\textbf{loss} &
\textbf{DM\_stat} &
\textbf{p\_value} &
\textbf{classical\_mean\_loss} &
\textbf{quantum\_mean\_loss} &
\textbf{delta\_mean\_loss} \\ \hline
(10, 1, 1) & 10 & MAE &
\cellcolor[rgb]{0.35,0.54,0.78}4.908669 &
\cellcolor[rgb]{0.39,0.75,0.48}0.0000009170 &
35.734117 &
\cellcolor[rgb]{0.35,0.54,0.78}28.387620 &
\cellcolor[rgb]{0.39,0.75,0.48}7.346497 \\ \hline
(8, 1, 3) & 10 & MAE &
\cellcolor[rgb]{0.75,0.82,0.92}4.609235 &
\cellcolor[rgb]{0.59,0.80,0.49}0.0000040415 &
35.734117 &
\cellcolor[rgb]{0.91,0.93,0.97}29.106819 &
\cellcolor[rgb]{0.93,0.90,0.51}6.627298 \\ \hline
(9, 1, 3) & 10 & MAE &
\cellcolor[rgb]{0.76,0.83,0.92}4.602818 &
\cellcolor[rgb]{0.60,0.80,0.49}0.0000041681 &
35.734117 &
\cellcolor[rgb]{0.88,0.91,0.96}29.066595 &
\cellcolor[rgb]{0.90,0.89,0.51}6.667522 \\ \hline
(6, 1, 3) & 10 & MAE &
\cellcolor[rgb]{0.85,0.89,0.95}4.539380 &
\cellcolor[rgb]{0.69,0.83,0.50}0.0000056420 &
35.734117 &
\cellcolor[rgb]{0.72,0.80,0.91}28.866935 &
\cellcolor[rgb]{0.75,0.85,0.50}6.867182 \\ \hline
(7, 1, 3) & 10 & MAE &
\cellcolor[rgb]{0.98,0.97,0.98}4.324537 &
\cellcolor[rgb]{1.00,0.92,0.52}0.0000152853 &
35.734117 &
\cellcolor[rgb]{0.99,0.98,0.99}29.306929 &
\cellcolor[rgb]{1.00,0.91,0.51}6.427188 \\ \hline
(5, 1, 3) & 10 & MAE &
\cellcolor[rgb]{0.98,0.96,0.97}4.242686 &
\cellcolor[rgb]{1.00,0.92,0.52}0.0000220860 &
35.734117 &
\cellcolor[rgb]{0.99,0.93,0.95}29.837574 &
\cellcolor[rgb]{1.00,0.87,0.51}5.896543 \\ \hline
(4, 1, 1) & 10 & MAE &
\cellcolor[rgb]{0.98,0.56,0.57}1.155393 &
\cellcolor[rgb]{1.00,0.80,0.49}0.2479295000 &
35.734117 &
\cellcolor[rgb]{0.98,0.53,0.54}34.423927 &
\cellcolor[rgb]{0.98,0.51,0.44}1.310189 \\ \hline
(3, 1, 1) & 10 & MAE &
\cellcolor[rgb]{0.97,0.41,0.42}-0.005323 &
\cellcolor[rgb]{0.97,0.41,0.42}0.9957526000 &
35.734117 &
\cellcolor[rgb]{0.97,0.41,0.42}35.739900 &
\cellcolor[rgb]{0.97,0.41,0.42}-0.005784 \\ \hline
\end{tabular}%
}%
\end{table}
\Needspace{0.34\textheight}
\subsection{Woolyarn Tables }
The following tables report the complete MSE- and MAE-based Diebold--Mariano comparisons for the Woolyarn dataset.
\begin{table}[H]
\centering
\caption{Woolyarn Classical VS Quantum - DM Stats MSE}
\label{tab:woolyarn_dm_mse}
\small
\setlength{\tabcolsep}{3pt}
\resizebox{\textwidth}{!}{%
\begin{tabular}{|l|c|c|r|r|r|r|r|}
\hline
\rowcolor[RGB]{119,48,156}
\multicolumn{8}{|c|}{\textcolor{white}{\textbf{Woolyarn Classical VS Quantum \quad DM Stats MSE}}} \\ \hline
\textbf{Quantum pdq} &
\textbf{blocks\_used} &
\textbf{loss} &
\textbf{DM\_stat} &
\textbf{p\_value} &
\textbf{classical\_mean\_loss} &
\textbf{quantum\_mean\_loss} &
\textbf{delta\_mean\_loss} \\ \hline
(2, 1, 1) & 1 & MSE &
\cellcolor[rgb]{0.97,0.41,0.42}-7.968394 &
\cellcolor[rgb]{0.90,1.00,0.40}0.0000000000 &
934889.823 &
\cellcolor[rgb]{1.00,0.80,0.80}1136324 &
\cellcolor[rgb]{1.00,0.70,0.70}-201434.0761 \\ \hline
(1, 1, 1) & 1 & MSE &
\cellcolor[rgb]{0.97,0.41,0.42}-7.952058 &
\cellcolor[rgb]{0.90,1.00,0.40}0.0000000000 &
934889.823 &
\cellcolor[rgb]{1.00,0.78,0.78}1185477 &
\cellcolor[rgb]{1.00,0.66,0.66}-250586.7362 \\ \hline
(3, 1, 1) & 1 & MSE &
\cellcolor[rgb]{0.97,0.43,0.44}-7.245462 &
\cellcolor[rgb]{0.92,1.00,0.60}0.0000000000 &
934889.823 &
\cellcolor[rgb]{1.00,0.81,0.81}1131699 &
\cellcolor[rgb]{1.00,0.72,0.72}-196809.3958 \\ \hline
(4, 1, 1) & 1 & MSE &
\cellcolor[rgb]{0.99,0.51,0.50}-6.427643 &
\cellcolor[rgb]{0.95,1.00,0.75}0.0000000001 &
934889.823 &
\cellcolor[rgb]{0.98,0.90,0.90}1042081 &
\cellcolor[rgb]{1.00,0.80,0.80}-107190.8507 \\ \hline
(7, 1, 1) & 1 & MSE &
\cellcolor[rgb]{0.99,0.69,0.63}-4.500098 &
\cellcolor[rgb]{1.00,0.95,0.60}0.0000067922 &
934889.823 &
\cellcolor[rgb]{1.00,0.52,0.52}1343679 &
\cellcolor[rgb]{1.00,0.54,0.54}-408789.1725 \\ \hline
(10, 1, 1) & 1 & MSE &
\cellcolor[rgb]{0.97,0.73,0.68}-4.071137 &
\cellcolor[rgb]{1.00,0.97,0.70}0.0000467843 &
934889.823 &
\cellcolor[rgb]{0.98,0.95,0.95}1026709 &
\cellcolor[rgb]{1.00,0.83,0.83}-91818.85786 \\ \hline
(8, 1, 1) & 1 & MSE &
\cellcolor[rgb]{0.97,0.74,0.69}-4.023652 &
\cellcolor[rgb]{1.00,0.97,0.72}0.0000573025 &
934889.823 &
\cellcolor[rgb]{0.99,0.96,0.96}1024751 &
\cellcolor[rgb]{1.00,0.84,0.84}-89861.61058 \\ \hline
(9, 1, 1) & 1 & MSE &
\cellcolor[rgb]{0.98,0.77,0.72}-3.937144 &
\cellcolor[rgb]{1.00,0.98,0.75}0.0000824572 &
934889.823 &
\cellcolor[rgb]{1.00,0.97,0.97}1009798 &
\cellcolor[rgb]{1.00,0.87,0.87}-74908.25687 \\ \hline
(6, 1, 1) & 1 & MSE &
\cellcolor[rgb]{0.99,0.87,0.84}-2.992495 &
\cellcolor[rgb]{0.91,1.00,0.50}0.0027670760 &
934889.823 &
\cellcolor[rgb]{0.86,0.94,0.98}942247.1 &
\cellcolor[rgb]{0.93,0.98,0.90}-7357.235698 \\ \hline
(5, 1, 1) & 1 & MSE &
\cellcolor[rgb]{0.99,0.91,0.87}-2.548627 &
\cellcolor[rgb]{0.83,1.00,0.45}0.0108148000 &
934889.823 &
\cellcolor[rgb]{0.89,0.95,0.99}948237.4 &
\cellcolor[rgb]{0.91,0.98,0.88}-13347.5792 \\ \hline
\end{tabular}%
}%
\end{table}
\begin{table}[H]
\centering
\caption{Woolyarn Classical VS Quantum - DM Stats MAE}
\label{tab:woolyarn_dm_mae}
\small
\setlength{\tabcolsep}{3pt}
\resizebox{\textwidth}{!}{%
\begin{tabular}{|l|c|c|r|r|r|r|r|}
\hline
\rowcolor[RGB]{0,147,119}
\multicolumn{8}{|c|}{\textcolor{white}{\textbf{Woolyarn Classical VS Quantum \quad DM Stats MAE}}} \\ \hline
\textbf{Quantum pdq} &
\textbf{blocks\_used} &
\textbf{loss} &
\textbf{DM\_stat} &
\textbf{p\_value} &
\textbf{classical\_mean\_loss} &
\textbf{quantum\_mean\_loss} &
\textbf{delta\_mean\_loss} \\ \hline
(3, 1, 1)  & 1 & MAE &
\cellcolor[rgb]{0.97,0.41,0.42}-7.36933 &
\cellcolor[rgb]{0.90,1.00,0.40}0.0000000000 &
826.718879 &
\cellcolor[rgb]{0.98,0.90,0.90}904.906818 &
\cellcolor[rgb]{0.98,0.77,0.52}-78.187938 \\ \hline
(4, 1, 1)  & 1 & MAE &
\cellcolor[rgb]{0.97,0.43,0.43}-7.148529 &
\cellcolor[rgb]{0.90,1.00,0.40}0.0000000000 &
826.718879 &
\cellcolor[rgb]{0.99,0.95,0.95}865.442915 &
\cellcolor[rgb]{0.99,0.88,0.57}-38.724036 \\ \hline
(2, 1, 1)  & 1 & MAE &
\cellcolor[rgb]{0.98,0.47,0.46}-6.434141 &
\cellcolor[rgb]{0.93,1.00,0.65}0.0000000001 &
826.718879 &
\cellcolor[rgb]{0.98,0.91,0.91}902.821988 &
\cellcolor[rgb]{0.98,0.80,0.54}-76.103109 \\ \hline
(1, 1, 1)  & 1 & MAE &
\cellcolor[rgb]{0.99,0.56,0.52}-5.220583 &
\cellcolor[rgb]{0.95,1.00,0.75}0.0000001784 &
826.718879 &
\cellcolor[rgb]{1.00,0.86,0.86}926.307162 &
\cellcolor[rgb]{0.99,0.76,0.50}-99.588282 \\ \hline
(7, 1, 1)  & 1 & MAE &
\cellcolor[rgb]{0.99,0.67,0.61}-4.272585 &
\cellcolor[rgb]{1.00,0.95,0.60}0.0000193220 &
826.718879 &
\cellcolor[rgb]{1.00,0.74,0.74}994.530279 &
\cellcolor[rgb]{1.00,0.66,0.46}-167.811399 \\ \hline
(10, 1, 1) & 1 & MAE &
\cellcolor[rgb]{0.99,0.77,0.71}-3.251013 &
\cellcolor[rgb]{0.91,1.00,0.50}0.0011499450 &
826.718879 &
\cellcolor[rgb]{0.98,0.97,0.98}869.822437 &
\cellcolor[rgb]{0.99,0.91,0.63}-43.103558 \\ \hline
(8, 1, 1)  & 1 & MAE &
\cellcolor[rgb]{0.99,0.78,0.71}-3.194489 &
\cellcolor[rgb]{0.91,1.00,0.48}0.0014007860 &
826.718879 &
\cellcolor[rgb]{0.98,0.97,0.98}869.053863 &
\cellcolor[rgb]{0.99,0.92,0.64}-42.334983 \\ \hline
(9, 1, 1)  & 1 & MAE &
\cellcolor[rgb]{0.99,0.80,0.73}-3.039556 &
\cellcolor[rgb]{0.92,1.00,0.55}0.0023693500 &
826.718879 &
\cellcolor[rgb]{0.99,0.98,0.99}862.313859 &
\cellcolor[rgb]{0.99,0.94,0.68}-35.594980 \\ \hline
(6, 1, 1)  & 1 & MAE &
\cellcolor[rgb]{0.85,0.90,0.95}-1.778833 &
\cellcolor[rgb]{0.95,1.00,0.82}0.0752672000 &
826.718879 &
\cellcolor[rgb]{0.80,0.88,0.96}829.852377 &
\cellcolor[rgb]{0.86,0.97,0.83}-3.133498 \\ \hline
(5, 1, 1)  & 1 & MAE &
\cellcolor[rgb]{0.66,0.79,0.88}0.767760 &
\cellcolor[rgb]{1.00,1.00,0.80}0.4426297000 &
826.718879 &
\cellcolor[rgb]{0.68,0.83,0.93}823.707948 &
\cellcolor[rgb]{0.76,0.95,0.74}3.010931 \\ \hline
\end{tabular}%
}%
\end{table}
\Needspace{0.34\textheight}
\subsection{Mauna Loa CO2 Tables}
The following tables report the complete MSE- and MAE-based Diebold--Mariano comparisons for the Mauna Loa CO$_2$ dataset.

\begin{table}[H]
\centering
\caption{CO$_2$ Classical vs Quantum - DM Stats (MSE)}
\label{tab:co2_dm_mse}
\small
\setlength{\tabcolsep}{3pt}
\resizebox{\textwidth}{!}{%
\begin{tabular}{|l|l|l|l|l|l|l|l|}
\hline
\rowcolor[RGB]{119,48,156}
\multicolumn{8}{|l|}{\textcolor{white}{\textbf{CO$_2$ Classical vs Quantum - DM Stats (MSE)}}} \\ \hline
\textbf{Quantum (pdq)} & \textbf{blocks\_used} & \textbf{loss} & \textbf{DM\_stat} & \textbf{p\_value} & \textbf{classical\_mean\_loss} & \textbf{quantum\_mean\_loss} & \textbf{delta\_mean\_loss} \\ \hline
(8, 1, 1)  & 174 & MSE &
{\cellcolor[rgb]{0.392,0.569,0.792}}8.681193 &
{\cellcolor[rgb]{0.388,0.745,0.482}}0.00E+00 &
9.080382 &
{\cellcolor[rgb]{0.408,0.576,0.792}}3.599858 &
{\cellcolor[rgb]{0.443,0.761,0.486}}5.480524 \\
\hline
(9, 1, 1)  & 172 & MSE &
{\cellcolor[rgb]{0.369,0.553,0.784}}8.991140 &
{\cellcolor[rgb]{0.388,0.745,0.482}}0.00E+00 &
9.099370 &
{\cellcolor[rgb]{0.365,0.549,0.780}}3.282792 &
{\cellcolor[rgb]{0.400,0.749,0.486}}5.816578 \\
\hline
(10, 1, 1) & 174 & MSE &
{\cellcolor[rgb]{0.353,0.541,0.776}}9.194297 &
{\cellcolor[rgb]{0.388,0.745,0.482}}0.00E+00 &
9.080382 &
{\cellcolor[rgb]{0.353,0.541,0.776}}3.192490 &
{\cellcolor[rgb]{0.388,0.745,0.482}}5.887893 \\
\hline
(5, 1, 1)  & 174 & MSE &
{\cellcolor[rgb]{0.973,0.412,0.420}}-6.729980 &
{\cellcolor[rgb]{0.882,0.886,0.510}}1.70E-11 &
9.080382 &
{\cellcolor[rgb]{0.976,0.420,0.427}}9.236257 &
{\cellcolor[rgb]{0.973,0.416,0.420}}-0.155875 \\
\hline
(6, 1, 1)  & 174 & MSE &
{\cellcolor[rgb]{0.973,0.416,0.424}}-6.673577 &
{\cellcolor[rgb]{1.000,0.922,0.518}}2.50E-11 &
9.080382 &
{\cellcolor[rgb]{0.973,0.412,0.420}}9.247513 &
{\cellcolor[rgb]{0.973,0.412,0.420}}-0.167131 \\
\hline
(4, 1, 1)  & 174 & MSE &
{\cellcolor[rgb]{0.973,0.478,0.486}}-5.947968 &
{\cellcolor[rgb]{1.000,0.914,0.518}}2.71E-09 &
9.080382 &
{\cellcolor[rgb]{0.976,0.427,0.435}}9.217262 &
{\cellcolor[rgb]{0.973,0.420,0.420}}-0.136880 \\
\hline
(3, 1, 1)  & 174 & MSE &
{\cellcolor[rgb]{0.973,0.533,0.541}}-5.288334 &
{\cellcolor[rgb]{0.980,0.553,0.447}}1.23E-07 &
9.080382 &
{\cellcolor[rgb]{0.976,0.420,0.427}}9.234463 &
{\cellcolor[rgb]{0.973,0.416,0.420}}-0.154080 \\
\hline
(7, 1, 1)  & 174 & MSE &
{\cellcolor[rgb]{0.627,0.733,0.875}}5.230577 &
{\cellcolor[rgb]{0.973,0.412,0.420}}1.69E-07 &
9.080382 &
{\cellcolor[rgb]{0.808,0.859,0.933}}6.563634 &
{\cellcolor[rgb]{0.827,0.875,0.510}}2.516748 \\
\hline
\end{tabular}%
}%
\end{table}
\begin{table}[H]
\centering
\caption{CO$_2$ Classical VS Quantum - DM Stats MAE}
\label{tab:co2_dm_mae}
\small
\setlength{\tabcolsep}{3pt}
\resizebox{\textwidth}{!}{%
\begin{tabular}{|l|l|l|l|l|l|l|l|}
\hline
\rowcolor[RGB]{0,147,119}
\multicolumn{8}{|c|}{\textcolor{white}{\textbf{CO$_2$ Classical VS Quantum \quad DM Stats MAE}}} \\ \hline
\textbf{Quantum pdq} & \textbf{blocks\_used} & \textbf{loss} & \textbf{DM\_stat} & \textbf{p\_value} & \textbf{classical\_mean\_loss} & \textbf{quantum\_mean\_loss} & \textbf{delta\_mean\_loss} \\ \hline
(8, 1, 1)  & 174 & MAE &
\cellcolor[rgb]{0.35,0.54,0.78}9.252065 &
\cellcolor[rgb]{0.39,0.75,0.48}0.00E+00 &
2.315727 &
\cellcolor[rgb]{0.69,0.82,0.91}1.406047 &
\cellcolor[rgb]{0.71,0.91,0.51}0.909681 \\ \hline
(9, 1, 1)  & 172 & MAE &
\cellcolor[rgb]{0.33,0.52,0.78}9.561353 &
\cellcolor[rgb]{0.39,0.75,0.48}0.00E+00 &
2.314338 &
\cellcolor[rgb]{0.63,0.80,0.90}1.348458 &
\cellcolor[rgb]{0.70,0.91,0.52}0.965881 \\ \hline
(10, 1, 1) & 174 & MAE &
\cellcolor[rgb]{0.31,0.51,0.77}9.857719 &
\cellcolor[rgb]{0.39,0.75,0.48}0.00E+00 &
2.315727 &
\cellcolor[rgb]{0.60,0.78,0.89}1.324859 &
\cellcolor[rgb]{0.69,0.91,0.52}0.990868 \\ \hline
(3, 1, 1)  & 174 & MAE &
\cellcolor[rgb]{0.97,0.41,0.42}-5.639472 &
\cellcolor[rgb]{0.97,0.41,0.42}1.71E-08 &
2.315727 &
\cellcolor[rgb]{0.97,0.41,0.42}2.338754 &
\cellcolor[rgb]{0.97,0.41,0.42}-0.023027 \\ \hline
(7, 1, 1)  & 174 & MAE &
\cellcolor[rgb]{0.55,0.74,0.85}5.486284 &
\cellcolor[rgb]{0.97,0.87,0.41}4.10E-08 &
2.315727 &
\cellcolor[rgb]{0.82,0.88,0.94}1.887741 &
\cellcolor[rgb]{0.82,0.95,0.60}0.427986 \\ \hline
(4, 1, 1)  & 174 & MAE &
\cellcolor[rgb]{0.96,0.54,0.54}-5.417399 &
\cellcolor[rgb]{0.99,0.83,0.47}6.05E-08 &
2.315727 &
\cellcolor[rgb]{0.96,0.52,0.53}2.332440 &
\cellcolor[rgb]{0.97,0.52,0.46}-0.016713 \\ \hline
(5, 1, 1)  & 174 & MAE &
\cellcolor[rgb]{0.96,0.58,0.58}-4.536355 &
\cellcolor[rgb]{1.00,0.90,0.55}5.72E-06 &
2.315727 &
\cellcolor[rgb]{0.96,0.55,0.56}2.328392 &
\cellcolor[rgb]{0.96,0.53,0.46}-0.012664 \\ \hline
(6, 1, 1)  & 174 & MAE &
\cellcolor[rgb]{0.96,0.60,0.60}-4.474334 &
\cellcolor[rgb]{1.00,0.91,0.55}7.66E-06 &
2.315727 &
\cellcolor[rgb]{0.96,0.56,0.57}2.329375 &
\cellcolor[rgb]{0.96,0.54,0.47}-0.013647 \\ \hline
\end{tabular}%
}%
\end{table}
\Needspace{0.34\textheight}
\subsection{Australian Beer Production Tables}
The following tables report the complete MSE- and MAE-based Diebold--Mariano comparisons for the Australian beer production dataset.

\begin{table}[H]
\centering
\caption{AusBeer Classical VS Quantum - DM Stats (MSE)}
\label{tab:ausbeer_dm_mse}
\small
\setlength{\tabcolsep}{3pt}
\resizebox{\textwidth}{!}{%
\begin{tabular}{|l|c|c|r|r|r|r|r|}
\hline
\rowcolor[RGB]{97,51,135}
\multicolumn{8}{|c|}{\textcolor{white}{\textbf{AusBeer Classical VS Quantum \quad DM Stats MSE}}} \\ \hline
\textbf{Quantum pdq} &
\textbf{blocks\_used} &
\textbf{loss} &
\textbf{DM\_stat} &
\textbf{p\_value} &
\textbf{classical\_mean\_loss} &
\textbf{quantum\_mean\_loss} &
\textbf{delta\_mean\_loss} \\ \hline
(1, 1, 6)  & 84 & MSE &
\cellcolor[rgb]{0.98,0.60,0.61}11.878028 &
\cellcolor[rgb]{0.39,0.75,0.48}0.00E+00 &
2559.328583 &
\cellcolor[rgb]{0.98,0.80,0.90}1296.194617 &
\cellcolor[rgb]{0.93,0.98,0.90}1263.133966 \\ \hline
(2, 1, 6)  & 84 & MSE &
\cellcolor[rgb]{0.97,0.49,0.50}15.353045 &
\cellcolor[rgb]{0.39,0.75,0.48}0.00E+00 &
2559.328583 &
\cellcolor[rgb]{0.74,0.88,0.95}564.606641 &
\cellcolor[rgb]{0.78,0.93,0.77}1994.721942 \\ \hline
(7, 1, 1)  & 84 & MSE &
\cellcolor[rgb]{0.97,0.47,0.48}15.626542 &
\cellcolor[rgb]{0.39,0.75,0.48}0.00E+00 &
2559.328583 &
\cellcolor[rgb]{0.55,0.76,0.89}388.120011 &
\cellcolor[rgb]{0.73,0.93,0.75}2171.208572 \\ \hline
(6, 1, 1)  & 84 & MSE &
\cellcolor[rgb]{0.97,0.47,0.48}15.626973 &
\cellcolor[rgb]{0.39,0.75,0.48}0.00E+00 &
2559.328583 &
\cellcolor[rgb]{0.64,0.81,0.90}412.991120 &
\cellcolor[rgb]{0.75,0.93,0.75}2146.337463 \\ \hline
(5, 1, 1)  & 84 & MSE &
\cellcolor[rgb]{0.97,0.46,0.48}15.708202 &
\cellcolor[rgb]{0.39,0.75,0.48}0.00E+00 &
2559.328583 &
\cellcolor[rgb]{0.63,0.81,0.90}408.258903 &
\cellcolor[rgb]{0.74,0.93,0.74}2151.069680 \\ \hline
(10, 1, 1) & 84 & MSE &
\cellcolor[rgb]{0.97,0.46,0.48}15.730669 &
\cellcolor[rgb]{0.39,0.75,0.48}0.00E+00 &
2559.328583 &
\cellcolor[rgb]{0.58,0.78,0.89}381.998708 &
\cellcolor[rgb]{0.72,0.93,0.74}2177.329875 \\ \hline
(9, 1, 1)  & 84 & MSE &
\cellcolor[rgb]{0.97,0.45,0.47}15.775209 &
\cellcolor[rgb]{0.39,0.75,0.48}0.00E+00 &
2559.328583 &
\cellcolor[rgb]{0.57,0.78,0.89}378.526593 &
\cellcolor[rgb]{0.71,0.93,0.74}2180.801990 \\ \hline
(3, 1, 3)  & 84 & MSE &
\cellcolor[rgb]{0.96,0.44,0.47}16.062654 &
\cellcolor[rgb]{0.39,0.75,0.48}0.00E+00 &
2559.328583 &
\cellcolor[rgb]{0.63,0.81,0.90}410.665446 &
\cellcolor[rgb]{0.74,0.93,0.75}2148.663137 \\ \hline
\end{tabular}%
}%
\end{table}
\begin{table}[H]
\centering
\caption{AusBeer Classical VS Quantum - DM Stats (MAE)}
\label{tab:ausbeer_dm_mae}
\small
\setlength{\tabcolsep}{3pt}
\resizebox{\textwidth}{!}{%
\begin{tabular}{|l|c|c|r|r|r|r|r|}
\hline
\rowcolor[RGB]{0,147,119}
\multicolumn{8}{|c|}{\textcolor{white}{\textbf{AusBeer Classical VS Quantum \quad DM Stats MAE}}} \\ \hline
\textbf{Quantum pdq} &
\textbf{blocks\_used} &
\textbf{loss} &
\textbf{DM\_stat} &
\textbf{p\_value} &
\textbf{classical\_mean\_loss} &
\textbf{quantum\_mean\_loss} &
\textbf{delta\_mean\_loss} \\ \hline
(1, 1, 6)  & 84 & MAE &
\cellcolor[rgb]{0.97,0.55,0.55}11.772301 &
\cellcolor[rgb]{0.39,0.75,0.48}0.00E+00 &
41.666725 &
\cellcolor[rgb]{0.99,0.87,0.92}29.306057 &
\cellcolor[rgb]{0.98,0.80,0.52}12.360668 \\ \hline
(2, 1, 6)  & 84 & MAE &
\cellcolor[rgb]{0.97,0.45,0.47}18.632163 &
\cellcolor[rgb]{0.39,0.75,0.48}0.00E+00 &
41.666725 &
\cellcolor[rgb]{0.74,0.88,0.95}18.651992 &
\cellcolor[rgb]{0.93,0.98,0.90}23.014733 \\ \hline
(3, 1, 3)  & 84 & MAE &
\cellcolor[rgb]{0.97,0.44,0.47}19.375578 &
\cellcolor[rgb]{0.39,0.75,0.48}0.00E+00 &
41.666725 &
\cellcolor[rgb]{0.62,0.80,0.91}16.011428 &
\cellcolor[rgb]{0.88,0.96,0.83}25.655297 \\ \hline
(6, 1, 1)  & 84 & MAE &
\cellcolor[rgb]{0.97,0.43,0.47}19.859260 &
\cellcolor[rgb]{0.39,0.75,0.48}0.00E+00 &
41.666725 &
\cellcolor[rgb]{0.62,0.81,0.91}16.104594 &
\cellcolor[rgb]{0.89,0.96,0.84}25.562131 \\ \hline
(7, 1, 1)  & 84 & MAE &
\cellcolor[rgb]{0.97,0.43,0.47}20.302789 &
\cellcolor[rgb]{0.39,0.75,0.48}0.00E+00 &
41.666725 &
\cellcolor[rgb]{0.59,0.79,0.90}15.588372 &
\cellcolor[rgb]{0.86,0.95,0.81}26.078353 \\ \hline
(5, 1, 1)  & 84 & MAE &
\cellcolor[rgb]{0.97,0.43,0.47}20.327477 &
\cellcolor[rgb]{0.39,0.75,0.48}0.00E+00 &
41.666725 &
\cellcolor[rgb]{0.62,0.81,0.91}16.109335 &
\cellcolor[rgb]{0.89,0.96,0.84}25.557390 \\ \hline
(9, 1, 1)  & 84 & MAE &
\cellcolor[rgb]{0.97,0.42,0.47}20.668863 &
\cellcolor[rgb]{0.39,0.75,0.48}0.00E+00 &
41.666725 &
\cellcolor[rgb]{0.55,0.76,0.89}15.037503 &
\cellcolor[rgb]{0.83,0.93,0.75}26.629222 \\ \hline
(10, 1, 1) & 84 & MAE &
\cellcolor[rgb]{0.97,0.42,0.47}20.854476 &
\cellcolor[rgb]{0.39,0.75,0.48}0.00E+00 &
41.666725 &
\cellcolor[rgb]{0.57,0.77,0.89}15.106466 &
\cellcolor[rgb]{0.84,0.94,0.77}26.560259 \\ \hline
\end{tabular}%
}%
\end{table}
\Needspace{0.34\textheight}
\subsection{Sydney weather 2024 tables}
The following tables report the complete MSE- and MAE-based Diebold--Mariano comparisons for the Sydney summer temperature dataset.

\begin{table}[H]
\centering
\caption{Sydney 2024 Summer Temp - Classical vs.\ Quantum \quad DM Stats (MSE)}
\label{tab:syd_summer_dm_mse}
\small
\setlength{\tabcolsep}{3pt}
\resizebox{\textwidth}{!}{%
\begin{tabular}{|l|c|c|r|r|r|r|r|}
\hline
\rowcolor[RGB]{97,51,135}
\multicolumn{8}{|c|}{\textcolor{white}{\textbf{Sydney 2024 Summer Temp Classical VS Quantum \quad DM Stats MSE}}} \\ \hline
\textbf{Quantum pdq} &
\textbf{blocks\_used} &
\textbf{loss} &
\textbf{DM\_stat} &
\textbf{p\_value} &
\textbf{classical\_mean\_loss} &
\textbf{quantum\_mean\_loss} &
\textbf{delta\_mean\_loss} \\ \hline

(4, 1, 1)  & 160 & MSE &
\cellcolor[HTML]{C6EFCE}-3.41981 &
\cellcolor[HTML]{C6EFCE}6.27E-04 &
8.1677 &
9.023377 &
\cellcolor[HTML]{FFE699}-0.855678 \\ \hline

(3, 1, 1)  & 161 & MSE &
\cellcolor[HTML]{D9EAD3}-2.943498 &
\cellcolor[HTML]{D9EAD3}3.25E-03 &
8.177383 &
\cellcolor[HTML]{F4CCCC}19.809067 &
\cellcolor[HTML]{EA9999}-11.631684 \\ \hline

(5, 1, 1)  & 161 & MSE &
\cellcolor[HTML]{E2F0D9}-2.456801 &
\cellcolor[HTML]{E2F0D9}1.40E-02 &
8.177383 &
8.713587 &
\cellcolor[HTML]{FFE699}-0.536204 \\ \hline

(6, 1, 1)  & 161 & MSE &
\cellcolor[HTML]{F2F2F2}-1.544826 &
\cellcolor[HTML]{F2F2F2}1.22E-01 &
8.177383 &
8.668534 &
\cellcolor[HTML]{FFE699}-0.491151 \\ \hline

(10, 1, 1) & 155 & MSE &
\cellcolor[HTML]{F4CCCC}0.610649 &
\cellcolor[HTML]{FCE4D6}5.41E-01 &
8.230546 &
\cellcolor[HTML]{C6EFCE}7.906464 &
\cellcolor[HTML]{C6EFCE}0.324082 \\ \hline

(7, 1, 1)  & 161 & MSE &
\cellcolor[HTML]{FCE4D6}-0.577497 &
\cellcolor[HTML]{FCE4D6}5.64E-01 &
8.177383 &
8.416235 &
\cellcolor[HTML]{FFE699}-0.238851 \\ \hline

(9, 1, 1)  & 159 & MSE &
\cellcolor[HTML]{F4CCCC}0.555303 &
\cellcolor[HTML]{FCE4D6}5.79E-01 &
8.258326 &
7.989777 &
\cellcolor[HTML]{C6EFCE}0.268549 \\ \hline

(8, 1, 1)  & 161 & MSE &
\cellcolor[HTML]{F2F2F2}0.175505 &
\cellcolor[HTML]{F2F2F2}8.61E-01 &
8.177383 &
8.099677 &
\cellcolor[HTML]{C6EFCE}0.077707 \\ \hline

\end{tabular}%
}%
\end{table}
\begin{table}[H]
\centering
\caption{Sydney 2024 Summer Classical VS Quantum \quad DM Stats (MAE)}
\label{tab:syd_summer_dm_mae}
\small
\setlength{\tabcolsep}{3pt}
\resizebox{\textwidth}{!}{%
\begin{tabular}{|l|c|c|r|r|r|r|r|}
\hline
\rowcolor[RGB]{0,153,0}
\multicolumn{8}{|c|}{\textcolor{white}{\textbf{Sydney 2024 Summer Classical VS Quantum \quad DM Stats MAE}}} \\ \hline
\textbf{Quantum pdq} &
\textbf{blocks\_used} &
\textbf{loss} &
\textbf{DM\_stat} &
\textbf{p\_value} &
\textbf{classical\_mean\_loss} &
\textbf{quantum\_mean\_loss} &
\textbf{delta\_mean\_loss} \\ \hline

(3, 1, 1)  & 161 & MAE &
\cellcolor[HTML]{C6EFCE}-4.038151 &
\cellcolor[HTML]{C6EFCE}5.40E-05 &
2.183719 &
\cellcolor[HTML]{BDD7EE}2.875055 &
\cellcolor[HTML]{FFE699}-0.691336 \\ \hline

(4, 1, 1)  & 160 & MAE &
\cellcolor[HTML]{D9EAD3}-2.975823 &
\cellcolor[HTML]{D9EAD3}2.92E-03 &
2.180798 &
2.280019 &
\cellcolor[HTML]{FFE699}-0.099221 \\ \hline

(5, 1, 1)  & 161 & MAE &
\cellcolor[HTML]{E2F0D9}-1.775833 &
\cellcolor[HTML]{E2F0D9}7.58E-02 &
2.183719 &
2.245571 &
\cellcolor[HTML]{FFE699}-0.061852 \\ \hline

(10, 1, 1) & 155 & MAE &
\cellcolor[HTML]{F4CCCC}1.205682 &
\cellcolor[HTML]{FCE4D6}2.28E-01 &
2.193426 &
\cellcolor[HTML]{C6EFCE}2.106318 &
\cellcolor[HTML]{C6EFCE}0.087108 \\ \hline

(6, 1, 1)  & 161 & MAE &
\cellcolor[HTML]{F2F2F2}-1.158140 &
\cellcolor[HTML]{F2F2F2}2.47E-01 &
2.183719 &
2.237213 &
\cellcolor[HTML]{FFE699}-0.053494 \\ \hline

(9, 1, 1)  & 159 & MAE &
\cellcolor[HTML]{F4CCCC}1.088950 &
\cellcolor[HTML]{FCE4D6}2.76E-01 &
2.198211 &
2.124441 &
\cellcolor[HTML]{C6EFCE}0.073770 \\ \hline

(8, 1, 1)  & 161 & MAE &
\cellcolor[HTML]{F2F2F2}0.601969 &
\cellcolor[HTML]{F2F2F2}5.47E-01 &
2.183719 &
2.145549 &
\cellcolor[HTML]{C6EFCE}0.038169 \\ \hline

(7, 1, 1)  & 161 & MAE &
\cellcolor[HTML]{F2F2F2}-0.164042 &
\cellcolor[HTML]{F2F2F2}8.70E-01 &
2.183719 &
2.193769 &
\cellcolor[HTML]{FFE699}-0.010050 \\ \hline

\end{tabular}%
}%
\end{table}
\endgroup
\FloatBarrier
\section*{Acknowledgment}
The authors express gratitude to the IBM Quantum Experience platform and its team for creating the Qiskit platform and granting free access to their simulators for executing quantum circuits and conducting the experiments detailed in this study. The authors also acknowledge the Centre for Quantum Software and Information (CQSI) for supporting quantum computing research.

\FloatBarrier
\section*{Competing Interests}
The authors have no financial or non-financial competing interests to declare.

\FloatBarrier
\section*{Author Contributions}
Study conception and design: N.M., B.K.B., B.M., and P.D.; data collection: N.M.; analysis and interpretation of results: N.M., B.K.B., B.M., and P.D.; draft manuscript preparation: N.M., B.K.B., and B.M. All authors reviewed the results and approved the final manuscript.

\FloatBarrier
\section*{Data Availability}
All datasets used in this study are publicly available benchmark time-series datasets obtained from standard forecasting repositories and publicly accessible sources.

\bibliographystyle{IEEEtran}
\bibliography{main}

\end{document}